\documentclass[notitlepage,nofootinbib,superscriptaddress,10pt,aps,prd,twocolumn]{revtex4-1}
\pdfoutput=1
\usepackage[utf8]{inputenc}

\usepackage{amssymb}
\usepackage{physics}
\usepackage{graphicx}
\usepackage{color}
\usepackage{hyperref}
\usepackage{tikz}
\usepackage{slashed}
\usetikzlibrary{decorations.pathmorphing}
\usetikzlibrary{decorations.markings}
\usetikzlibrary{shapes.misc}
\usetikzlibrary{positioning, shapes, arrows}
\usetikzlibrary{backgrounds}
\usepackage{bbm}

\usepackage{enumitem}

\newcommand{\be}{\begin{equation}}
\newcommand{\ee}{\end{equation}}
\newcommand{\bea}{\begin{eqnarray}}
\newcommand{\eea}{\end{eqnarray}}

\tikzset{
scalar/.style={
  dashed,
  },
scalarV/.style={
  dashed,
  },
scalarD/.style={
  dash dot,
  },
fermion/.style={
  postaction={decorate},
  decoration={markings,mark=at position .5 with {\arrow[#1]{latex}}}
  },
graviton/.style={
 decorate,
 decoration={coil,amplitude=2pt, segment length=2pt,  pre length=.1cm, post length=.1cm}
 },
regulator/.style={
  draw,
  circle,
  preaction={fill, white},
  minimum width=0.2cm,
  path picture={ 
   \draw[black] (path picture bounding box.south east) -- (path picture bounding box.north west) (path picture bounding box.south west) -- (path picture bounding box.north east);
   }
  }
}

\begin{document}
 \title{
Constraining power of asymptotic safety for scalar fields
 }
 \author{Astrid Eichhorn}
 \email{eichhorn@cp3.sdu.dk}
 \affiliation{CP3-Origins,  University  of  Southern  Denmark,  Campusvej  55,  DK-5230  Odense  M,  Denmark}
 \author{Martin Pauly}
 \email{m.pauly@thphys.uni-heidelberg.de}
 \affiliation{Institut f\"ur Theoretische Physik, Ruprecht-Karls-Universit\"at Heidelberg, \\
Philosophenweg 16, 69120 Heidelberg, Germany}

\begin{abstract}
 We explore the phenomenology of scalar fields coupled to asymptotically safe quantum gravity, in light of their potential significance for dark matter, for the inflaton as well as dynamical dark energy, and in the Higgs sector in and beyond the Standard Model. This work is a step towards delineating the boundaries of the asymptotically safe swampland by exploiting the constraining power of the asymptotic-safety paradigm. Firstly, we strengthen indications that quantum gravitational fluctuations could drive scalar potentials towards flatness, with intriguing potential implications for inflation and dark energy. Secondly, we explore how asymptotic safety could rule out large parts of the parameter space in models for scalar dark matter. Thirdly, we discover hints that at an asymptotically safe fixed point with finite top quark mass, the non-minimal Higgs-curvature coupling could be constrained.
  Finally, by combining the constraining power of asymptotic safety in particle physics and cosmology, we find hints that Higgs inflation lies in the asymptotically safe swampland.\\
 In summary, we strengthen previous indications for the constraining power of asymptotic safety and the resulting large extent of the asymptotically safe swampland.
\end{abstract}

\maketitle
\tableofcontents

\section{Motivation: Scalars in particle physics and cosmology: in search of a UV completion}
 The experimental discovery of the Higgs particle \cite{Aad:2012tfa,Chatrchyan:2012ufa} has shown that at least one fundamental scalar field might exist in nature. Its production and subsequent decay into various Standard-Model (SM) particles is consistent with the expected values in the Standard Model \cite{Zyla:2020zbs}. Yet, this impressive success exposes one of the open problems of the Standard Model, namely a transplanckian Landau pole that is expected in the Higgs-Yukawa sector. It is expected to limit the validity of the theory and require new physics. Given the scale at which  it appears, quantum gravity constitutes a prime candidate for new physics to remedy this Landau pole\footnote{This is the reason why the measurement of 125 GeV for the mass of the Higgs particle is a far-reaching result of the LHC -- at a not too different mass, vacuum instability would imply a much lower scale of new physics \cite{Degrassi:2012ry,Buttazzo:2013uya,Andreassen:2017rzq,Bezrukov:2014ina}, as would low-scale Landau poles if the Higgs particle was somewhat heavier \cite{Cabibbo:1979ay,Lindner:1985uk}.}. Within the asymptotic-safety approach to quantum gravity, the Higgs field could become a truly fundamental field, as there are indications for an asymptotically safe fixed point in the Higgs sector of the Standard Model under the impact of quantum-gravity fluctuations, see, e.g., \cite{Shaposhnikov:2009pv, Wetterich:2016uxm, Hamada:2017rvn, Eichhorn:2017ylw, Eichhorn:2018whv,Pawlowski:2018ixd, Wetterich:2019zdo, deBrito:2019umw} and \cite{Eichhorn:2017als, Kwapisz:2019wrl,Reichert:2019car,Eichhorn:2019dhg,Domenech:2020yjf,Hamada:2020vnf} for beyond-SM settings. For earlier work on gravity-scalar systems, see, e.g., \cite{Narain:2009fy,Narain:2009gb,Zanusso:2009bs,Vacca:2010mj,Eichhorn:2012va,Percacci:2015wwa,Labus:2015ska}. Asymptotic safety generalizes the success story of asymptotic freedom from non-Abelian gauge theories to a quantum gravitational setting. It corresponds to a quantum-scale-invariant fixed-point regime of the Renormalization Group (RG) flow. This enables an ultraviolet (UV) completion or extension of an effective field theory such as the Standard Model. 
 Requiring the enhancement of the symmetry to quantum scale symmetry in the UV implies relations between the couplings of the theory that hold in the infrared (IR). This results in a potential enhancement of predictive power and could open the door to observational tests of the asymptotic-safety paradigm. 
 \\
  Beyond the Higgs sector, fundamental scalar fields could  play a variety of roles in cosmology. In the late universe, they could  constitute a part or all of the dark matter \cite{Silveira:1985rk,McDonald:1993ex,Burgess:2000yq,Boehm:2003hm}. 
 The dark-matter sector might become asymptotically safe under the impact of quantum gravitational fluctuations. There are tentative hints that this could entail an enhancement of predictive power \cite{Eichhorn:2017als}, see also \cite{Reichert:2019car,Hamada:2020vnf,Eichhorn:2020kca}, strongly restricting the parameter space of the dark matter.\\
 Further, scalars could be relevant to drive the early- and/or late-time expansion of the universe. The potential for such scalars is also expected to be subject to constraints if coupled to asymptotically safe quantum gravity, see, e.g.,  \cite{Narain:2009fy,Henz:2013oxa,Henz:2016aoh,Rubio:2017gty,Wetterich:2019rsn}.  At the same time, observational data already constrains the form of the potential for a dark-energy scalar \cite{Abbott:2017wau,Scolnic:2017caz,Aghanim:2018eyx}.  This raises the question whether cosmological observations, such as, e.g., upcoming surveys aimed at restricting the dark-energy equation of state \cite{Abell:2009aa,Amendola:2012ys,Aghamousa:2016zmz,Bacon:2018dui,Bull:2018lat}, could provide insights into the viability of the asymptotic-safety paradigm. Similarly, Planck constraints on inflationary potentials \cite{Aghanim:2018eyx,Akrami:2018odb} might enable observational tests of the paradigm.
\\

 Taking these manifold physics applications of fundamental scalars within the Standard Model, as dark matter, dark energy and as the inflaton as our motivation, we explore the interplay of asymptotically safe gravity with scalars and Yukawa systems, building on and extending previous work in \cite{Shaposhnikov:2009pv,Narain:2009fy,Narain:2009gb,Zanusso:2009bs,Vacca:2010mj,Eichhorn:2012va,Henz:2013oxa,Percacci:2015wwa,Labus:2015ska,Henz:2016aoh,Wetterich:2016uxm,Rubio:2017gty,Hamada:2017rvn,Eichhorn:2017ylw,Eichhorn:2017als, Eichhorn:2018whv,Pawlowski:2018ixd,Wetterich:2019rsn,deBrito:2019umw,Kwapisz:2019wrl,Reichert:2019car,Wetterich:2019zdo, Eichhorn:2019dhg,Hamada:2020vnf,Eichhorn:2020kca}. \\

 This paper is structured as follows: In Sec.~\ref{sec:setup} we introduce the method and setup we use.  In Sec.~\ref{sec:single_scalar} we study how quantum gravity impacts the potential of a single scalar and discuss potential implications for inflation and dark energy. In Sec.~\ref{sec:yukawa_model} we extend the purely scalar model to a Yukawa system representing the Higgs-top sector of the Standard Model. We study the impact of the non-minimal coupling on potential bounds on the top mass \cite{Eichhorn:2017ylw} and explore the viability of asymptotically safe Higgs inflation. Sec.~\ref{sec:portal} discusses two Yukawa systems coupled via a portal coupling, and explores how one of these two sectors could be a dark sector \cite{Eichhorn:2020kca}, accounting for dark matter.\\
 We refer \emph{readers interested in potential cosmological implications} to Sec.~\ref{sec:slow_roll} on slow-roll-like parameters in asymptotic safety; Sec.~\ref{sec:phenosinglescalar} on asymptotically safe inflation and dark energy, Sec.~\ref{sec:noHiggsinflationinAS} on the viability of Higgs inflation in asymptotic safety as well as Sec.~\ref{sec:portal} on potential dark-matter candidates.\\
Additionally, we refer \emph{readers interested in potential particle-physics implications}, to Sec.~\ref{subsubsec:globalsyms} for a discussion of global symmetries under the impact of asymptotically safe gravity and Sec.~\ref{sec:topmass} on the impact of the non-minimal coupling on the bound on the top mass \cite{Eichhorn:2017ylw}.\\
Finally, we highlight that \emph{readers interested in formal aspects of functional Renormalization Group setups} 
will find a study of auxiliary background-scalar field dependencies in App.~\ref{sec:methods}.

 \section{Asymptotic safety: conceptual and technical framework}
 \label{sec:setup}
  \emph{...where we first introduce some concepts related to an asymptotically safe fixed point and then elucidate the method we use to explore the scale-dependence of quantum field theories. Finally, we provide specifics (field content and interaction structure) of the systems we explore.}
  \subsection{Predictivity and effective asymptotic safety}\label{sec:effAS}
\emph{...where we review how an interacting fixed point imposes predictivity on a quantum field theory. Further, we revisit the concept of effective asymptotic safety and how it can result in predictivity in an effective-field-theory setup with a finite new-physics scale.}\\

In a quantum field theory (QFT), scale-dependent couplings encode the effect of quantum fluctuations. By integrating out quantum fluctuations with momenta larger than a momentum scale $k$, one obtains the scale dependence of a dimensionless coupling $g_i(k)$, encoded in its beta function $\beta_{g_i}$.
At a Renormalization Group (RG) fixed point, all beta functions of a system vanish, realizing quantum scale symmetry. If RG trajectories are picked for which the UV regime is determined by quantum scale symmetry, one refers to the resulting model as asymptotically safe, and the fixed point itself as an asymptotically safe fixed point.
Several dynamical mechanisms can be at the heart of such a fixed point, see \cite{Eichhorn:2018yfc} for a review.
As a consequence, one can consider the mathematical limit in which the RG scale is taken to arbitrarily high values, and obtain a path integral that is well-defined in this limit, i.e., a fundamental theory. Firstly, such a limit is physically challenging to access; and secondly, it does not imply that new physics might not exist at some high scale. In this paper, our main focus will therefore not be on the UV completeness that follows as a consequence of asymptotic safety. Instead, we will focus on the predictivity that it implies -- even in settings where new physics does indeed exist at high scales.

The predictivity of a fixed point is encoded in its critical exponents. The critical exponents are defined as
\be
\theta^{(I)} = -{\rm eig} \left(\frac{\partial \beta_{g_i}}{\partial g_{j}} \right)\Big|_{\vec{g} = \vec{g}_{\ast}}.
\ee
They govern the linearized flow around the fixed point: Directions in the space of couplings associated to a positive $\theta$ are relevant, i.e., IR repulsive. A tiny deviation from the fixed-point value continues to grow towards the IR, i.e., quantum fluctuations associated to such a direction drive the system away from scale symmetry. The IR values of the corresponding couplings are not calculable from the fixed-point requirement; a range of IR values is compatible with asymptotic safety in the UV. \\
Directions associated with negative $\theta$ are irrelevant, i.e., IR attractive. Deviations from the fixed point decrease under the RG flow to the IR, i.e., quantum fluctuations associated to such a direction drive the system towards scale symmetry. Therefore, the IR values of such couplings are calculable from the fixed-point requirement. Essentially, for each $\theta<0$ there is a relation between the values of couplings, which also has to hold in the IR for quantum scale symmetry to be realized in the UV.

Asymptotic safety is typically introduced as an ultraviolet completion for a quantum field theory (QFT), with its main interest being the extension of the regime of validity of the theory up to arbitrarily high energy scales.
 A more pragmatic view on asymptotic safety originates in  the concept of ``effective asymptotic safety", discussed in \cite{deAlwis:2019aud,Held:2020kze}, see also \cite{Percacci:2010af}: even if the QFT description of matter-gravity systems breaks down at a transplanckian momentum scale $\Lambda_{\rm UV}> M_{\rm Pl}$, the predictivity of an asymptotically safe fixed point can continue to play a major role in determining the IR physics. In this case, the RG flow starts at initial conditions at $\Lambda_{\rm UV}$ that depend on the UV completion. From there, it is attracted by the asymptotically safe fixed point along all its (infinitely many) irrelevant directions, and repulsed along the finitely many relevant ones. A wide range of suitable initial conditions at $\Lambda_{\rm UV}$ thus results in RG trajectories exhibiting near-scale invariance close to the fixed point over a wide range of scales. In the deep IR, they therefore closely resemble ``true" fixed-point trajectories exactly satisfying the relation between couplings associated to negative critical exponents.
 Accordingly, predictions for low-energy physics do not necessarily depend on asserting that asymptotic safety provides a UV completion valid ``to arbitrarily high scales".
 Taking the fixed-point limit, where RG trajectories are required to emanate exactly out of the fixed point, can be considered as a mere mathematical convenience, as it allows to easily make ``sharp" statements about low-energy predictions. In a more general setting with effective asymptotic safety, the degree of predictivity a fixed point imposes on IR physics has been quantified in \cite{Held:2020kze}.\\

  \subsection{Functional Renormalization Group}\label{subsec:FRG}
 \emph{...where we provide an introduction to the Functional Renormalization Group, focusing on conceptual and technical challenges specific to the gravitational context.}\\

 We work in a Euclidean, Wilsonian framework, where quantum fluctuations in the path integral are taken into account in a momentum-shell wise fashion. This allows us to identify candidates for quantum scale invariant dynamics in the UV and connect them with physics in the IR, to derive the predictions that follow from quantum scale symmetry. Below, we first introduce the setup in momentum space, then discuss the generalization of a momentum cutoff to the gravitational setting and highlight its Euclidean nature, before reviewing the functional differential equation that underlies this formalism.\\

 In the gravitational context, the distinction of high-momentum and low-momentum modes is more involved than in QFT on flat space. In general, defining a local notion of a coarse-graining scale -- in other words, a suitable generalization of a momentum scale -- requires a background geometry. In standard QFT, the flat metric provides a background such that one can implement a cutoff in momentum space. In a gravitational setting, one can use the background-field formalism.
To this end, the full metric $g_{\mu\nu}$ is split into a background $\bar{g}_{\mu\nu}$ and a fluctuation $h_{\mu\nu}$,
\be
  g_{\mu\nu} = \bar{g}_{\mu\nu} + h_{\mu\nu}.\label{eq:metricsplit}
\ee 
 This enables a generalization of the notion of momentum, by considering the background-covariant Laplacian $-\bar{D}^2$, such that $-\bar{D}^2 \rightarrow p^2$ when one chooses $\bar{g}_{\mu\nu} = \delta_{\mu\nu}$.  We stress that despite the suggestive notation, metric fluctuations $h_{\mu\nu}$ need not be small.\\
The split in Eq.~\eqref{eq:metricsplit} also facilitates the introduction of a gauge-fixing term for the metric, which is necessary in order to define the metric propagator. While the gauge-fixing term breaks the full diffeomorphism symmetry, it can be chosen to respect an auxiliary invariance with respect to the background metric, as is standard in the background-field formalism \cite{Reuter:1996cp}. This is central for the recovery of diffeomorphism symmetry in the full effective action $\Gamma=\Gamma_{k \rightarrow 0}$.
 \\

 The comparison of $p^2$  (or $-\bar{D}^2$) to a momentum scale $k^2$  is a meaningful distinction of UV and IR modes in Euclidean signature. For QFTs on a flat background, a Wick rotation or analytic continuation provides a relation of the Euclidean theory to the Lorentzian one. For quantum gravity, the analytic continuation of a number of metric configurations is not straightforward \cite{Baldazzi:2018mtl} and the configuration spaces of Riemannian and Lorentzian metrics feature a different topology/connectivity \cite{Demmel:2015zfa}. For these reasons, it is not expected that a straightforward relation between Riemannian and Lorentzian quantum gravity exists beyond the perturbative regime. With this cautionary remark, let us proceed with the Riemannian calculation, which can be thought of as exploring the effect of quantum fluctuations of a four-dimensional space (in contrast to a 3+1 dimensional spacetime).\\

 Within the Functional Renormalization Group (FRG), one introduces an (unphysical) momentum scale $k$ that serves to distinguish UV modes (with momenta higher than $k$) from IR modes (with momenta lower than $k$). One then introduces a mass-like regulator term to implement the idea of Wilsonian  renormalization: 
 The regulator $R_k(p)$ suppresses fluctuations with momenta $p^2<k^2$. Fluctuations with momenta larger than the scale $k$ are integrated out.
 After introducing this term in the path integral, one obtains a generalization of the full effective action $\Gamma$ to a scale-dependent action $\Gamma_k$, which contains the effect of quantum fluctuations above the momentum scale $k$ only, see \cite{Wetterich:1992yh,Ellwanger:1993mw,Morris:1993qb}. The full effective action $\Gamma$, encoding the physics, is recovered by integrating over all $k$. In the settings we explore, a main focus is on finding well-defined initial conditions for this integration, i.e., a microscopic dynamics $\Gamma_{k\rightarrow \infty}$ which is defined by an asymptotically safe fixed point. The physics implications of asymptotic safety are then explored in a second step by following the RG flow to $k\rightarrow 0$.\\
The key advantage of introducing a quadratic (i.e., mass-like) regulator term in the path integral is the resulting functional differential equation for $\Gamma_k$, which is of an exact one-loop structure at the formal level
 \cite{Wetterich:1992yh,Ellwanger:1993mw,Morris:1993qb} 
  \be
 \label{eqn:wetterich}
 \partial_t \Gamma_k = k \partial_k\, \Gamma_k = \frac{1}{2} {\rm STr} \left( \left(\Gamma_k^{(2)} + R_k \right)^{-1} \partial_t R_k \right).
 \ee
 Here, $t= \ln k/k_0$, with a reference scale $k_0$ and $\Gamma_k^{(2)}$ is the second functional derivative of the effective action with respect to the fluctuation fields. In our case, it is a matrix in field space, and the supertrace includes a trace in field space, in addition to a summation/integration over the discrete/continuous eigenvalues of the  Laplacian $-\bar{D}^2$.  To evaluate the right-hand-side of the flow equation, we sum over eigenvalues of the appropriate Laplacians, see, e.g., \cite{Eichhorn:2009ah,Dona:2012am,Benedetti:2012dx,Ohta:2015efa,Ohta:2015fcu,Alkofer:2018fxj}. We employ the ``middle-of-the-staircase'' approximation discussed in \cite{Alkofer:2018fxj}. \\
 For Grassmann-valued fields, such as, e.g., Faddeev-Popov ghost fields and fermion fields, the supertrace includes an additional negative sign. For reviews of the method see, e.g., \cite{Pawlowski:2005xe,Gies:2006wv,Rosten:2010vm,Dupuis:2020fhh}, and \cite{Reuter:2012id,Percacci:2017fkn,Reuter:2019byg,Bonanno:2020bil} specifically for the gravitational context.

The flow equation Eq.~\eqref{eqn:wetterich} shifts the focus away from the classical/microscopic action $S$: Instead of postulating a form for $S$ and starting to integrate out quantum fluctuations in the path integral, where configurations are weighed by ${\rm exp}[-S]$, one can explore the scale-dependence of whole families of interactions. This enables a search for viable microscopic dynamics defined by RG fixed points. More specifically, from Eq.~\eqref{eqn:wetterich}, one can obtain the beta function for any coupling in the effective action. By projecting the right-hand side of Eq.~\eqref{eqn:wetterich} on the corresponding interaction monomial, one generates diagrams of a one-loop form (with the regulator insertion $\partial_t R_k$ sitting on each one of the internal propagators in turn). These analytic beta functions enable a fixed-point search, allowing to identify candidates for asymptotically safe dynamics.\\
 The relation of these beta functions to those derived in different schemes in perturbation theory lies in the agreement of their universal parts; in particular the one-loop contributions from perturbation theory are recovered directly from the FRG-beta functions. Higher-loop contributions which are not universal in perturbation theory beyond two loops, are encoded in contributions coming from higher-order terms in the dynamics $\Gamma_k$ \cite{Baldazzi:2020vxk}. 
 At the same time, the FRG beta functions feature nontrivial denominators from Eq.~\eqref{eqn:wetterich}, which result in threshold effects and the automatic decoupling of modes once the RG scale $k$ falls below their mass. This is in contrast to standard perturbative schemes, where modes typically have to be decoupled by hand.

 The scale-dependent effective action $\Gamma_k$ in principle contains all operators compatible with the symmetries of the system under consideration, including all higher-order interactions. 
For practical calculations, one needs to truncate this infinite tower of operators, thereby introducing a systematic uncertainty.\\
 Within asymptotically safe gravity-matter systems, the principle used to select the truncation is the conjectured near-perturbative nature of the fixed point. This is supported by studies of nontrivial symmetry-identities in gravity-matter systems \cite{Eichhorn:2018nda,Eichhorn:2018ydy,Eichhorn:2018akn}, studies of the scaling spectrum in extended pure-gravity truncations \cite{Falls:2013bv,Falls:2014tra,Falls:2017lst,Falls:2018ylp,Kluth:2020bdv} and critical exponents in matter-gravity systems \cite{Narain:2009fy,Narain:2009gb,Eichhorn:2011pc, Percacci:2015wwa,Eichhorn:2016esv,Eichhorn:2016vvy,Eichhorn:2017eht,Eichhorn:2017sok,Eichhorn:2018nda,Eichhorn:2019yzm,Daas:2020dyo},
indications for asymptotic safety from one-loop perturbation theory \cite{Niedermaier:2009zz,Niedermaier:2010zz}, and the conjecture that the asymptotic-safety mechanism \cite{Eichhorn:2018yfc} relies on a balance of canonical scaling and quantum scaling that is operative in $d=2+\epsilon$ dimensions \cite{Gastmans:1977ad,Christensen:1978sc,Kawai:1989yh,Kawai:1992np,Kawai:1993mb,Aida:1996zn} and can be extended continuously to $d=4$ \cite{Reuter:1996cp,Reuter:2001ag}.
This implies that residual interactions  shift the critical exponents away from their canonical value by an $\mathcal{O}(1)$ amount. Accordingly, canonically highly irrelevant operators are expected to remain irrelevant at the interacting fixed point and can be neglected to obtain a robust count of the number of relevant directions, i.e., the free parameters of the theory.

 While such a choice of truncation would automatically be robust at non-interacting fixed points, its self-consistency needs to be checked a posteriori at an interacting fixed point. 
 Anticipating the results of our study, the fixed points we explore exhibit near-canonical scaling, making our choice of truncation self-consistent.\\

 Further support for such functional RG results can in the future come from other techniques. Similar to the gravitational case, Gross-Neveu systems are a paradigmatic example for interacting fixed points. The underlying mechanism generating scale symmetry is the same as that conjectured for asymptotically safe quantum gravity. Thus, a choice of truncation based on canonical power counting  produces reliable results \cite{Rosa:2000ju,Hofling:2002hj,Braun:2010tt,Gies:2010st,Gehring:2015vja,Classen:2015mar,Vacca:2015nta,Knorr:2016sfs} in agreement with other approaches \cite{Parisi:1975im,Gawedzki:1985ed,Gracey:1990sx,Hikami:1976at,Rosenstein:1988pt,Gat:1991bf,Rosenstein:1993zf,Mihaila:2017ble,Zerf:2017zqi}.
These systems also provide a blueprint how the confluence of several methods can provide strong evidence for an interacting fixed point. In the case of asymptotically safe gravity, a similar degree of convergence from different techniques has not been achieved yet, but asymptotic safety is under active investigation both with lattice techniques \cite{Loll:2019rdj} as well as combinatorial tensor models \cite{Eichhorn:2018phj}. The causal-set approach \cite{Surya:2019ndm} could in the future provide direct access to a Lorentzian continuum limit, a.k.a.~Lorentzian asymptotic safety \cite{Eichhorn:2019xav}, as could spin foam studies \cite{Bahr:2016hwc}. 
 \subsection{Three systems: Single scalar, simple Yukawa, portal to simple Yukawa dark matter}
 \emph{...where we specify the matter field content of the three distinct systems we will explore and define the truncation of the dynamics for each case.}\\

We consider a real scalar field $\phi$, featuring a $\mathbb{Z}_2$ symmetry, $\phi \rightarrow -\phi$, with  the truncated effective action
 \bea
  \label{eqn:ansatz}
  \Gamma_k^\text{scal+grav} &=& \int \dd[4]{x} \sqrt{g} \Big(  \frac{1}{2} Z_{\phi} g^{\mu \nu} \partial_{\mu}\phi \partial_{\nu}\phi \nonumber \\
   & &- \left(\frac{1}{16\pi G_N} + \xi \phi^2\right) R + V(\phi^2) \Big).
 \eea
 Similar truncations for scalar-gravity system have been explored, e.g., in \cite{Narain:2009fy,Narain:2009gb,Eichhorn:2012va,Percacci:2015wwa,Labus:2015ska, Eichhorn:2017als,Pawlowski:2018ixd, Wetterich:2019zdo, deBrito:2019umw}. For work on a scalar field in the presence of a non-minimal coupling $\xi$ but without quantum gravitational effects with FRG techniques, see \cite{Merzlikin:2017zan}.
 The potential $V(\phi^2)$ includes a cosmological constant when expanded about $\phi=0$, such that our truncation includes the Einstein-Hilbert action. 
We will typically terminate this expansion at the quartic order,
 \be
  \label{eqn:truncation}
  V(\phi^2) = \frac{2 \bar{\Lambda}}{16\pi G_N} + \frac{\bar{m}^2}{2} \phi^2 + \frac{\lambda_4}{4} \phi^4
  ,\ee 
and always express our results in terms of the dimensionless couplings
  \be
 g = G_N\, k^2, \, \lambda = \bar{\Lambda}\, k^{-2}, \, m^2 = \bar{m}^2\, k^{-2}. \label{eq:dimlesscouplings}
 \ee
 Dimensionless couplings are best suited to search for asymptotic safety in the UV. On the other hand, dimensionful couplings, such as dimensionful masses can be of interest in the IR. The rescaling  Eq.~\eqref{eq:dimlesscouplings} could beg the question how such dimensionful couplings can be finite in the IR, where $k \rightarrow 0$. For the limit $ k \rightarrow 0$, we focus on the case in which every field in the theory comes with a mass (this includes the massless graviton, where the Planck-scale serves as the mass-scale for metric fluctuations). In this situation, quantum fluctuations are suppressed, once k has fallen below the mass scale. All dimensionful quantities are then constant, since their dimensionless counterparts scale according to the canonical dimension. For instance, the beta function for a dimensionless mass parameter $m^2$ becomes $\beta_m^2 = -2 m^2$, once $k^2 < m^2$, such that all loop contributions are suppressed. Thus, $m^2(k) \sim k^{-2}$. Accordingly, the dimensionful mass $\bar{m}^2 = m^2 k^2 \rightarrow \rm const$. This extends to all other couplings.

 In part of what follows, the real scalar field $\phi$ will serve as a toy model for the SM Higgs field. Both have in common that there are no massless Goldstone modes after spontaneous symmetry breaking, making a $\mathbb{Z}_2$-symmetric real scalar  a suitable toy model to explore aspects of Higgs physics in the FRG setup, see \cite{Gies:2009hq,Gies:2009sv,Eichhorn:2015kea}.

 As a representative for the top quark  we additionally incorporate a single Dirac fermion in Sec.~\ref{sec:yukawa_model} by including the terms
 \be
  \label{eqn:action_fermion}
  \Gamma_k^\text{ferm} = \int \dd[4]{x} \sqrt{g}\left(i\bar{\psi}\slashed{\nabla}\psi + i y\, \phi \,\bar{\psi}\psi\right)
 \ee
 into our truncation for the effective action, where $\slashed{\nabla}$ is the Dirac operator which contains the spin connection in the presence of gravity.  Earlier work on simple Yukawa systems coupled to quantum gravity can be found in \cite{Zanusso:2009bs,Vacca:2010mj,Oda:2015sma,Eichhorn:2016esv,Hamada:2017rvn,Eichhorn:2017eht}.\\

 In Sec.~\ref{sec:portal} we first consider two real scalars coupled to gravity, labeled by $v/d$ for a visible/dark sector. 
In that section, we first work with
\bea
\label{eqn:action_two_scalars}
\Gamma_k^{\rm 2-scal + grav}&=&\int d^4\sqrt{g}\Bigl(  \frac{Z_v}{2}g^{\mu\nu}\partial_{\mu}\phi_v \partial_{\nu}\phi_v+\frac{\bar{m}_v^2}{2}\phi_v^2\nonumber\\
&{}& +\frac{\lambda_v}{8}\phi_v^4 - \xi_v \phi_v^2 R - \frac{1}{16\pi G_N} \left(R-2 \bar{\Lambda} \right)\nonumber\\
&{}&+ \frac{Z_d}{2}g^{\mu\nu}\partial_{\mu}\phi_d \partial_{\nu}\phi_d+\frac{\bar{m}_d^2}{2}\phi_d^2\nonumber\\
&{}&+  \frac{\lambda_d}{8}\phi_d^4 - \xi_d \phi_d^2 R\Bigr).
\eea
 The two sectors are coupled via a portal coupling
 \be
  \label{eqn:action_portal}
  \Gamma_k^\text{portal} = \int \dd[4]{x} \sqrt{g} \frac{\lambda_\text{HP}}{4} \phi_v^2 \phi_d^2.
  \ee

In the second part of Sec.~\ref{sec:portal}, the visible scalar $\phi_v$ and the dark scalar $\phi_d$ are additionally coupled to a visible fermion $\psi_v$ and a dark fermion $\psi_d$, respectively, by a flowing action of the type \eqref{eqn:action_fermion}.  There, we explore the same truncation as put forward in \cite{Eichhorn:2020kca}.

 While obtaining the beta functions from the effective action, one encounters technical choices related to the non-minimal coupling. One can choose both, the gauge fixing action $S_\text{gf}$ as well as the regulator $R_k$ to depend on the scalar background field. In case of a scalar-background dependent gauge fixing, one needs to consider an appropriately chosen ghost term. We discuss these technical aspects and give the explicit form of our gauge fixing and regulator terms in App.~\ref{sec:methods}. The (artificial) dependence on various technical choices is relatively mild, and does not qualitatively alter our results as we investigate in detail in that appendix.

 \section{Single scalar}
 \label{sec:single_scalar}
  \emph{...where we search for asymptotic safety and characterize its predictivity in a scalar-gravity system. We tie our results for the flattening of the scalar potential under the impact of quantum gravity into discussions of global symmetries, as well as slow-roll parameters. Finally we highlight potential phenomenological consequences in a cosmological context, with a focus on inflation with and without non-minimal coupling as well as a short discussion of dynamical dark energy and the cosmological constant in asymptotic safety.}\\

 In this section we concentrate on the pure scalar gravity system, with its truncated flowing action given by
 \be
  \Gamma_k = \Gamma_k^\text{scal+grav} + S_{\rm gf} + S_{\rm gh}.\label{eq:truncationpurescalar}
 \ee

 \subsection{Fixed-point results}
 \emph{...where we show fixed-point results for the scalar potential and non-minimal coupling. 
 We compare the case of the  scalar-gravity system to a case with additional, purely gravitationally coupled SM-type fields. Finally, we treat the gravitational couplings as free parameters, to obtain a comprehensive overview of the predictivity of asymptotic safety in this context.}\\

\begin{figure}
   \includegraphics[width=0.45\textwidth]{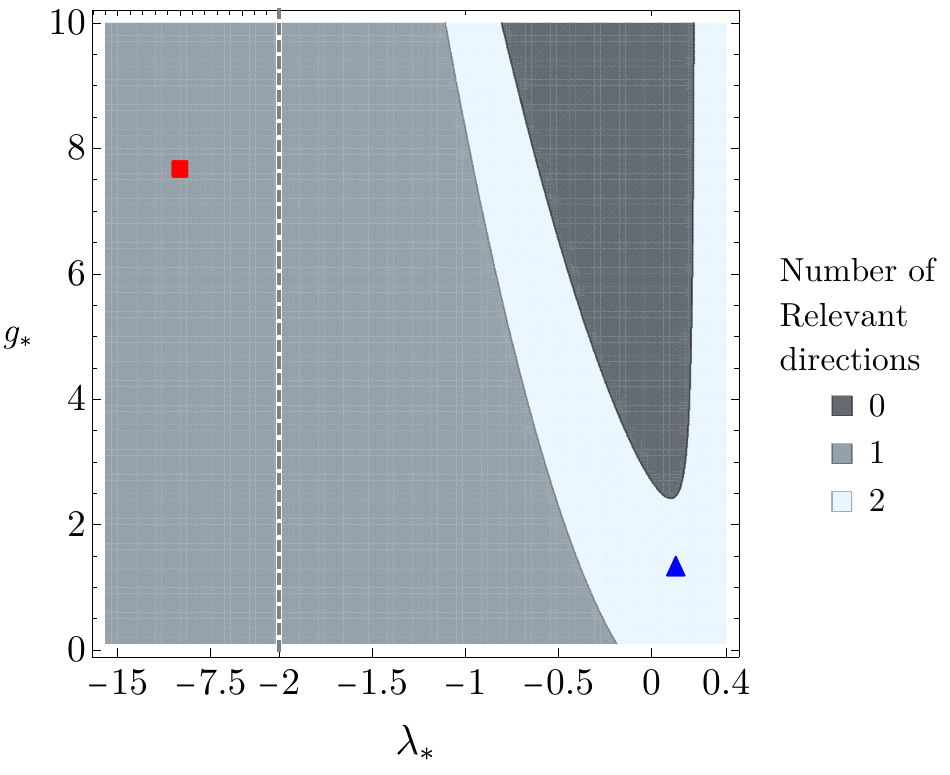}
   \caption{\label{fig:fv_nr_relv_dir}
   Number of relevant directions at the free fixed point as a function of the gravitational couplings $\lambda_\ast$ and $g_\ast$ respectively, for the three scalar couplings $(m^2, \lambda_4, \xi)$. 
   The blue triangle/red square marks the position of the fixed point Eq.~\eqref{eq:FPpurescalar}/\eqref{eqn:fp_grav_canonical}. 
   For larger values of the gravitational coupling the fixed point is fully irrelevant. For (large) negative values of the cosmological constant one relevant direction turns irrelevant.
   The vertical dashed line highlights a change in the scaling of the $\lambda_\ast$ axis at $\lambda_\ast=-2$.
   }
 \end{figure}

 Without additional matter degrees of freedom, the only viable fixed point present in the truncation \eqref{eq:truncationpurescalar} with the expansion \eqref{eqn:truncation} for the scalar potential lies at 
 \be
  \lambda_\ast = 0.13, \quad g_\ast =1.32, \quad
   m^2_{\ast} = \lambda_{4\,\ast} = \xi_\ast = 0 
 \label{eq:FPpurescalar}
 .\ee
 There are indications that the inclusion of additional matter degrees of freedom significantly alters the gravitational fixed-point values \cite{Dona:2013qba}. In fact, within the background-field approximation we employ here, an increasing number of fermions shifts the fixed-point values towards negative microscopic cosmological constant. An effective strength of gravity can be defined as $g_{\rm eff}=g_{\ast}/(1-2 \lambda_{\ast})^{\#}$, with $\#>0$, as this quantity enters the beta functions for matter couplings \cite{Eichhorn:2017eht,Eichhorn:2018nda}. As a function of an increasing number of fermions, the effective strength of gravity decreases. Similar results have been found for fluctuation calculations \cite{Eichhorn:2018nda}.
 Under the impact of
 the full Standard Model particle content, one obtains the gravitational fixed-point values
 \be
  \label{eqn:fp_grav_canonical}
  \lambda_\ast = -9.97, \quad g_\ast = 7.63
 .\ee
 To assess the effect of such a shift in the gravitational couplings, we study the critical exponents of the couplings $m^2,\, \lambda_4$ and $\xi$ at the fixed point $m_{\ast}^2=0, \, \lambda_{4\, \ast}=0,\, \xi_{\ast}=0$ at varying values for the gravitational couplings $\lambda_\ast, g_\ast$.
 Fig.~\ref{fig:fv_nr_relv_dir} shows the number of relevant directions as a function of the gravitational parameters.
 For the fixed-point values in Eq.~\eqref{eq:FPpurescalar}, there are two relevant directions, which can be associated with the scalar mass and non-minimal coupling. In contrast, there is only one relevant direction, associated to the mass parameter, at the fixed point in Eq.~\eqref{eqn:fp_grav_canonical}. A third region, with no relevant direction, would also in principle be available. The determination of the gravitational fixed-point values is subject to systematic uncertainties. Therefore, we analyze the entire $\lambda-g$ plane instead of focusing on fixed-point values in a particular truncation. We focus on $g>0$ throughout the paper, as $g<0$ cannot be connected to a regime with attractive gravity in the IR.

 To understand why the different regions of the $\lambda-g$ plane show different degrees of predictivity in terms of number of relevant directions, we will first discuss the critical exponents in a simple approximation, where we assume a diagonal stability matrix, neglecting the mixing of operators.\\
Then, the gravitational contribution to the anomalous dimension for the scalar potential is always \emph{towards} irrelevance. Given the vanishing canonical dimension of $\lambda_4$, there is accordingly at least one irrelevant direction everywhere in the $\lambda-g$ plane. Where $g_{\rm eff}=g/(1-2 \lambda)^{\#}$ (with $\#>0$) is largest, the gravitational effects can even shift the mass parameter into irrelevance. In fact, $g_{\rm eff}$ is largest for large $g$ and $0<\lambda<1/2$ and decreases, as $g$ and $\lambda$ are lowered. Thence, the gravitational contribution to the scaling of the scalar mass parameter is larger than its canonical dimension of 2 for 
\be
g_{\rm crit} > \frac{12\pi (1 -2 \lambda)^2 (3-4\lambda)^2}{159-460\lambda+352\lambda-16\lambda^3},
\ee
see also the analysis in \cite{Eichhorn:2017als,Pawlowski:2018ixd}. 
\\
Finally, the gravitational contribution to $\beta_{\xi}$ is towards irrelevance for $\lambda<-0.17$, as follows from an inspection of $\beta_{\xi}$ at the free matter fixed point (with $m_{\ast}^2=0$, $\lambda_{4\, \ast}=0$, $\xi_{\ast}=0$),
 \bea
 \beta_{\xi}&=& -g\frac{99+318 \lambda-1464 \lambda^2+1232 \lambda^3-96 \lambda^4}{18 \pi (1-2\lambda)^3(3-4\lambda)^2}\xi\nonumber\\
 &{}&+ 4\,g \frac{21-8 \lambda}{\pi(3-4\lambda)^2}\xi^2+ 54\, g\frac{5-8\lambda}{\pi(3-4\lambda)^2}\xi^3.
 \eea
 The linear coefficient changes sign at sufficiently negative $\lambda<-0.17$. This is associated to a degeneracy of zeros of $\beta_{\xi}$ (which are not fixed points of the full system, except at the degenerate point corresponding to the free fixed point.) 

These two boundaries, namely $\lambda<-0.17$ and $g>g_{\rm crit}$, become deformed once we lift the simple approximation and consider off-diagonal contributions in the stability matrix. Nevertheless, a ``strong-gravity" island with no relevant directions (the deformation of the $g>g_{\rm crit}$ boundary), and a region with only one relevant direction for sufficiently negative $\lambda$ (the deformation of the $\lambda<-0.17$ line) are still visible in Fig.~\ref{fig:fv_nr_relv_dir}. 

If the gravitational coupling values fall in the strong-gravity regime, then all of the couplings in the scalar sector are irrelevant at the fixed point. Along the full flow towards the IR , i.e., at all $k$, their value is predicted. In particular, the scalar potential is predicted to remain flat in this scenario. Outside of the strong-gravity regime, every relevant direction can start to deviate from its fixed point value at some finite scale $k_{\rm tr}$. The corresponding (linear combination of) coupling(s) can take a range of values in the IR and needs to be determined by observation. Once all relevant couplings are fixed the theory becomes fully predictive.

 \subsection{Three perspectives on flat scalar potentials in  quantum gravity}
 \emph{...where we interpret the fixed-point results of the previous subsection in light of the more general question about global symmetries in quantum gravity, explain how our results provide a hint for a mechanism of gravitational flattening of scalar potentials and finally rewrite our scale-dependent potential in terms of scale-dependent slow-roll-inspired parameters, in order to provide a more straightforward link of our results to cosmological settings.}

 \subsubsection{Global symmetries}\label{subsubsec:globalsyms}
The fate of global symmetries in quantum gravity is an intriguing open question. Arguments suggesting the violation of global symmetries through quantum gravitational effects have been put forward in \cite{Banks:1988yz,Kamionkowski:1992mf,Kallosh:1995hi,Banks:2010zn} and substantiated to some extent in the context of the AdS/CFT conjecture \cite{Harlow:2018jwu,Harlow:2018tng}, in particular in relation to the weak-gravity conjecture  \cite{ArkaniHamed:2006dz}, see also \cite{Palti:2019pca} for a review. In several other approaches to quantum gravity, less is known about this question -- in part as a  consequence of the general difficulty to account for matter degrees of freedom in many approaches to quantum gravity. As one example, conflicting results on the realization of chiral symmetry in a fermion sector coupled to Loop Quantum Gravity exist in the literature \cite{Barnett:2015ara, Gambini:2015nra}.
In an asymptotically safe context, the situation is different:
many different gravity-matter systems that feature global symmetries in the matter sector have been explored, see, e.g., \cite{Narain:2009fy,Narain:2009gb,Eichhorn:2011pc, Eichhorn:2016vvy, Eichhorn:2017eht,Eichhorn:2017sok}. All results in truncations of the Euclidean RG flow support the conclusion that quantum gravitational fluctuations at fixed topology do not lead to a violation of global symmetries, see also the discussion in \cite{Eichhorn:2020mte}. 
Let us highlight two potential caveats of these results:
Firstly,  Euclidean and Lorentzian quantum gravity could differ when it comes to the fate of global symmetries. Secondly, it could be the expansion to finite orders in the fluctuation field $h_{\mu\nu}$ that prevents global-symmetry-violating terms to become visible in the RG flow. Yet, within Euclidean Dynamical Triangulations, one also finds that shift-symmetry is left intact by quantum gravitational fluctuations \cite{Jha:2018xjh}. This suggests that the non-violation of global symmetries in the FRG setting is not a consequence of truncations to finite order in the gravitational fluctuation field at fixed topology. A first result in \cite{Hamada:2020mug} suggests that the presence of topological fluctuations might change the picture. Here, we remain conservative in that topological fluctuations are not taken into account.

We discuss the fixed point in Eq.~\eqref{eq:FPpurescalar} from the perspective of global symmetries: Firstly, the $\mathbb{Z}_2$ symmetry of the scalar field is respected by quantum gravitational fluctuations. This follows, as the rhs of the flow equation vanishes when projected onto field monomials with an uneven power of $\phi$. Secondly, the shift symmetry of the scalar kinetic term, $\phi \rightarrow \phi+a$, is respected by the flow.
This is an example of how all global symmetries manifested in the kinetic term of a matter field are respected by quantum gravity, see \cite{Eichhorn:2017eht}. Hence, the  $\mathbb{Z}_2$- and shift- symmetric hypersurface in theory space is a \emph{fixed hypersurface} under the RG flow. 
This does not automatically imply that there must be a \emph{fixed point in this hypersurface}, as there are derivative-interactions compatible with shift-symmetry and $\mathbb{Z}_2$ symmetry. These are  generically nonzero in the presence of asymptotically safe gravity \cite{Eichhorn:2012va, Eichhorn:2017eht}. They are however not guaranteed to feature a fixed point. Since these are neglected in our truncation, there must be a fixed point respecting $\mathbb{Z}_2$- and shift symmetry in our truncated theory space, i.e., a fixed point at vanishing scalar potential\footnote{The inclusion of derivative-interactions is generically determined by the following pattern of quantum-gravity induced interactions: As long as gravitational interactions are sufficiently weak, there is a real fixed point for such interactions, which is the shifted Gau\ss{}ian fixed point, at which they remain irrelevant \cite{Eichhorn:2011pc, Eichhorn:2017eht}. At strong enough gravitational interactions, some of the induced interactions no longer feature a real fixed point, see, e.g., \cite{Eichhorn:2016esv, Eichhorn:2017eht, Christiansen:2017gtg}. }, as found in Eq.~\eqref{eq:FPpurescalar} and previously in \cite{Narain:2009fy,Narain:2009gb,Percacci:2015wwa,Oda:2015sma,Labus:2015ska,Wetterich:2016uxm,Hamada:2017rvn,Eichhorn:2017als,Pawlowski:2018ixd,Wetterich:2019rsn}. 
Beyond, additional fixed points at nonvanishing potential might in principle also exist, but turn out to require the presence of additional matter fields, e.g., fermions coupled through a Yukawa term, see Sec.~\ref{sec:yukawa_model}.

 \subsubsection{The gravity-induced flow towards flatness and predictivity}\label{sec:QGflattens}

 In our truncation, shift symmetry guarantees the existence of a fixed point corresponding to a flat potential and vanishing non-minimal coupling.
 The subsequent pressing question is whether combined fluctuations of gravity and matter drive the system away from this fixed point or towards it. 
 This information is encoded in the number of relevant directions, shown as a function of the gravitational parameters in Fig.~\ref{fig:fv_nr_relv_dir}.
Among the three couplings $m^2, \lambda_4$ and $\xi$, at least one combination is always irrelevant. 
 This implies that flowing from the fixed point towards the IR, at least one of the three couplings is fixed in terms of the others. In fact, under the assumption of a diagonal stability matrix, the quartic coupling is required to vanish at all scales, preventing a $\phi^4$-type deviation from flatness.

Additionally, there are regions in the $\lambda-g$ plane, where two or even all three of the couplings $m^2, \lambda_4$ and $\xi$ are irrelevant, cf. Fig.~\ref{fig:fv_nr_relv_dir}. In these regions, the gravity-induced flow is towards flatness; all perturbations away from flatness are irrelevant and therefore excluded. In the regime with no relevant directions, the ``resurgence mechanism" for the scalar mass becomes applicable, which could provide a dynamical mechanism for the mass hierarchy between the electroweak scale and Planck scale \cite{Wetterich:2016uxm}.

 Each irrelevant direction also restricts settings in which the initial conditions at $k_\text{UV}>M_\text{Pl}$ are chosen away from the fixed point as in effective asymptotic safety, see Sec.~\ref{sec:effAS}: flowing towards $M_\text{Pl}$, an irrelevant direction is IR-attractive, the RG flow is driven towards the fixed-point value.  The simplest case is that of no relevant directions in which case the flat potential and vanishing non-minimal coupling are fully IR attractive: quantum fluctuations drive the RG flow towards $m^2(M_{\rm Pl})\approx0,\, \lambda_4(M_{\rm Pl})\approx0,\, \xi(M_{\rm Pl})\approx0$. With increasing ratio $k_{\rm UV}/M_{\rm Pl}$, this prediction becomes sharper; see \cite{Held:2020kze} for a quantitative measure of predictivity in such a setting. As an upshot, quantum-gravity fluctuations could drive scalar potentials towards flatness also in effective asymptotic safety, i.e., in presence of a UV scale of new physics.

 \subsubsection{Small slow-roll parameters from asymptotic safety}
 \label{sec:slow_roll}
\begin{figure}
   \includegraphics[width=0.3\textwidth]{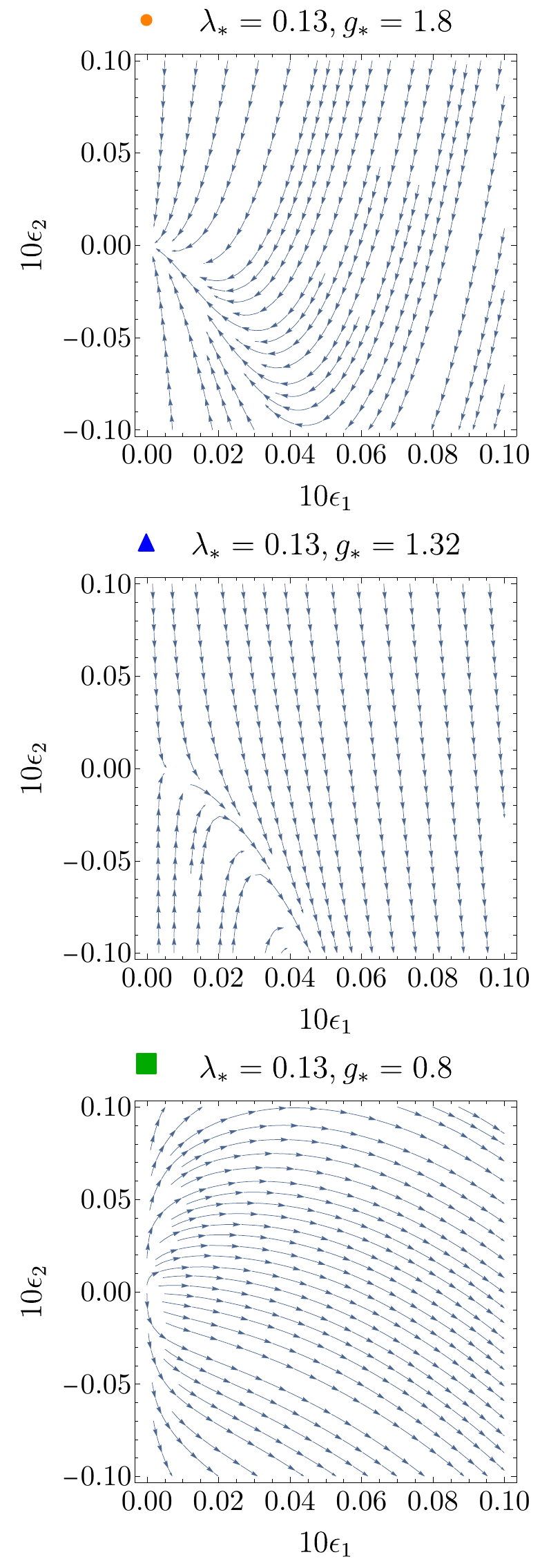}
   \includegraphics[width=0.4\textwidth]{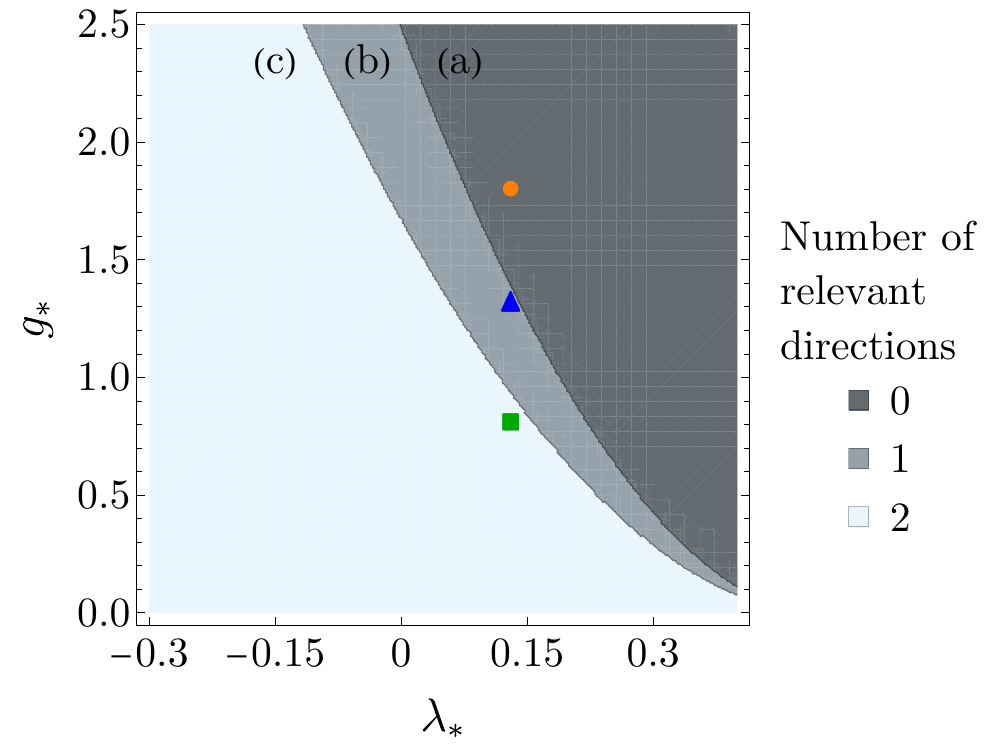}
   \caption{
   \label{fig:slow_roll_flow_epseta}
  (top) The flow of $\epsilon_1$ and $\epsilon_2$ at  various values of $g_\ast$, with $\lambda_\ast$ given by the fixed point \eqref{eq:FPpurescalar} and $\phi/k=1$. For larger values of $g_{\ast}$, the fixed point exhibits fewer relevant directions. (bottom) The number of relevant direction as a function of $\lambda_\ast$ and $g_\ast$. The three markers indicate the values for $(\lambda_\ast, g_\ast)$ represented in the panels at the top of this figure, and (a)-(c) label the three regions explained in more detail in the text.
   }
 \end{figure}

 The flattening of scalar potentials could be of relevance for scalar fields in cosmological settings, such as inflation. 
 In this context, a description in terms of the slow-roll parameters is commonly applied. 
 To relate to that, we re-express our results in a similar terminology
by rewriting the flow equation for the scalar potential in terms of the dimensionless quantities
 \be
  \label{eqn:slow_roll_parameters}
  \epsilon_i = \frac{1}{k^{4-i}}\pdv[i]{V(\phi^2)}{\phi},
 \ee
 which are inspired by the standard slow-roll parameters. Their physical values are those at $k=0$, when all quantum fluctuations are integrated out. At $k\approx M_\text{Pl}$, quantum gravitational fluctuations decouple from the flow, and the remaining evolution is determined by quantum fluctuations of the scalar field, only. 
We work within the approximation $\xi = 0$, which is self-consistent in the vicinity of the fixed point $\xi_\ast = 0$  and additionally neglect the scalar anomalous dimension.
 The scale dependence of the potential  is given in App.~\ref{app:flow_potential} and determines the flow of the $\epsilon_i$ by the appropriate derivatives with respect to the scalar field.
The beta functions for the $\epsilon_i$ are explicitly field dependent, since they can be understood as the flowing coefficients of a local Taylor expansion of the potential at finite values of the field.
 The flow of $\epsilon_i$ depends on $\epsilon_{i+1}$ and $\epsilon_{i+2}$. We truncate this hierarchy by setting $\epsilon_i = 0\,\, \forall\,\, i>2$.
 As expected, the system exhibits a fixed point with flat potential, $\epsilon_{i, \, \ast} = 0$ that can be traced back to the fixed point \eqref{eq:FPpurescalar}, and no other fixed points.  The critical exponents encode whether the low-energy limit of this setting has any free parameters which allow the potential to deviate from flatness. For $\phi/k \sim \mathcal{O}(1)$ 
the corresponding RG flows are shown in Fig.~\ref{fig:slow_roll_flow_epseta}. 
  At a fixed point with vanishing Newton coupling $g_\ast=0$, both $\epsilon_{1,2}$ would be relevant, as expected from their canonical mass dimension. The leading term in $g$ in all critical exponents is 
 \be
  \theta^{(i)} = - \frac{2(51-144\lambda+104\lambda^2)}{3\pi(3-4\lambda)^2(1-2\lambda)^2} g
 ,\ee 
 which is negative for all $\lambda \in (-\infty, 0.5)$.
 Hence, as $g_\ast$ increases, the $\epsilon_i$ one by one become irrelevant and are thereby required to be small as a consequence of quantum gravity fluctuations.

 \subsection{Outlook on potential phenomenological consequences}\label{sec:phenosinglescalar}
 \emph{...where we discuss some of the potential implications our findings might have for inflation and (dynamical) dark energy in an asymptotically safe context. While these results are obtained in a truncation of the Euclidean dynamics, they tentatively hint at the potential constraining/predictive power of asymptotic safety in cosmology.}\\

 Scalar fields appear in many cosmological applications and our results might entail certain features that could arise in these settings as a consequence of asymptotic safety. These types of study are relevant for various reasons.
Results of such studies can serve as a theoretical principle that distinguishes between effective models (e.g., in the context of dark matter or dark energy) with or without a quantum-field theoretic UV completion\footnote{A related idea is being explored within the effective field theory setting, where, e.g., certain definitions of locality and/or unitarity are demanded of a viable UV completion and serve to restrict the effective field theories, see, e.g., \cite{deRham:2017zjm,Melville:2019wyy}.}. The availability of such an asymptotically safe UV completion is expected to come with an enhanced predictive power.
On the one hand, this can connect asymptotically safe gravity to observations, making observational consistency tests or even the search for ``smoking-gun" signatures possible. On the other hand, the potential constraining power of a given UV completion, e.g., asymptotic safety, might provide guidance for observational searches, as it could (dis)favor certain regions of the EFT-parameter space.

Let us highlight that, as opposed to an effective-field-theory setting, transplanckian excursions of the inflaton are not expected to be problematic in an asymptotically safe context. In particular, transplanckian imprints on inflationary observables could arise. Asymptotic safety might therefore not be restricted by the transplanckian censorship conjecture \cite{Martin:2000xs,Bedroya:2019snp,Bedroya:2019tba}.

 \subsubsection{Towards asymptotically safe consequences for the inflaton}\label{sec:ASinfl}
 \emph{...where we use results on the behavior of the derivatives of the potential from Sec.~\ref{sec:slow_roll} in the context of inflation. We first focus on the possibility that quantum-gravity fluctuations might yield a flattened inflaton potential. In a second step, we explore whether the predictive power of asymptotic safety might make the observed amplitude of scalar fluctuations a critical observational test of the asymptotic-safety scenario for a gravity-inflaton system.}\\

 The near-scale-invariance of the spectrum of primordial scalar perturbations might be interpreted as a hint towards scale invariance as a guiding principle for inflationary model building \cite{Shaposhnikov:2008xb, GarciaBellido:2011de,Bezrukov:2012hx,GarciaBellido:2012zu,Khoze:2013uia,Rubio:2014wta,Ferreira:2016wem,Kannike:2016wuy,Karananas:2016kyt,Ferreira:2016vsc,Rubio:2017gty}.
Within the quantum-scale invariant framework of asymptotic safety, the scalar degree of freedom that arises in $f(R)$ theories and is present in corresponding truncations of the gravitational action, has been explored as a driver of inflation \cite{Bonanno:2010bt,Bonanno:2015fga,Kofinas:2016lcz,Bonanno:2018gck,Platania:2019qvo}. An RG-improvement procedure within a toy- model has been used to calculate inflationary parameters from this setting, see \cite{Platania:2020lqb} for a review as well as \cite{Bonanno:2020bil} for a critical discussion of RG improvement. 

 Here, we follow an alternative route, and include an additional scalar as the inflaton.
 We consider the ``vanilla'' model of inflation, a scalar that does not couple to any of the (B)SM degrees of freedom except for a non-minimal coupling to gravity. Potential derivative-couplings of the inflaton to (B)SM fields that are expected in an asymptotically safe setting \cite{Eichhorn:2012va,Eichhorn:2017eht,Eichhorn:2017sok} will be neglected here, motivated by their canonical irrelevance. 
Asymptotic safety fixes several of the parameters determining the inflaton potential, as well as potentially the non-minimal coupling. In fact, the non-minimal coupling $\xi$ vanishes at the asymptotically safe UV fixed point. We first explore the approximation that $\xi$ remains negligible at all scales, before exploring the case with nonvanishing $\xi$ below.

 \emph{Inflation with vanishing non-minimal coupling?}\\
From observations, the deviation from scale-invariance of the spectrum of primordial scalar fluctuations, $n_s - 1$, the tensor-to-scalar ratio $r$ and the amplitude of  scalar fluctuations $A_s$ can be inferred or constrained, respectively, see, e.g., \cite{Akrami:2018odb} for the most recent results. In turn, these quantities are calculable in a given inflationary setting, and depend on the potential as well as its first and second derivative.  For vanishing non-minimal coupling, our study of the slow-roll-inspired parameters $\epsilon_i$, see Sec.~\ref{sec:slow_roll}, is applicable. The value of the potential (typically evaluated at vanishing field) always corresponds to a relevant direction, i.e., its IR value can be chosen freely, and corresponds to a cosmological constant. 
Within our truncation, there are three regimes to distinguish, cf.~Fig.~\ref{fig:slow_roll_flow_epseta}, where a) the first and second, b) the second, c) neither the first nor the second derivates of the inflaton potential are fixed in the IR by demanding asymptotic safety in the UV. In regime a), where all derivatives of the scalar potential are fixed by asymptotic safety, the potential stays completely flat and inflation does not exit gracefully. In the second regime, b), the first derivative of the scalar potential remains a free parameter, as it corresponds to a relevant direction of the fixed point. The dimensionless second derivative is predicted as a function of the first, cf.~separatrix in the second panel of  Fig.~\ref{fig:slow_roll_flow_epseta}.  Interestingly, the second derivative, $\epsilon_2$ is driven to negative values, favoring concave potentials as are indeed also favored by the data, cf.~Fig.~8 in \cite{Akrami:2018odb}. At this point, we do not embark on a quantitative comparison with actual data, due to the systematic limitations of our study, and simply stress that the predictive power of this setting could in the future allow for powerful comparison with data. In addition to an investigation of the flatness of the potential, the amplitude of the primordial perturbations provides a further test (see below).

Additionally, the least predictive of the three regimes, c), remains, where the fixed point is IR repulsive in both the first and second derivative of the potential and no predictions about their IR values can be derived from asymptotic safety. In this regime, the observed values of $r$ and $n_s$ can potentially be accommodated, leaving us with the amplitude $A_s$. It is fixed by the value of the potential, measured at the value of the field  corresponding to CMB scales. Within asymptotic safety, this quantity corresponds to a relevant direction and can therefore be chosen freely. However, late-time observations can constrain it: In the absence of an additional scalar field, i.e., dynamical dark energy, the free parameter has to be chosen such as to provide the correct value of the cosmological constant. 

In the inflationary context, it is particularly worthwhile to consider the scenario of \emph{effective asymptotic safety}, cf.~Sec.~\ref{sec:effAS}, see, e.g., \cite{deAlwis:2019aud,Held:2020kze}, see also \cite{Percacci:2010af}: In this scenario, a finite UV scale of new physics, $\Lambda_{\rm UV}> M_{\rm Pl}$ exists, beyond which a (non quantum field theoretic) UV completion is valid. Below $\Lambda_{\rm UV}$, the gravity-matter RG flow provides an effective-field-theory description of the microphysics. In this case, the initial value of the potential and its derivatives is specified at $\Lambda_{\rm UV}$ by the UV completion. The gravity-matter RG flow then drives the derivatives of the scalar potential towards smaller values, when the UV values of the Newton coupling and cosmological constant fall into the dark gray region in Fig.~\ref{fig:slow_roll_flow_epseta}.
While quantitative statements rely on the details of the UV completion, we conclude that asymptotic safety could provide a mechanism that drives potentials towards flatness, and thus potentially towards an agreement with the data. Therefore, asymptotic safety might provide a mechanism to lift models out of the string-theoretic swampland into the string-theoretic landscape, see also \cite{deAlwis:2019aud}. 

We also point out that this might imply that criteria that have been proposed as indications for the possibility of eternal inflation \cite{Rudelius:2019cfh} might be realized in an asymptotically safe context.

 \emph{Inflation with nonvanishing non-minimal coupling?}\\
 Going beyond the approximation $\xi=0$ in the previous paragraph, we also explore inflation with a nonzero value of the non-minimal coupling. In this case, a conformal transformation from the Jordan frame (which features a non-minimal $\phi^2 R$ term) to the Einstein frame (which is minimally coupled) is typically made to analyze the system, and our study of the derivatives of the potential does not apply since we worked at $\xi=0$ in that analysis. Instead, we now assume that the potential can be parameterized fully in terms of $\xi$, $m$ and $\lambda_4$, including at large field values. Within our truncation, the following description then holds.
 Observationally, the case of a purely polynomial potential of order two or four without non-minimal coupling is ruled out by the exceedingly large tensor-to-scalar ratio \cite{Aghanim:2018eyx}. Therefore, a finite value of the non-minimal coupling is required.  
We first neglect the scalar mass, setting $m^2 = 0$.
 To match the observed amplitude of primordial scalar fluctuations, one needs to adjust the magnitude of the Einstein frame potential. In a simple approximation; we require a ratio of $\lambda_4/\xi^2 \sim O(10^{-9})$ to do so.
 As we find no indication for the existence of an interacting fixed point with finite $\xi$, the finite value of $\xi$ would need to be generated along the flow away from the fixed point \eqref{eq:FPpurescalar}.

\begin{figure}
\includegraphics[width=\linewidth]{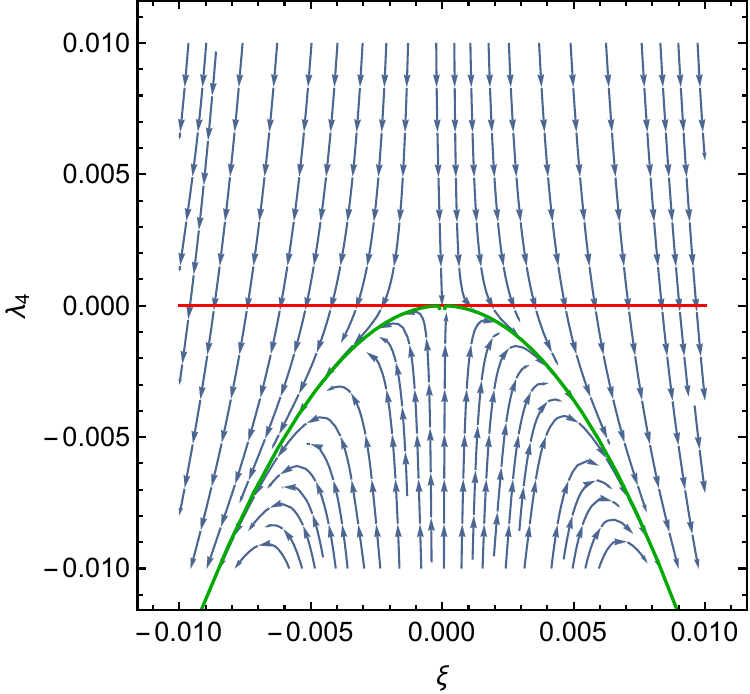}\\
\includegraphics[width=\linewidth]{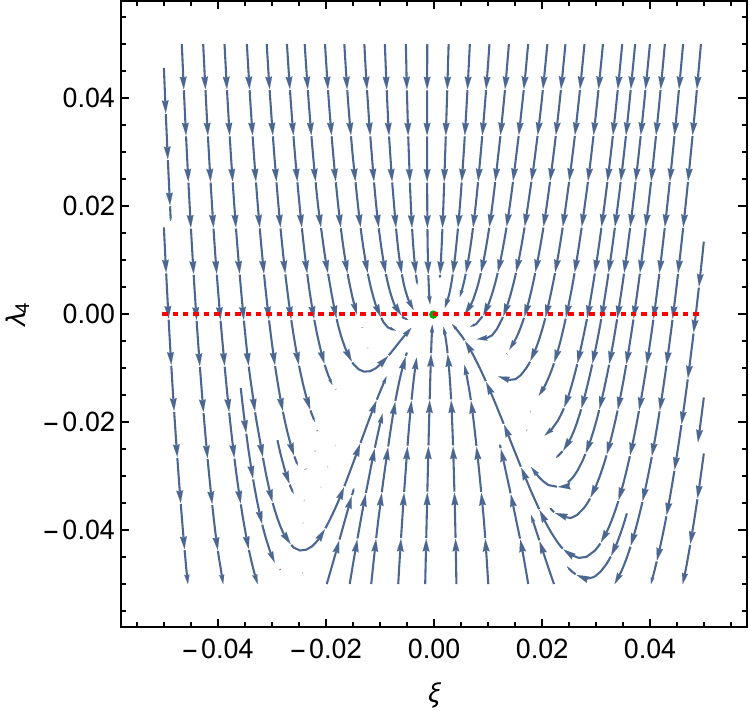}
\caption{\label{fig:xilambdaflow} We show the flow towards the IR in the $\xi-\lambda_4$ plane at $g=1$ for $ \lambda=0$ (upper panel) and $\lambda=-1$ (lower panel) and vanishing scalar mass. The thick green lines in the upper panel indicate the two fixed-point trajectories which result in negative $\lambda_4$ as $\xi$ grows only touch the relation $\lambda_4 \sim 10^{-9} \xi^2$, indicated in red dotted, in the fixed point. The green dot in the lower panel indicates the fixed point, which is IR attractive in the $\xi-\lambda_4$ plane. Accordingly, the phenomenological relation $\lambda_4 \sim 10^{-9} \xi^2$ cannot be satisfied in this case.}
\end{figure}

 There are two distinct regimes of the RG flow:
 \begin{enumerate}
 \item[a)] transplanckian, matter-gravity-driven flow away from the fixed point,
 \item[b)] subplanckian, purely matter-driven flow.
 \end{enumerate}
 For the regime a), there are two cases to distinguish: \\
 For a sufficiently large fixed-point value of the cosmological constant $\lambda$, the non-minimal coupling has overlap with a relevant direction, cf.~light blue region in Fig.~\ref{fig:fv_nr_relv_dir}. 
 As the upper panel in Fig.~\ref{fig:xilambdaflow} highlights, the fixed-point trajectory that reaches a finite $\xi$, does not safisfy the relation $ \lambda_4/\xi^2 \sim 10^{-9}$.  Instead, a linearization of the critical surface around the fixed point approximately yields $\lambda_4/\xi^2 \sim \mathcal{O}(-10^{-2})$.\\
 For sufficiently negative $\lambda$, there are two irrelevant, i.e., IR attractive directions that overlap with $\lambda_4$ and $\xi$, cf.~light gray region in Fig.~\ref{fig:fv_nr_relv_dir}. In this case, the values of $\lambda_4$ and $\xi$ remain at zero, i.e., quantum-gravity fluctuations drive $\lambda_4$ and $\xi$ back to zero, even if they deviate from it, cf.~lower panel in Fig.~\ref{fig:xilambdaflow}.

\begin{figure}
  \includegraphics[width=\linewidth]{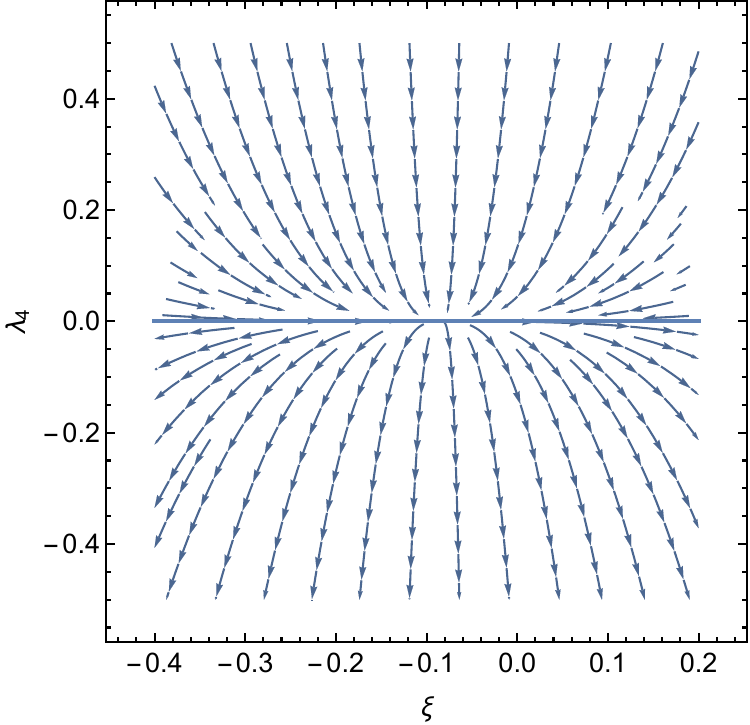}\\
  \caption{\label{fig:xilambdaflowwograv} We show the flow towards the IR in the $\xi-\lambda_4$ plane at $g=0$ for  vanishing scalar mass. On the $\lambda_4=0$ line, the flow vanishes identically.}
 \end{figure}

 In both cases, this transplanckian, matter-gravity-driven flow is followed by the second regime b). In this matter-driven regime, there is a fixed line at finite $\xi$ and vanishing $\lambda_4$. It is a consequence of the fact that at vanishing $\lambda_4$, the beta function for $\xi$ vanishes identically, as
 \be
 \beta_{\xi}\Big|_{\rm no\, gravity}= \frac{\lambda_4}{64\pi^2} (1 + 12\xi).
 \ee
 The fixed line itself does not satisfy the phenomenological relation $\lambda_4 \sim 10^{-9} \xi^2$, as it lies at vanishing $\lambda_4$.  The flow away from this fixed line leads to negative $\lambda_4$, cf.~Fig.~\ref{fig:xilambdaflowwograv}, whereas the fixed line is IR attractive from the regime of positive $\lambda_4$. While no conclusions about global stability can be drawn from a polynomial expansion of the potential around vanishing field, a destabilization around the origin is indicated by such a flow. \\

 Overall, we conclude that the phenomenological relation $\lambda_4 \sim 10^{-9} \xi^2$ that results in an amplitude of scalar perturbations in agreement with observations appears to be difficult to reconcile with an asymptotically safe UV completion with gravity within our truncation. The limitations of our study, lying mainly in its Euclidean nature and choice of truncation, should be kept in mind when interpreting this result.   

Having highlighted the potential difficulties with a potential which features a quartic coupling, let us instead consider the case of a purely quadratic potential, $\lambda_4 = 0$.  The two remaining parameters need to take on values $m^2 \sim \mathcal{O}(10^{-11})$ and $\xi \sim \mathcal{O}(10^{-3})$ in order to be compatible with observations \cite{Tenkanen:2017jih}. However, these  values are incompatible within our set of beta functions: the non-vanishing value of $\xi$ induces a value of $\lambda_4$ large enough to jeopardize the shape of the potential, as Fig.~\ref{fig:xilambdaflowwograv} highlights that one cannot flow to finite $\xi$ while remaining at $\lambda_4=0$, at least in the presence of quantum-gravity fluctuations.

We tentatively conclude that the situation does not appear promising for a ``vanilla" model of inflation with a single scalar field. In general it appears to be challenging to reconcile the potential predictive power of asymptotic safety with a given value of the amplitude of primordial scalar perturbations.
This situation might change in two-field models, and, as emphasized before, our tentative conclusions should be interpreted within the limitations of our study.  Nevertheless, these results might provide further motivation to instead focus on inflation driven by the $R^2$ coupling within asymptotic safety \cite{Weinberg:2009wa,Bonanno:2012jy,Copeland:2013vva,Bonanno:2015fga,Bonanno:2018gck}. 

\subsubsection{Towards asymptotically safe consequences for dynamical dark energy}
  \emph{...where we explore potential asymptotically safe consequences for dark energy; highlighting the role of the cosmological constant and emphasizing the possibility that quantum gravity fluctuations could drive scalar potentials towards flatness.}\\

 A scalar might be responsible for the accelerated expansion of the late universe, see, e.g., \cite{Wetterich:1987fk,Wetterich:1987fm}. Scale symmetry is a valuable guiding principle to construct viable dark-energy models \cite{Shaposhnikov:2008xb,Blas:2011ac,GarciaBellido:2011de,Bezrukov:2012hx,Rubio:2014wta,Casas:2017wjh,Casas:2018fum}. Potential implications of quantum scale symmetry, i.e., asymptotic safety, for dynamical dark energy have been explored, e.g., in \cite{Wetterich:2014gaa,Henz:2016aoh,Rubio:2017gty,Anagnostopoulos:2018jdq,Wetterich:2019qzx}.
 The notion that different scales in physics decouple could be taken to suggest that the late-time acceleration of the observable universe is completely decoupled from  the physics of quantum gravity. Various quantum-gravity approaches have contested that statement.  For instance, the everpresent-Lambda proposal \cite{Ahmed:2002mj}, rooted in causal sets \cite{Bombelli:1987aa,Surya:2019ndm}, is a proposal that has been related to observations \cite{Zwane:2017xbg} and the swampland conjectures, see \cite{Palti:2019pca} for a review, in string theory and their implications for late-time cosmology are a topic of intense discussions \cite{Obied:2018sgi,Agrawal:2018own,Garg:2018reu} with the potential to put observational pressure \cite{Abell:2009aa,Amendola:2012ys,Aghamousa:2016zmz,Bacon:2018dui} on the underlying theory \cite{Heisenberg:2018yae}. 
 The swampland conjectures, if realized, would exclude a cosmological constant by constraining the derivative of scalar potentials to be larger than an $\mathcal{O}(1)$ constant divided by an appropriate power of  the Planck scale \cite{Obied:2018sgi,Agrawal:2018own}. 
 This indicates a preference for steeper potentials.\\
 The situation appears to be somewhat different in asymptotic safety, judging by results within truncations of the full RG flow: Firstly, a positive cosmological constant of the correct size to agree with observations can be reached along RG trajectories emanating from the asymptotically safe fixed point in truncations \cite{Reuter:2004nx,Gubitosi:2018gsl}. Secondly, scalar potentials tend to become flattened under the impact of asymptotically safe quantum gravity, as discussed above, see Sec.~\ref{sec:QGflattens}. The fixed-point potential for a scalar that is coupled only to gravity and not to any other fields is flat within our truncation; furthermore the non-minimal coupling vanishes. The shape of the full effective potential at $k\approx 0$, which encodes the physics of asymptotic safety, depends on the number of relevant directions associated to this fixed point: 
 Depending on the strength of the gravitational effects, all couplings associated to the potential as well as the non-minimal coupling are irrelevant, cf.~Fig.~\ref{fig:fv_nr_relv_dir}. In this region (the dark grey region in Fig.~\ref{fig:fv_nr_relv_dir}), the potential remains flat at all scales, and the full effective potential is constant. This would imply a cosmological behavior for the scalar dominated by a cosmological constant.\\
 In a second region, the light grey region in Fig.~\ref{fig:fv_nr_relv_dir}, which contains the gravitational fixed point under the impact of minimally coupled matter \cite{Dona:2013qba} in our truncation, only the scalar mass is relevant, allowing a quadratic effective potential.\\
 In a third region, the light blue region in Fig.~\ref{fig:fv_nr_relv_dir}, in which gravitational fixed-point values tend to lie when the impact of matter is not accounted for, the mass and the non-minimal coupling are relevant; therefore a quadratic effective potential together with a nonvanishing non-minimal coupling can be achieved at $k\approx 0$.\\
  Therefore we tentatively conclude that the asymptotic-safety scenario could a priori be compatible with different effective potentials for dark energy, depending on  the gravitational fixed-point values. In all cases, a flat effective potential is compatible with the fixed-point scenario within our truncation, together with a small cosmological constant. Therefore, deviations from $w=-1$, where $w$ is the equation-of-state-parameter, can potentially be accommodated by using relevant directions of the fixed point. Nevertheless, $w=-1$ appears to be preferred in that a cosmological constant is always present in asymptotic safety. In all regions of the gravitational parameter space in our truncation, deviations from $w=-1$, linked to various terms in the scalar effective potential, are actually constrained due to the predictive power of the asymptotically safe fixed point.
 Again, we highlight that these conclusions regarding phenomenology should be viewed in light of the Euclidean signature of our studies.

\begin{figure}
    \includegraphics[width=\linewidth]{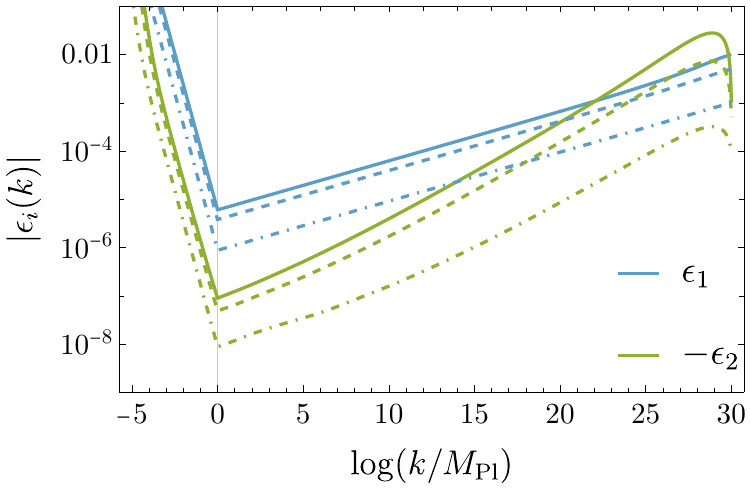}\\
    \caption{\label{fig:slow_roll_flow} We show the flow towards the IR for $\epsilon_1, \epsilon_2$ for various starting values at trans-planckian scales at $\phi/k=1$. The solid/dashed/dashed-dotted lines correspond to different initial conditions in the UV. The values of the gravitational couplings $g = 1.5$ and $\lambda=0.13$ lie in the dark grey region in Fig.~\ref{fig:slow_roll_flow_epseta}, i.e., region a). They are held constant here for simplicity.
    As long as gravitational contributions are active, the flow heads towards flatter scalar potentials. 
    For large $\epsilon_i$, our truncation of setting $\epsilon_i=0, i \geq 3$, is expected to break down. 
    }
  \end{figure}

Besides a fixed-point scenario, we can also use our calculation of the gravitational effects in an effective-field-theory setting, where a finite UV cutoff $\Lambda_{\rm UV}$ exists at a transplanckian scale. Such an "effective-asymptotic safety" scenario, cf.~Sec.~\ref{sec:effAS}, has been discussed, e.g., in \cite{Percacci:2010af,deAlwis:2019aud}. At $\Lambda_{\rm UV}$, the underlying microscopic theory provides the initial conditions for the RG flow. In the range of scales between $\Lambda_{\rm UV}$ and $M_{\rm Pl}$, gravitational fluctuations drive the flow. Fig.~\ref{fig:slow_roll_flow} showcases how gravitational fluctuations drive the slow-roll-inspired parameters $\epsilon_1$ and $\epsilon_2$ (cf.~Sec.~\ref{sec:slow_roll}) to zero, as soon as the gravitational couplings $g$ and $\lambda$ fall into the dark grey region in Fig.~\ref{fig:fv_nr_relv_dir}. Below the Planck scale, where gravitational fluctuations decouple dynamically, scalar fluctuations then drive the flow of $\epsilon_1$ and $\epsilon_2$ towards larger values again. This part of the flow is also present in a theory in which $\Lambda_{\rm UV} = M_{\rm Pl}$, i.e., where quantum gravitational fluctuations do not impact the scalar potential. We observe that the larger the ratio $\Lambda_{\rm UV}/M_{\rm Pl}$, the smaller the $\epsilon_i$ at $M_{\rm Pl}$ and thus in the IR, ultimately resulting in a flatter effective potential. 
Therefore a scenario as proposed in \cite{deAlwis:2019aud} might have the added consequence of flattening the potential for dynamical dark energy coming out of string theory. It would be interesting to explore this quantitatively and compare the ratio $V'/V$ from string theory with and without an intermediate asymptotically safe QFT regime with gravitational fluctuations.

\section{Simple Yukawa model for the Higgs-top system}
\label{sec:yukawa_model}
\emph{...where we focus on a simple Yukawa system as a toy model for the Higgs-top sector under the impact of quantum gravity. We investigate the stability of previous results, indicating a potential mechanism to fix the top-quark mass from asymptotic safety \cite{Eichhorn:2017ylw}, under the inclusion of the non-minimal coupling. Further, we explore whether within the present truncation, Higgs inflation could be compatible with asymptotic safety.}\\

 As a step towards a more SM-like scalar sector, we include a single Dirac fermion that captures the qualitative effects of the top quark. This amounts to studying the flowing action
\be
  \Gamma_k = \Gamma_k^\text{scal+grav} + \Gamma_k^\text{ferm} + S_{\rm gf} + S_{\rm gh}.
\ee

Herein, $\Gamma_k^{\rm scal+grav}$ stands for the single-scalar-gravity action in Eq.~\eqref{eqn:ansatz} and $\Gamma_k^{\rm ferm}$ is given by Eq.~\eqref{eqn:action_fermion}.  
  The above model neglects the chiral  SU(2) structure that is present in the coupling of the top and bottom quark to the Higgs, resulting in numerical differences of various factors in the beta functions. Nevertheless, as a precursor to a more realistic chiral Higgs-top-bottom model as analyzed with the FRG in \cite{Gies:2014xha} and under the impact of quantum gravity in \cite{Eichhorn:2017ylw,Eichhorn:2018whv,Alkofer:2020vtb}, the model with a single real scalar and a Dirac fermion can already provide valuable insights into potential asymptotic-safety mechanisms.

  Various aspects of the interplay of quantum gravity with fermions have previously been studied in \cite{Eichhorn:2011pc, Dona:2012am,Gies:2013noa,Meibohm:2016mkp,Eichhorn:2016vvy,Eichhorn:2018nda,deBrito:2019epw,Daas:2020dyo,Kowalska:2020gie}. Yukawa systems under the impact of asymptotically safe gravity have been explored in \cite{Zanusso:2009bs,Vacca:2010mj,Oda:2015sma,Eichhorn:2016esv,Hamada:2017rvn,Eichhorn:2017eht,Eichhorn:2017ylw,Eichhorn:2018whv,Held:2019vmi,Alkofer:2020vtb}. In \cite{Eichhorn:2017eht}, it was first proposed that asymptotic safety might allow to calculate the top quark mass; the proposal was extended to a chiral top-bottom system in \cite{Eichhorn:2018whv} and to three generations of quarks with CKM mixing in \cite{Alkofer:2020vtb}.
\subsection{Effects of Fermions}\label{sec:effofferm}
 \emph{...where we provide additional contributions to beta functions in the literature arising from the non-minimal coupling and focus on the number of relevant directions in the simple Yukawa system. This provides an indication for the predictivity that quantum gravity generates in this matter system.}\\

 To compute fermion contributions, we apply similar techniques as before: 
 we evaluate traces either in momentum space or using explicit summation over the spectrum, depending on our interest in the curvature dependence of the result. For the relevant expressions regarding the evaluation with spectral methods we again refer to \cite{Benedetti:2012dx,Alkofer:2018fxj} for the corresponding expressions, while stressing the importance to use an appropriate cutoff term for the fermions, as discussed in \cite{Dona:2012am}. In particular, for the fermionic fluctuations, the regulator is chosen as a function of the Dirac operator, cf. Eq.~\eqref{eqn:reg_fermion}.

 A closed fermion loop contributes
 \be
  \label{eqn:flow_potential_fermion}
  \Delta \partial_t \left(V(\phi)/k^4 \right)= \frac{-1}{8\pi^2(1+y^2 \phi^2/k^2)},
 \ee
 to the flow of the potential $V(\phi)$ and a factor
 \be
  \label{eqn:flow_non-minimal_fermion}
  \Delta\partial_t \left(\frac{1}{16\pi g} + \xi \phi^2/k^2 \right) = \frac{-1}{48\pi^2(1+y^2 \phi^2/k^2)},
 \ee
 to the flow of the prefactor of the Ricci scalar, respectively. Herein, the notation signifies that the full flow of these quantities is given by the previous result from the single-scalar section, plus the contribution \eqref{eqn:flow_potential_fermion}, or \eqref{eqn:flow_non-minimal_fermion}, respectively.
 These results agree with the ones reported in \cite{Hamada:2017rvn,Wetterich:2019zdo}.  Additionally, the Yukawa coupling contributes to the scalar anomalous dimension via a fermionic loop, 
 as well as to the fermion anomalous dimension via a scalar sunset-type diagram. 
 We adopt the expressions for these universal contributions from Ref.~\cite{Eichhorn:2017eht}.

\begin{figure}
  \begin{tikzpicture}[scale=3.0]
    \coordinate (in) at (0.0, 0);
    \coordinate (l) at (0.25,0);
    \coordinate (center) at (0.5,0);
    \coordinate (rup) at (0.68,0.18);
    \coordinate (rdown) at (0.68,-0.18);
    \coordinate (rupout) at (0.86, 0.36);
    \coordinate (rdownout) at (0.86, -0.36);
    \coordinate (reg) at (0.375, 0.125);
    \begin{scope}[on background layer]
      \draw [scalar] (in) -- (l);
      \draw [scalar] (l) -- (center);
      \draw [graviton] (l) arc(180:90:0.125) -- (reg);
      \draw [graviton] (reg) arc(90:0:0.125) -- (center);
      \draw [fermion] (rdownout) -- (rdown);
      \draw [fermion] (rdown) -- (center);
      \draw [fermion] (center) -- (rup);
      \draw [fermion] (rup) -- (rupout);
      \node [regulator] at (reg) {};
    \end{scope}
  \end{tikzpicture}
  \qquad
  \begin{tikzpicture}[scale=3.0]
    \coordinate (in) at (0.0, 0);
    \coordinate (l) at (0.25,0);
    \coordinate (center) at (0.5,0);
    \coordinate (rup) at (0.68,0.18);
    \coordinate (rdown) at (0.68,-0.18);
    \coordinate (rupout) at (0.86, 0.36);
    \coordinate (rdownout) at (0.86, -0.36);
    \coordinate (reg) at (0.404, 0.231);
    \begin{scope}[on background layer]
      \draw [scalar] (in) -- (l);
      \draw [scalar] (l) -- (center);
      \draw [graviton] (l) arc(180:112.5:0.25) -- (reg);
      \draw [graviton] (reg) arc(112.5:45:0.25) -- (rup);
      \draw [fermion] (rdownout) -- (rdown);
      \draw [fermion] (rdown) -- (center);
      \draw [fermion] (center) -- (rup);
      \draw [fermion] (rup) -- (rupout);
      \node [regulator] at (reg) {};
    \end{scope}
  \end{tikzpicture}
  \caption{\label{fig:feynman_diags_yukawa} Diagrams that contain contributions from the non-minimal coupling, contributing to the Yukawa coupling. Curly lines represent gravitons, dashed lines represent scalars, and the crossed circle represents a regulator insertion.}
 \end{figure}
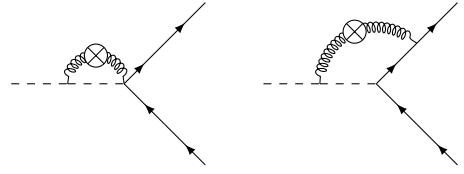

 In our truncation, the Yukawa beta function at vanishing scalar mass $m^2=0$ takes the form
 \be
  \beta_y = \frac{5}{16\pi^2} y^3 
  - f_y\, y,
  \ee
 with $f_y$ capturing the effect of gravitational contributions and
 \be
  f_y = f_y^{\rm min} + \Delta f_y^{(1)} + \Delta f_y^{(2)}.
 \ee
 Herein $f_y^{\rm min}$ agrees with \cite{Eichhorn:2017eht}, given by
 \be 
 f_y^{\rm min}=-\frac{g(96-235\lambda+103\lambda^2+56\lambda^3)}{12\pi(3-4\lambda)^2(1-2\lambda)^2}.
 \ee
 Additionally, 
 the two diagrams depicted in Fig.~\ref{fig:feynman_diags_yukawa} contain parts proportional to the non-minimal coupling.
 The corresponding contributions to $f_y$ read
 \be
  \Delta f_y^{(1)} =-g \frac{24 (1 - \frac{\eta_h}{8}) \xi}{\pi(3-4 \lambda )^2} - g \frac{8 (1-\frac{\eta_\phi}{8}) \xi}{\pi  (3-4 \lambda)},
 \ee
 and 
 \bea
  \Delta f_y^{(2)} &=& g \frac{108(1-\frac{\eta_h}{9})\xi}{7\pi (3-4\lambda)^2}+g \frac{36(1-\frac{\eta_\phi}{9})\xi}{7\pi(3-4\lambda)} \nonumber \\
  &+&g\frac{18 (1-\frac{\eta_\psi}{8})\xi}{7\pi (3-4\lambda)}.
 \eea
All corresponding expressions including the scalar mass are lengthier and can be found in App.~\ref{app:beta_functions}.

 For $f_y<0$, the only fixed point for the Yukawa coupling at finite gravitational fixed-point values is the free one, $y_\ast=0$. Indeed, $f_y <0$ is realized for the values of the gravitational couplings  in the vicinity of 
  the fixed point \eqref{eq:FPpurescalar}.
  Therefore, this fixed point persists in the larger theory space extended by an additional fermion, and enforces
   $y_\ast=0$, see also \cite{Eichhorn:2016esv}. The gravitational coupling values at that fixed point are
   \be
   \label{eqn:fp_grav_with_yukawa}
    g_\ast = 3.2 \quad \lambda_\ast = -0.146
   .\ee
   The critical exponents of the combined system are
 \begin{align}
  \theta^{(1)}&=2.92 & \theta^{(2)}&=0.728 \nonumber \\
  \theta^{(3/4)}&=0.0921\pm 1.58i & \theta^{(5)}&=-0.524 \nonumber \\
  \theta^{(6)}&= -1.94. &
  \end{align}
  The first two critical exponents are associated with the gravitational couplings, the next two relevant directions are associated with the mass parameter and the non-minimal coupling, the last two with the quartic coupling and the Yukawa coupling. The small changes compared to the values in \eqref{eq:FPpurescalar} arise due to the impact of the additional Dirac fermion.
 \\

\begin{figure}
\includegraphics[width=\linewidth]{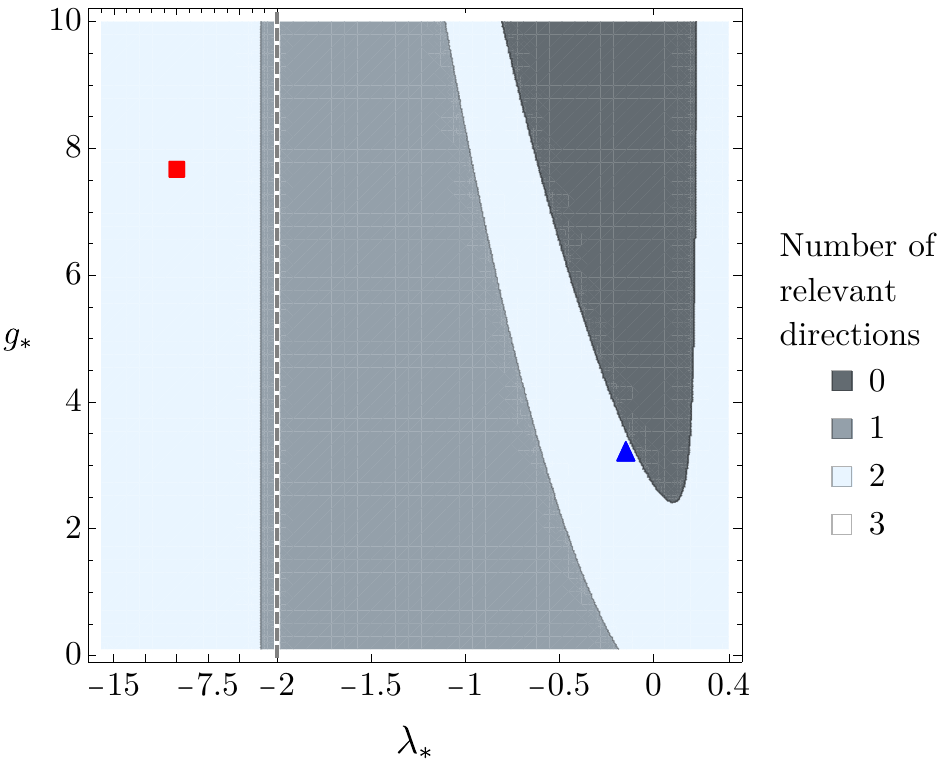}
\caption{\label{fig:glambda_GFP_rel_with_yukawa}We show the number of relevant directions in the $y, \lambda_4, m^2, \xi$-subspace as a function of the gravitational couplings $g, \lambda$ at the fixed point with vanishing matter couplings. There are no relevant directions in the dark gray area, one relevant direction in the light gray area, and two relevant directions in both light blue area. In the lighter blue area at negative $\lambda$, the Yukawa coupling is relevant at the GFP, indicating the existence of an interacting fixed point at which it is irrelevant, as in \cite{Eichhorn:2017ylw}. The dashed gray line at $\lambda_\ast=-2$ marks a difference in the scaling of the $\lambda_\ast$-axis. The blue triangle/red square indicate the fixed points \eqref{eqn:fp_grav_with_yukawa}/\eqref{eqn:fp_grav_canonical}, i.e., without/with the effect of minimally coupled SM fields.}
\end{figure}

 Varying the gravitational fixed-point values, for moderately negative values of $-3.3 < \lambda_\ast < -0.6$ (at $g_\ast$ given by \eqref{eqn:fp_grav_canonical}) the Gau\ss{}ian matter fixed point exhibits one relevant direction associated with the scalar mass, implying a vanishing Yukawa coupling at all scales, cf. Fig.~\ref{fig:glambda_GFP_rel_with_yukawa}. This is challenging to reconcile with the presence of finite fermion masses, generated through the Higgs mechanism, in the SM, cf.~\cite{Eichhorn:2016esv,Eichhorn:2017eht,Held:2019vmi}. \\
 For $\lambda_\ast \lesssim -3.3$, the gravity-induced anomalous dimension of the Yukawa coupling becomes  positive, $f_y  > 0$.  
 The gravitational fixed-point values incorporating all SM degrees of freedom \eqref{eqn:fp_grav_canonical} fall into this regime in the truncation studied in \cite{Dona:2013qba}.
 The Gau\ss{}ian fixed point features  two relevant directions for the matter couplings, associated with the scalar mass parameter and  a relevant direction in the canonically marginal couplings. The fixed point has  critical exponents 
 \begin{align}
  \theta^{(1)}&=3.98 & \theta^{(2)}&=1.97 \nonumber \\
  \theta^{(3)}&=1.92 & \theta^{(4)}&=0.0107 \nonumber \\ 
  \theta^{(5)}&=-0.0185 & \theta^{(6)}&=-0.038
  \label{eq:critexpGFPYuk}
 .\end{align}
 At the same time, a new fixed point emerges that will be the focus of the next subsection.

\subsection{Comparison to the case without non-minimal coupling: Impact on predictions of the fermion mass}\label{sec:topmass}
\emph{...where we investigate the robustness of results from \cite{Eichhorn:2017ylw} that indicate that (an upper bound on) the top quark mass might be calculable from asymptotic safety, and study the impact of the non-minimal coupling on the fermion-mass prediction.}\\

For $\lambda_\ast < -3.3$ and correspondingly $f_y > 0$, and in the approximation where the non-minimal coupling is set to zero, $\xi = 0$, two fixed points at $y_{\ast}>0$ exist in addition to the fixed point at $y_{\ast}=0$, see Eq.~\eqref{eq:critexpGFPYuk}. One of these is characterized by $\lambda_{4\,\ast}<0$, the other by $\lambda_{4\,\ast} > 0$.
 For values of the gravitational couplings given by Eq.~\eqref{eqn:fp_grav_canonical}, the former is located at 
 \be
  \label{eqn:fp_without_xi_neg_lam4}
  \lambda_{4\,\ast} = -1.15, \quad y_\ast = 0.58,
 \ee
 whereas the latter is located at 
 \be
  \label{eqn:fp_without_xi}
  \lambda_{4\,\ast} = 0.18, \quad y_\ast = 0.58.
 \ee
 As first observed in \cite{Eichhorn:2016esv} and generalized in \cite{Eichhorn:2017ylw,Eichhorn:2018whv,Held:2019vmi,Alkofer:2020vtb} to a realistic quark sector, the latter fixed point is infrared attractive in both these couplings, thus also generalizing the scenario proposed in \cite{Shaposhnikov:2009pv}. Naturally, the interacting nature of the fixed point implies a mixing of couplings in the relevant/irrelevant directions, yet, as the fixed point is near-perturbative, this mixing is small. 
 Consequently, initializing the model at the fixed point  in Eq.~\eqref{eqn:fp_without_xi} in the UV, the value of the Yukawa coupling is fixed at all energy scales and its ``sibling" in a top-bottom-Higgs system allows to calculate the top mass and bottom mass as a consequence of asymptotic safety \cite{Eichhorn:2017ylw,Eichhorn:2018whv,Alkofer:2020vtb}.  The difference in the fixed-point values in Eq.~\eqref{eqn:fp_without_xi} compared to the ones reported in \cite{Eichhorn:2017ylw} arises, as we work in a simple Yukawa system here instead of a chiral Yukawa system, and our fixed-point values of the gravitational couplings differ. The shift in the gravitational fixed-point values is due to a different treatment of gravitational fluctuations; the gravitational beta functions used in \cite{Eichhorn:2017ylw} and first computed in \cite{Eichhorn:2016vvy} are obtained (a) without decomposing the graviton according to \eqref{eqn:york_decomposition} and (b) employing a regulator function that features $\eval{\Gamma_k^{(2)}}_{\lambda=0}$ as its argument. On the one hand the shift in fixed-point values highlights the uncertainty in the determination of fixed-point values, cf.~the estimated lower bound on the systematic uncertainty in \cite{Eichhorn:2017ylw}, emphasizing the need to study the full gravitational parameter space. On the other hand, the stability of the qualitative features of the interacting fixed point is encouraging.

As discussed in the previous Sec.~\ref{sec:effofferm}, in addition to the interacting fixed points, the system also features a fixed point at which all matter couplings vanish. The interplay of these fixed points results in an upper bound on the fermion mass. This upper bound is the value that is reached along the unique trajectory connected to the interacting fixed point, see \cite{Eichhorn:2017ylw}.

 In the following, we will study the fate of both fixed points in Eq.~\eqref{eqn:fp_without_xi_neg_lam4} and \eqref{eqn:fp_without_xi}  under the inclusion of the non-minimal coupling $\xi$. 

\begin{figure}[!t]
 \includegraphics[width=\linewidth]{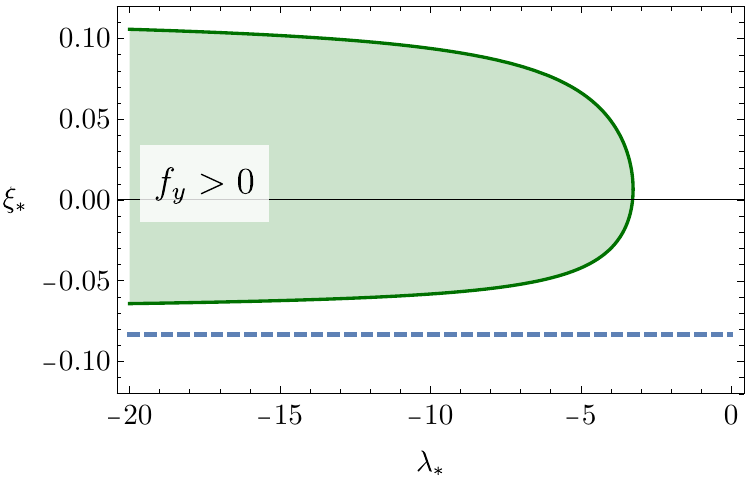}
 \caption{\label{fig:xiinterval} 
 Region of positive $f_y$ marked in green at vanishing mass $m^2=0$. The dashed line marks the conformal fixed point $\xi=-1/12$. The region is independent of the value of $g$.}
 \end{figure}

 To qualitatively understand the effect of a non-minimal coupling on these fixed points, we first solve the $\beta$-functions for the Yukawa and the quartic coupling with the non-minimal coupling $\xi$ as a free parameter.  The gravity-induced anomalous dimension $f_y$ depends on the non-minimal coupling $\xi$, and  for gravitational coupling values given by \eqref{eqn:fp_grav_canonical}, $f_y$ is positive only for
 \be
  \label{eqn:window_xi}
  \xi \in [-0.06, 0.09],
 \ee
 for more general values of $\lambda$, see Fig.~\ref{fig:xiinterval}.

 Let us now compare this interval to the actually realized fixed-point value for $\xi$.
 Crucially, a non-vanishing Yukawa coupling breaks shift symmetry and therefore causes a nonvanishing non-minimal coupling.
 At order $\mathcal{O}(\xi)$, which suffices for an analysis within the above window, the $\beta$-function of the non-minimal coupling at vanishing mass parameter, $m^2=0$, is given by
 \bea
  \label{eqn:beta_xi}
  \beta_\xi &=& \left(\xi + \frac{1}{12}\right) \left( \frac{y^2}{4\pi^2}+\frac{3\lambda_4}{16\pi^2} \right) \nonumber \\
  &+&g \,\xi\, \left(\frac{2(3-\lambda)}{3\pi(3-4\lambda)^2} - \frac{5(3+10\lambda)}{18\pi(1-2\lambda)^3} \right).
  \eea

\begin{figure}
\includegraphics[width=\linewidth]{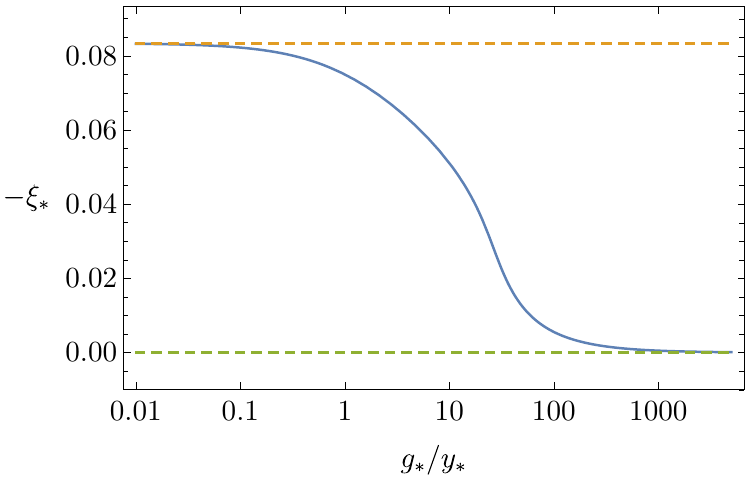}
\caption{\label{fig:zero_beta_xi}We show the value of $\xi_\ast$, leaving $y_\ast$ and $g_\ast$ as free parameters  with $\lambda_\ast$ given by \eqref{eqn:fp_grav_canonical}. The Yukawa and gravitational effects compete. For small gravitational effects the fixed-point value approaches $\xi_\ast = -1/12$, for large gravitational effects it approaches the free fixed point. }
\end{figure}

 A non-vanishing Yukawa and quartic coupling generate a fixed-point value $\xi_{\ast} = -1/12 < -0.06$ if $g=0$. 
 The direct gravitational contribution, i.e., the second line in Eq.~\eqref{eqn:beta_xi}, is therefore crucial. 
 The Yukawa and the gravitational contribution  compete in shifting the fixed point between the conformal value $\xi_\ast = -1/12$ and the free fixed point, as can be seen in Fig.~\ref{fig:zero_beta_xi}.
 The key question is, whether at the fixed point \eqref{eqn:fp_without_xi} the gravitational contributions are sufficiently strong for $\xi_\ast$ to fall into the interval in \eqref{eqn:window_xi}. Only then can the fixed point from \cite{Eichhorn:2017ylw,Eichhorn:2018whv,Alkofer:2020vtb} continue to exist in the larger truncation we study here.

The gravitational contribution in Eq.~\eqref{eqn:beta_xi} is positive for sufficiently negative $\lambda_\ast$. 
 At the fixed point \eqref{eqn:fp_without_xi_neg_lam4} with negative $\lambda_{4\,\ast}$, the coefficent of the term $(\xi + 1/12)$ is negative; the resulting would-be fixed point value for $\xi$ is positive and too large to fall into the interval \eqref{eqn:window_xi}. The inclusion of $\xi$ hence rules out this previous fixed-point candidate with negative $\lambda_{4\, \ast}$.\\
 In contrast, at the fixed point \eqref{eqn:fp_without_xi} with positive $\lambda_{4\, \ast}$, the gravitational effects are strong enough 
to shift $\xi_{\ast}$ into the interval in Eq.~\eqref{eqn:window_xi}. This fixed point, the generalization of which underlies \cite{Eichhorn:2017ylw,Eichhorn:2018whv,Alkofer:2020vtb} hence persists under the inclusion of the non-minimal coupling.\\
 It is rather nontrivial that the region of sufficiently negative $\lambda_{\ast}$ is actually the region which is already selected by requiring $f_y>0$. Now, we find that this is the region in which it appears to be a qualitatively viable approximation to neglect $\xi$ in the analysis of the fixed-point structure of Higgs-Yukawa systems.\\

\begin{figure*}
   \includegraphics[width=0.48\textwidth]{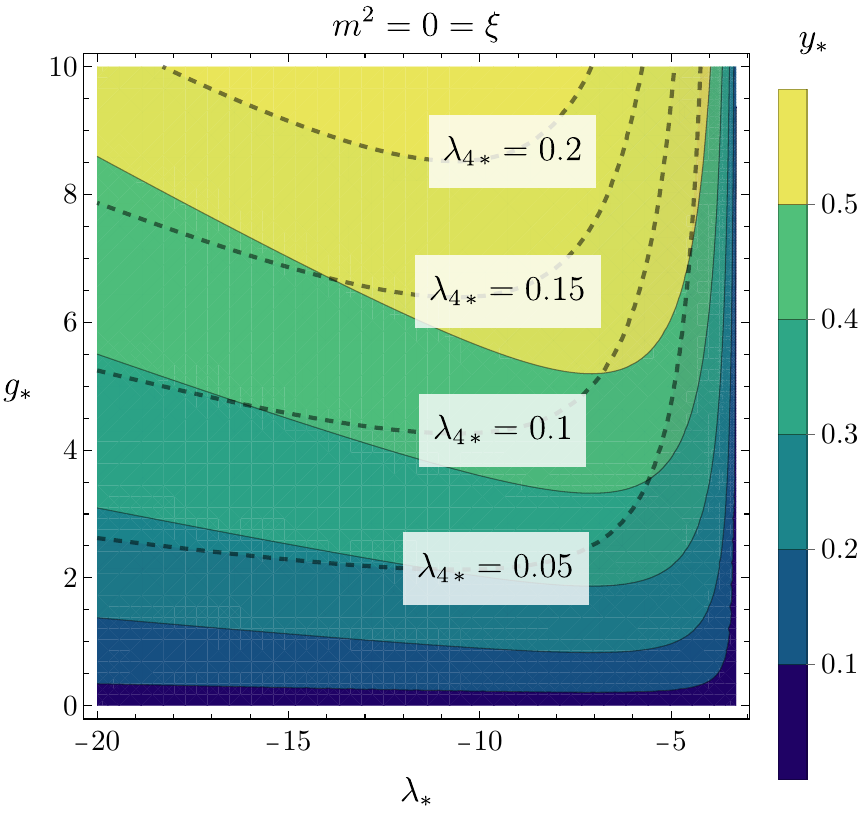}
   \includegraphics[width=0.48\textwidth]{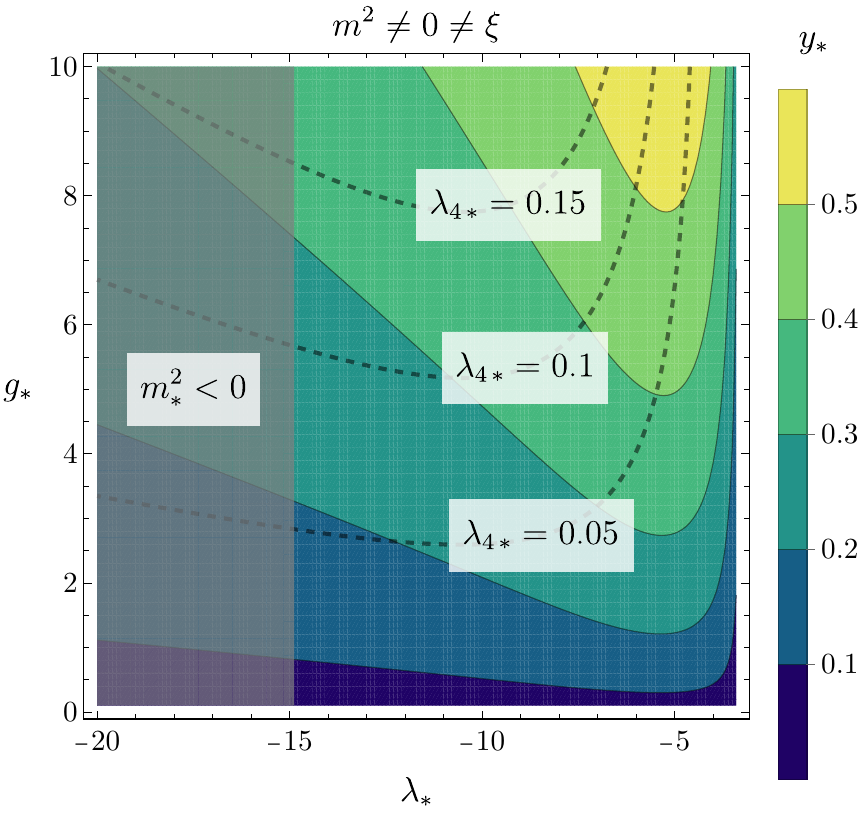}
   \caption{
   \label{fig:yukawa_contours}
 Fixed point-value of the Yukawa coupling $y_\ast$ (color) and the quartic coupling $\lambda_{4\, \ast}$ (dashed lines) as a function of the gravitational fixed-point values without (with) a non-minimal coupling on the left (right). 
   The left plot is obtained setting $m^2 = 0$, the right one considers the full mass dependence.
   }
 \end{figure*}

 In our full system of beta-functions, we find a fixed point at
 \begin{align}
   \lambda_\ast&= -9.49 & g_\ast &= 7.26 \nonumber \\
   \lambda_{4\,\ast} &= 0.139 & y_\ast &= 0.381 \nonumber \\
   \xi_\ast &= -0.04 &
   m^2_{\ast} &= 0.00086,
 \end{align}
 with three relevant directions and critical exponents
 \begin{align}
  \theta^{(1)} &= 3.98 & \theta^{(2)} &= 1.99 \nonumber \\
  \theta^{(3)} &= 1.91 & \theta^{(4)} &= -0.0364 \nonumber \\
  \theta^{(5/6)} &= -0.00855 \pm 0.0108 i
 .\end{align}
 In comparison to the fixed point \eqref{eqn:fp_without_xi}, both $\lambda_{4\, \ast}$ as well as $y_{\ast}$ are considerably lowered due to the presence of the non-minimal coupling $\xi$.
 This can also be seen in Fig.~\ref{fig:yukawa_contours} which shows contours of constant Yukawa coupling in the gravitational parameter space.
 Under the inclusion of the non-minimal coupling, the curves are shifted upwards considerably. 
  Assuming that a similar effect will hold in the full SM, this aids to realize the scenario presented in \cite{Eichhorn:2017ylw}. 
 Neglecting the non-minimal coupling $\xi$, in that reference the Higgs self-coupling fixed-point value was determined to be $\lambda_{4\,\ast} = 0.01$ at a point with a roughly realistic top mass, rendering the Higgs boson slightly too heavy. 
 Our results indicate that the inclusion of a non-minimal coupling could reduce this overestimation of the Higgs mass. \\
 As one of the nontrivial results of our paper, we highlight that the non-minimal coupling, which is canonically marginal, actually corresponds to an irrelevant coupling at the asymptotically safe fixed point. Therefore, its value is fixed at all scales as a unique function of the relevant couplings.

 Let us briefly mention that, depending on the value of the top mass, the Higgs potential in the SM might be metastable \cite{Degrassi:2012ry,Buttazzo:2013uya,Andreassen:2017rzq}. In that case, the presence of a non-minimal coupling can stabilize or destabilize the Higgs potential in the early universe, where the background curvature is large enough, see \cite{Herranen:2014cua,Herranen:2015ima,Rajantie:2016hkj,Joti:2017fwe,Figueroa:2017slm,Markkanen:2018pdo,Lalak:2020dao}. At present, it is not clear whether or not the SM Higgs potential remains stable when coupled to asymptotically safe gravity. To answer this question, the gauge degrees of freedom of the SM are important, as they affect the Higgs potential; therefore we leave the question of vacuum stability, as well as a comparison with the corresponding cosmological bounds on $\xi$, for future work. We highlight that if the fixed-point properties of the simple Yukawa system persist under this extension, the non-minimal coupling takes on a uniquely calculable value. Therefore, there might be a unique answer to the question whether asymptotic safety aids or prevents vacuum stability through the non-minimal coupling in the early universe.

\subsection{Spontaneous symmetry breaking in the quantum-gravity regime}
\emph{ ...where we explore the emergence of a symmetry-broken fixed-point potential in parts of the gravitational parameter space.}\\

\begin{figure}[!t]
\includegraphics[width=0.48\textwidth]{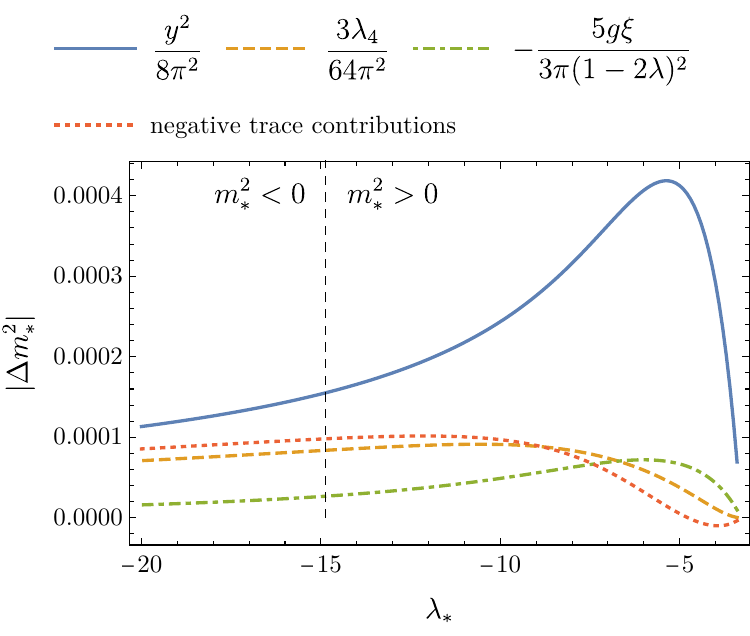}
\caption{\label{fig:UVSSB} The leading contributions to $\beta_{m^2}$ for $g=1$, cf.~Eq.~\eqref{eq:msqFP}. The dotted contribution stems from three diagrams that all include the gravitational trace mode, i.e., the three terms with $(3-4\lambda)$ in the denominator in Eq.~\eqref{eq:msqFP}. The dashed and dotted contributions appear with a negative sign in the beta function for the mass. The vertical dashed line marks the onset of the symmetry-broken regime.}
\end{figure}

Spontaneous symmetry breaking, in our case of the $\mathbb{Z}_2$ symmetry that maps $\phi$ to $-\phi$, is signalled by a negative value for the squared mass parameter, $m^2$. An asymptotically safe fixed point could a priori lie at any sign of $m^2$, i.e., in principle spontaneous symmetry breaking might already occur in the UV. As explained also in \cite{Eichhorn:2019dhg}, in the presence of a Yukawa coupling, the fixed-point value for the mass remains positive in the strong-gravity regime -- which lies at $\lambda \gtrsim 0$ in our truncation. Therefore, we focus on the weak-gravity regime, into which the gravitational fixed-point value is pushed under the impact of the SM degrees of freedom \cite{Dona:2013qba}. In this regime, fixed-point values for matter couplings become perturbatively small, and we can focus on a simple approximation of the beta function for the mass parameter. In the limit $\lambda \ll 0$, it can be approximated as
\bea
\beta_{m^2}&\approx&-2 m^2 + \frac{1}{4\pi^2}y^2 - \frac{3}{32\pi^2}\lambda_4 -\frac{18}{\pi(3-4\lambda)}g\, \xi^2 \nonumber \\
 &{}& - \frac{6 g \xi}{\pi (3-4\lambda)^2} - \frac{10 g \xi}{3\pi (1-2\lambda)^2} - \frac{54 g \xi^2}{\pi (3-4\lambda)^2} \nonumber\\
&{}& + \,\mathcal{O}(m^2 y, m^2 \lambda_4, m^2 \xi).
\eea
The resulting fixed-point value is
\bea
\label{eq:msqFP}
m^2_{\ast} &=& \frac{1}{8\pi^2}y^2 - \frac{3}{64\pi^2}\lambda_4 -\frac{9}{\pi(3-4\lambda)}g\, \xi^2 \\
&{}& - \frac{3 g \xi}{\pi (3-4\lambda)^2} - \frac{5 g \xi}{3\pi (1-2\lambda)^2} - \frac{27 g \xi^2}{\pi (3-4\lambda)^2} \nonumber
.\eea
In the absence of the non-minimal coupling, this results in a positive mass at all values of $g$ and $\lambda$, since the $\lambda_4$-contribution turns out to be subleading compared to the $y^2$-contribution. In the presence of a finite non-minimal coupling, the combined pure scalar and trace mode contributions $\sim \lambda_4$ and $\sim \frac{1}{(3-4\lambda)^{\#}}$ are negative (at negative $\lambda$) and start to dominate over the positive $y^2$-contribution for sufficiently negative $\lambda$, as shown in Fig.~\ref{fig:UVSSB}. Accordingly, the fixed point lies in the spontaneously symmetry- broken regime for $\lambda< -14.9$.\\
We leave the exploration of the resulting phenomenology and the investigation of the viability of this region to future work.

\subsection{Indications against the viability of Higgs inflation in asymptotic safety}\label{sec:noHiggsinflationinAS}
\emph{...where we explore whether a ratio of quartic coupling to non-minimal coupling as it would be required for Higgs inflation could be achieved  in our toy model. In particular, we investigate whether the regime of gravitational parameter space that appears preferred by Higgs inflation in our truncation is also the regime that could allow a realistically large value of the Yukawa coupling to accommodate the measured top quark mass.}\\

\begin{figure}
\includegraphics[width=\linewidth]{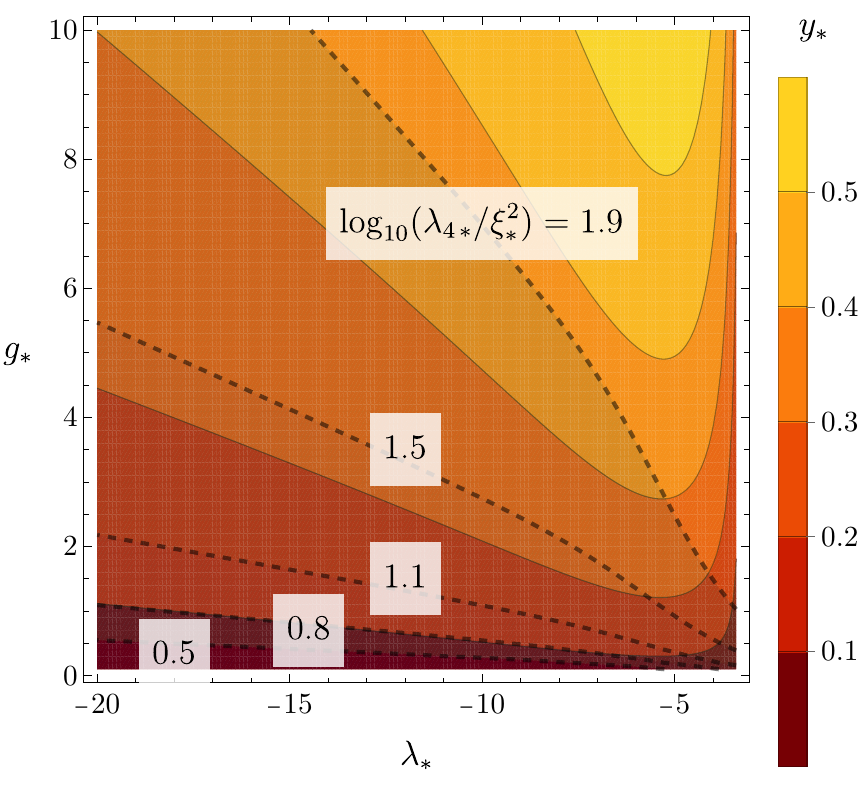}
\caption{\label{fig:xi_g_lambda} Fixed-point values for $\lambda_4/\xi^2$ (dashed lines) at the interacting fixed point as a function of $g_{\ast}$ and $\lambda_{\ast}$, together with contours showing the fixed-point value of $y$. The fixed-point value for $y$ translates into an upper bound on the fermion mass compatible with asymptotic safety. 
In order to achieve fixed-point values for $\lambda_4/\xi^2$ compatible with  those suggested by Higgs inflation, the lower bound on the Yukawa coupling generically has to be very small. Assuming that these results carry over to the SM Higgs-top sector, Higgs inflation would not be compatible with a viable top mass within asymptotic safety.}
\end{figure}

Higgs inflation is an attractive theoretical idea that ties together particle physics and cosmology and does not invoke any BSM physics to generate an inflationary regime \cite{Bezrukov:2007ep}, see \cite{Rubio:2018ogq} for a review. The standard questions regarding the status of the initial conditions \cite{Ijjas:2013vea} remain, but in contrast to inflationary settings with a new scalar degree of freedom as the inflaton, Higgs-inflation in principle only has one additional free parameter, namely the non-minimal coupling of the Higgs field to gravity. \\
Additional free parameters, linked to higher-order corrections in the Higgs potential, start to play a role if the Higgs potential is metastable, or close to metastability \cite{Bezrukov:2014ina,Bezrukov:2014ipa}.
 In principle, such corrections can be evaluated in an asymptotically safe setting. \\
 Here, we ask a simpler question, namely, whether the non-minimal coupling is available as a free parameter in this setting, and we work under the assumption that higher-order terms in the potential are negligible. We find that within our truncation, the non-minimal coupling is predicted by the asymptotic-safety requirement in that region of parameter space which could admit a finite Yukawa coupling in the IR, i.e., a finite top quark mass. In other words, within our truncation the non-minimal coupling is only available as a free parameter at the cost of fixing the top quark mass to zero. \\
We therefore focus on the region of parameter space with $\lambda<-3.3$. In this region, an upper bound on the fermion mass exists \cite{Eichhorn:2017ylw,Eichhorn:2018whv,Alkofer:2020vtb}, in addition to a prediction of the non-minimal coupling. 
The presence of the finite Yukawa coupling distinguishes the scenario considered here from the one discussed in Sec.~\ref{sec:phenosinglescalar} as well as in  \cite{Saltas:2015vsc}. \\
Within our toy model, defined by our choice of degrees of freedom and our truncation, asymptotic safety does not appear to be compatible with Higgs inflation. 
Unless $y_{\ast}$ is made very tiny, $y_\ast \lesssim 10^{-6}$,
 the ratio $\lambda_{4\,\ast}/\xi_\ast^2$ at the interacting fixed point is of $\mathcal{O}(1)$, cf.~Fig.~\ref{fig:xi_g_lambda}. 
 As all three couplings are irrelevant and only flow logarithmically towards the IR, it appears unlikely that this yields the observed amplitude of primordial scalar fluctuations, which would require $\lambda_4/\xi^2 \sim \mathcal{O}(10^{-9})$, cf.~Sec.~\ref{sec:phenosinglescalar}. This finding fits with with Ref.~\cite{Wetterich:2019rsn}, that found that based on the fixed point value for $\xi$ Higgs inflation is unlikely to be realized within asymptotic safety.\\
In conclusion, our results indicate that Higgs inflation might be hard to realize in asymptotic safety as it appears difficult to obtain the correct amplitude of primordial fluctuations while accommodating a viable top-quark-mass. Again, this result holds under an extrapolation from the Euclidean to the Lorentzian regime.

\section{Scalar portal to dark matter}
\label{sec:portal}
\emph{...where we focus on settings, where dark matter is coupled to the SM Higgs through a portal coupling. 
We study two coupled scalars as a toy model for the portal sector and additionally propose a setting with an additional dark fermion as the simplest dark-matter sector with an interacting fixed point in the portal coupling. We then explore to what extent asymptotic safety leads to an enhancement of the predictivity of the model over the effective-field-theory approach without gravity.}\\

The ongoing experimental searches for dark matter have not yielded a clear discovery yet \cite{Wagner:2010mi,Hooper:2010mq,Ahnen:2016qkx,Aartsen:2016pfc,Akerib:2016vxi,Amole:2017dex,Agnese:2017jvy,Cui:2017nnn,Agnes:2018fwg,Aprile:2018dbl,Abdelhameed:2019hmk,Leane:2019xiy,Aguilar-Arevalo:2019wdi,Braine:2019fqb,Arnaud:2020svb,Andrianavalomahefa:2020ucg,Aprile:2020tmw}.
While these searches have already excluded an impressive number of models and parameter combinations, an even larger space lies ahead of them. Novel theoretical ideas could change this situation.
 Classically, scale-invariance enhances the predictivity of Higgs portal models  \cite{AlexanderNunneley:2010nw,Khoze:2013uia,Holthausen:2013ota,Carone:2013wla,Gabrielli:2013hma,Hambye:2013dgv,Guo:2014bha,Karam:2015jta,Karam:2016rsz,Helmboldt:2016mpi,Ishida:2017ehu,Cosme:2018nly,YaserAyazi:2018lrv,Ishida:2019gri}. Quantum scale invariance \cite{Wetterich:2019qzx,Wetterich:2020cxq} additionally might allow to fix marginal couplings that are not constrained by classical scale invariance, including the portal coupling. 

In particular, a single, uncharged dark scalar remains an attractive model for dark matter, due to its conceptual simplicity as a thermal relic and due to its experimental accessibility, see \cite{Arcadi:2019lka} for a review. In \cite{Eichhorn:2017als}, it has been found that quantum-gravity fluctuations drive the Higgs portal coupling to zero, such that it vanishes at the Planck scale. If confirmed in extended theoretical studies, this has important phenomenological consequences, as it leads to the decoupling of this dark-matter candidate, and requires an extended dark sector to regenerate the portal coupling at \cite{Eichhorn:2020kca} or below \cite{Reichert:2019car,Hamada:2020vnf} the Planck scale.

In the first part of this section, \ref{sec:decouplingsingledm}, we work with two scalars coupled to gravity non-minimally and to each other through a portal coupling, described by the truncation
\be
\Gamma_k = \Gamma_k^{\rm 2-scal + grav} + \Gamma_k^{\rm portal} + S_\text{gf} + S_\text{gh}.
\ee
This provides a model for dark matter coupled to a visible scalar as a toy model for the SM Higgs, analyzed without the non-minimal couplings in \cite{Eichhorn:2017als}. We confirm that the decoupling-result from that paper persists under the inclusion of the non-minimal couplings in Sec.~\ref{sec:decouplingsingledm}. This motivates an extension of the dark sector in order to reconcile the asymptotic-safety paradigm with the idea that dark matter is a thermal relic. In contrast to \cite{Reichert:2019car,Hamada:2020vnf}, where new physics is added to regenerate the portal coupling below the Planck scale, we explore the setting in which an asymptotically safe fixed point lies at finite portal coupling, introduced in \cite{Eichhorn:2020kca}, see Sec.~\ref{sec:darkYukawa}. 
\subsection{Decoupling of a single dark scalar in asymptotic safety}\label{sec:decouplingsingledm}
 \emph{...where we confirm that  the result that the portal coupling between two purely scalar sectors vanishes at an asymptotically safe fixed point is robust under the inclusion of non-minimal couplings.}\\

Our first key result is that the inclusion of the non-minimal couplings does not alter the decoupling result from \cite{Eichhorn:2017als}, as $\lambda_{{\rm HP}\, \ast}=0$ and $\lambda_\text{HP}$ corresponds to an irrelevant direction. The corresponding beta function for the portal coupling can be found in App.~\ref{app:beta_functions}. 
Including the additional effect of 45 minimally coupled Weyl fermions, 12 gauge bosons, 3 minimally coupled scalars, as for the degrees of freedom of the Standard Model, we find a fixed point at
\bea
g_{\ast}=9.66, \, \lambda_{\ast}=-12.4, \label{eq:fpdarkmatter}
\eea
with all other couplings vanishing. The underlying reason for the flat scalar potential is shift symmetry in the two scalar sectors; the kinetic term for both scalars is invariant under $\phi \rightarrow \phi + C$, with $C$ a constant. Under the impact of gravity, that symmetry remains intact, cf. Sec.~\ref{subsubsec:globalsyms}.
The critical exponents read
\begin{align}
\theta^{(1)}&=3.99 & \theta^{(2/3)}&=1.97, & \theta^{(4)}&=1.92,\nonumber\\
\theta^{(5/6)}&=-0.0173, & \theta^{(7/8/9)}&=-0.0353.
\end{align}
Several of these come in identical pairs due to exchange symmetry between the visible and the dark scalar. The three smallest critical exponents show a three-fold degeneracy, as they are associated with the three quartic couplings.

The Higgs portal coupling is an irrelevant direction, just as in \cite{Eichhorn:2017als} and is protected by shift symmetry at all scales once set to zero at high scales. Thus, a departure from $\lambda_\text{HP}=0$ at low scales is incompatible with the fixed-point requirement at high scales.
Accordingly, the conclusions in \cite{Eichhorn:2017als} persist under the inclusion of non-minimal couplings; the dark scalar decouples from the visible one.\\

Within our truncation, asymptotic safety with gravity is therefore incompatible with the thermal production of a single dark scalar in a setting without additional fields in the dark sector, see \cite{Reichert:2019car,Hamada:2020vnf} for proposed extensions. \\
Whether alternative production mechanisms could be available, depends on the gravitational parameters: At the fixed point in Eq.~\eqref{eq:fpdarkmatter}, the dark-matter mass remains associated with a relevant direction and can therefore depart from its fixed-point value $m^2_{d\, \ast}=0$ to take any finite value in the infrared. For a massive, purely gravitationally coupled particle, gravitational particle production in an inflating universe might be viable \cite{Ema:2018ucl}, based on the classic results in \cite{Parker:1969au}. 
In particular, superheavy stable dark matter, minimally coupled to gravity, might be produced in the early universe \cite{Chung:1998zb}. The availability of such a mechanism also depends on the compatibility of the required inflationary potentials with asymptotic safety, see also Sec.~\ref{sec:ASinfl}.\\
Additionally, just as for a single scalar \cite{Eichhorn:2012va}, one expects momentum-dependent interactions between the dark scalar and the Higgs scalar to necessarily be present if an asymptotically safe fixed point exists, see also \cite{Eichhorn:2017eht}. These are present at high RG scales, but suppressed in the IR due to their canonical dimensionality. As discussed in \cite{Garny:2015sjg}, such planckian-interacting massive dark matter might be produced in the early universe.

\subsection{Portal coupling to a dark Yukawa system}\label{sec:darkYukawa}
\emph{...where we expand on results in \cite{Eichhorn:2020kca}, tentatively hinting at a highly predictive model of a scalar-fermion dark sector under the impact of quantum gravity, coupled through a portal to the visible Yukawa sector.}\\

 The results in the previous section, lending further support to the decoupling of a simple scalar as dark matter found in \cite{Eichhorn:2017als}, motivate us to ask whether thermal production of dark matter could be compatible with asymptotically safe gravity once the dark matter sector is extended. Thus,
 we investigate a model that illustrates how to generate a finite Higgs portal coupling at the Planck scale.
 As a consequence of asymptotic safety, we will find that this model features a very strongly enhanced predictive power compared to its leading-order effective-field-theory treatment.\\
  In our toy model, the first scalar $\phi_v$ again
 plays the role of the Higgs. The second scalar $\phi_d$ assumes the role of a dark scalar.
 To dynamically break shift symmetry in the dark sector, we additionally include a dark fermion. This generates a non-vanishing portal, as we discovered in \cite{Eichhorn:2020kca}.
 We hence work with two copies of the action from Sec.~\ref{sec:yukawa_model}, a dark and a visible one. 
 Additionally, we include a portal coupling in the effective action which reads
 \be
  \Gamma_k = \Gamma_k^{2\rm -scal+grav} + \Gamma_k^{\text{ferm}, v}+ \Gamma_k^{\text{ferm}, d} + \Gamma_k^\text{portal} + S_\text{gf} + S_\text{gh},
 \ee 
 where $\Gamma_k^{\rm 2-scal+grav}$ is the action for two non-minimally coupled scalars given in Eq.~\eqref{eqn:action_two_scalars}, $\Gamma_k^\text{portal}$ is the portal coupling given in Eq.~\eqref{eqn:action_portal}, and $\Gamma_k^{{\rm ferm}, v}$ ($\Gamma_k^{{\rm ferm}, d}$) are the fermion action \eqref{eqn:action_fermion} for a visible (dark) fermion $\psi_v$ ($\psi_d$) coupled to the visible (dark) scalar $\phi_v$($\phi_d$), respectively.  The beta functions for this system are given in App.~\ref{app:beta_functions}. 

 \subsection{Fixed-point structures, predictivity and near-perturbativity}
 \emph{...where we list all possible fixed-point structures of the model with a dark and a visible Yukawa sector, explore their predictivity, and study whether the most predictive fixed point is near-perturbative in nature.}\\

 The action features a $\mathbb{Z}_2$ exchange symmetry between dark and visible sector. This exchange symmetry is also reflected in the fixed-point structure of the system and leads to degeneracies in the spectrum of critical exponents at the interacting fixed point that explicitly exhibits this symmetry.

The system features several partially Gau\ss{}ian fixed points, as soon as $\lambda_{\ast}<-3.3$. These arise as a consequence of the fact that under the impact of gravity, Yukawa systems could feature a free as well as an interacting fixed point \cite{Eichhorn:2016esv, Eichhorn:2017eht, Eichhorn:2017ylw}, see Sec.~\ref{sec:yukawa_model}. There are four fixed points in the system cf.~Tab.~\ref{tab:darkmatterFPs}. Once the fixed-point value of the Yukawa coupling is non-vanishing, the resulting explicit breaking of shift-symmetry results in a non-flat scalar potential and non-minimal coupling  in the corresponding sector.

\begin{table}
 \begin{tabular}{c||c|c|c||c|}
 FP &  \parbox[t]{1.45cm}{$y_{v\, \ast}, \xi_{v\, \ast},$\\$ m^2_{v\, \ast},\lambda_{4, v\, \ast}$} & \parbox[t]{1.45cm}{$ y_{d\, \ast}, \xi_{d\, \ast},$\\ $ m^2_{d\, \ast}, \lambda_{4, d\, \ast}$} & $\lambda_{{\rm HP}\,\ast}$ &  \parbox[t]{3cm}{number of\\relevant directions}\\\hline\hline
 GMFP& 0 & 0 & 0 & 6\\\hline
 FP1 & $\neq 0$ & 0 &0 & 5\\\hline
 FP2 & 0 &  $\neq 0$ & 0  & 5\\\hline
 FP3 & $\neq 0$& $\neq 0$ & $\neq 0$& 4
 \end{tabular}
 \caption{\label{tab:darkmatterFPs} Fixed points and their predictivity in the gravity-matter system. FP1 and FP2 are symmetric under the exchange of the dark and the visible sector. Two of the relevant directions approximately correspond to the gravitational couplings $g$ and $\lambda$, and two approximately correspond to the scalar masses $m_v^2$ and $m_d^2$. There is a total of eleven canonically relevant or marginal couplings in our truncation.}
 \end{table}

Fixed points FP1 and FP2 are connected by the exchange symmetry between dark and visible sector. This symmetry entails that fixed points which do not obey the symmetry, i.e., where, $y_{v\, \ast} \neq y_{d\, \ast}$ etc.~, must come in pairs that map into each other under the symmetry.

 At all four fixed points, both gravitational couplings are associated to relevant directions, as are the two mass parameters; in addition various of the marginal couplings correspond to relevant directions. 
 At the most predictive fixed point FP3, both sectors are interacting and all canonically marginal couplings are actually irrelevant, i.e., predicted. In the following we will concentrate on this fixed point. This can either be understood as a choice for the universality class underlying the UV completion, or can also be interpreted in view of effective asymptotic safety \cite{Held:2020kze}, see Sec.~\ref{sec:effAS}: If there is a finite UV cutoff, beyond which the QFT description for matter-gravity systems ceases to hold, then the microscopic model provides the initial conditions for the RG flow. 
 From there, the flow will be attracted by fixed points along their IR attractive directions. The fixed point with most irrelevant directions is therefore most likely to play a role as an ``effective" fixed point, generating trajectories with near-scale symmetry. \\
There are only two relevant directions in the matter sector at FP3, both of which set the mass scale in the corresponding sector. All matter interactions are predicted at this fixed point. This is in marked contrast to an effective-field theory setting, where all of seven canonically marginal couplings, ($\lambda_{4\, v}, \lambda_{4,d}, \lambda_{\rm HP}, \xi_v, \xi_d, y_v, y_d$), correspond to free parameters.

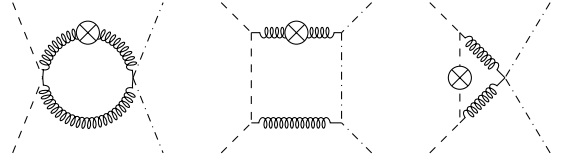
\begin{figure}
\begin{tikzpicture}[scale=2.0]
   \coordinate (tl) at (0.0, 1.0);
   \coordinate (tr) at (1.0, 1.0);
   \coordinate (bl) at (0.0, 0.0);
   \coordinate (br) at (1.0, 0.0);
   \coordinate (l) at (0.2, 0.5);
   \coordinate (r) at (0.8, 0.5);
   \coordinate (reg) at (0.5, 0.8);
   \begin{scope}[on background layer]
     \draw [scalarV] (tl) -- (l);
     \draw [scalarV] (bl) -- (l);
     \draw [graviton] (r)  arc(0:90:0.3) -- (reg);
     \draw [graviton] (reg)  arc(90:180:0.3) -- (l);
     \draw [graviton] (l)  arc(180:360:0.3) -- (r);
     \draw [scalarD] (r) -- (tr);
     \draw [scalarD] (r) -- (br);
     \node [regulator] at (reg) {};
   \end{scope}
 \end{tikzpicture}
 \qquad
 \begin{tikzpicture}[scale=2.0]
   \coordinate (ttl) at (0.0, 1.0);
   \coordinate (tl) at (0.2, 0.8);
   \coordinate (ttr) at (1.0, 1.0);
   \coordinate (tr) at (0.8, 0.8);
   \coordinate (bbl) at (0.0, 0.0);
   \coordinate (bl) at (0.2, 0.2);
   \coordinate (bbr) at (1.0, 0.0);
   \coordinate (br) at (0.8, 0.2);
   \coordinate (reg) at (0.5, 0.8);
   \begin{scope}[on background layer]
     \draw [scalarV] (ttl) -- (tl);
     \draw [scalarD] (ttr) -- (tr);
     \draw [scalarD] (bbr) -- (br);
     \draw [scalarV] (bbl) -- (bl);
     \draw [scalarV] (tl) -- (bl);
     \draw [scalarD] (tr) -- (br);
     \draw [graviton] (tr)  -- (reg);
     \draw [graviton] (reg)  -- (tl);
     \draw [graviton] (bl)  -- (br);
     \node [regulator] at (reg) {};
   \end{scope}
 \end{tikzpicture}
 \qquad
 \begin{tikzpicture}[scale=2.0]
   \coordinate (ttl) at (0.0, 1.0);
   \coordinate (tl) at (0.2, 0.8);
   \coordinate (tr) at (0.8, 1.0);
   \coordinate (bbl) at (0.0, 0.0);
   \coordinate (bl) at (0.2, 0.2);
   \coordinate (br) at (0.8, 0.0);
   \coordinate (r) at (0.5, 0.5);
   \coordinate (reg) at (0.2, 0.5);
   \begin{scope}[on background layer]
     \draw [scalarV] (ttl) -- (tl);
     \draw [scalarV] (bbl) -- (bl);
     \draw [scalarV] (tl) -- (reg);
     \draw [scalarV] (reg) -- (bl);
     \draw [graviton] (tl)  -- (r);
     \draw [graviton] (bl)  -- (r);
     \draw [scalarD] (tr) -- (r);
     \draw [scalarD] (br) -- (r);
     \node [regulator] at (reg) {};
   \end{scope}
 \end{tikzpicture}
   \caption{ \label{fig:diagrams_portal}
   These diagrams induce the Higgs portal coupling. The crossed circles correspond to regulator insertions, the wiggly lines represent gravitons, the dashed/dashed-dotted lines represent the two scalars. Additionally, the diagrams with regulator insertions on the other internal propagators contribute.
   }
 \end{figure}

 At the fixed point FP3, the Yukawa couplings break the shift symmetry in both sectors. This induces non-vanishing fixed-point values for $\xi_{v(d)}$ via the $\xi_{v(d)}$ independent terms in Eq.~\eqref{eqn:beta_xi}. These in turn induce a  nonvanishing fixed-point value for the portal coupling via the diagrams depicted in Fig.~\ref{fig:diagrams_portal}.
 These diagrams generate $\lambda_\text{HP}$ independent contributions in the beta function for $\lambda_{\rm HP}$, which reads
 \bea
  \label{eqn:beta_hp_indep}
  \beta_{\lambda_\text{HP}} &=& \frac{160 g^2 \xi_d \xi_v}{(1-2 \lambda)^3}
  +\frac{27648 g^2 \xi_d^2 \xi_v^2}{(3-4 \lambda )^2}
  +\frac{82944 g^2 \xi_d^2 \xi_v^2}{(3-4 \lambda)^3}\nonumber \\
  &+&\frac{864 g^2 \xi_d \xi_v}{(3-4 \lambda)^3}
  +\frac{6912 g^2 (9-4 \lambda) (\xi_d \xi_v^2+ \xi_d^2 \xi_v)}{5 (3-4 \lambda)^3} \nonumber\\
  &{}& + \mathcal{O}(\lambda_{\rm HP}).
 \eea
 Eq.~\eqref{eqn:beta_hp_indep} showcases that $\lambda_{{\rm HP}\, \ast}=0$ is not available, once $\xi_{v(d)\,\ast}\neq0$.
 In summary, the choice of a fixed point at finite Yukawas breaks shift symmetry in both scalar sectors, resulting in asymptotic safety at non-vanishing portal coupling.

\begin{figure}
   \includegraphics[width=\linewidth]{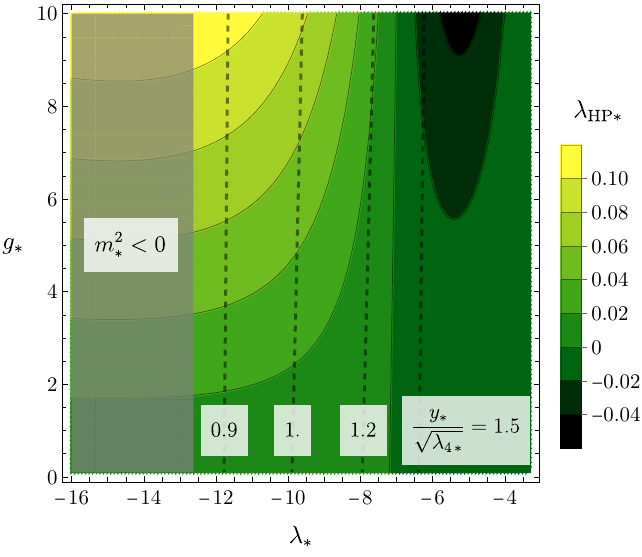}
   \caption{ \label{fig:portal_parameter_space}
   Fixed-point value of the portal coupling as a function of the gravitational fixed-point values. For $\lambda_\ast < -12.6$, the fixed point lies in the symmetry-broken regime. The ratio $\frac{y_\ast}{\sqrt{\lambda_{4\,\ast}}}$ is indicated by the dashed contours.
   }
 \end{figure}

 We analyze potential physics implications of the fixed point FP3 with non-vanishing portal coupling as a function of $\lambda_\ast$ and $g_\ast${, see Fig.~\ref{fig:portal_parameter_space}}. This completes our analysis presented in \cite{Eichhorn:2020kca}, where we focus more particularly on the fixed-point values obtained in a specific setup with the gravitational beta functions obtained in \cite{Eichhorn:2016vvy}.
  Interestingly, the fixed point FP3 can lie at negative $m_{\ast}^2<0$ for sufficiently negative $\lambda_{\ast}$, cf.~Fig.~\ref{fig:portal_parameter_space}, where both $\mathbb{Z}_2$ symmetries are spontaneously broken in the UV. \\
  Further, the portal coupling can become negative at the fixed point. Nevertheless, a stability analysis around the origin in field space indicates that the local stability criterion $\lambda_{4, v\, \ast} \lambda_{4, d\, \ast}- \lambda_{\rm HP\, \ast}^2>0$ is always satisfied. That there is a region in parameter space, in which an interacting fixed point not only exists, but also satisfies local stability and lies in the symmetric phase, is clearly a nontrivial result -- since both conditions can be violated in corresponding regimes of the parameter space. \\
  The ratio of dark-scalar and dark-fermion mass in a no-mixing approximation is $y(k_\text{IR})/\sqrt{\lambda_{4}(k_\text{IR})}$, where $k_\text{IR}$ is a sufficiently IR scale. We observe that the ratio $y_{\ast}/\sqrt{\lambda_{4\,\ast}}$ is larger than but of the order one, unless the fixed-point value $\lambda_\ast$ becomes smaller than $\lambda_\ast \approx -10$. Both couplings flow logarithmically towards the IR, therefore the ratio of fixed-point values contains indirect information about the mass ratio. Indeed, following the RG flow towards the IR indicates that, as long as the absolute value of the gravitational couplings is not too large, the dark (and visible) fermions are heavier than their scalar counterparts, making the dark scalar a potential candidate for a stable dark-matter particle. In all cases that we consider, the ratio of the two masses remains close to one.\\

 Our choice of truncation relies on assuming a near-perturbative nature of the fixed point, which needs to be checked a posteriori.\\
 In particular, FP3 has been obtained in an approximation of the beta functions which only takes into account the leading terms in the anomalous dimension and reproduces the universal one-loop results for marginal couplings, see also Sec.~\ref{sec:anomalous_dimension}. Going beyond this approximation, we find  only small quantitative changes:
 Taking into account the full dependence on the anomalous dimension $\eta_\phi$ and $\eta_\psi$, and hence considering non-polynomial terms in the beta functions, only alters both, the fixed-point values, as well as the critical exponents, at the percent level at generic values of $g$ and $\lambda$. The stability of our results under this change is an indication for the near-perturbative nature of the system.

 Additionally, the fixed-point values for $g_{\ast}$ and $\lambda_{\ast}$
 lie at roughly similar values, no matter which of the four fixed-point structures is chosen in the matter sector. This indicates that there is no huge ``back-reaction" from the interactions in the matter sector in the regime of parameter space of interest to us, i.e., for $\lambda_\ast<-3.3$ and $g_\ast$ not too large.\\ 

 \begin{figure*}
    \includegraphics[width=0.48\textwidth]{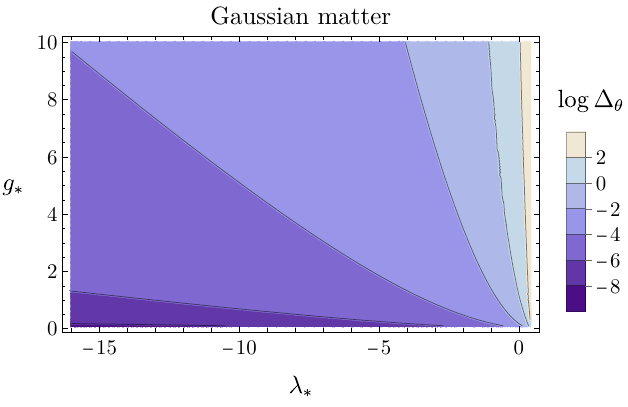}
    \includegraphics[width=0.48\textwidth]{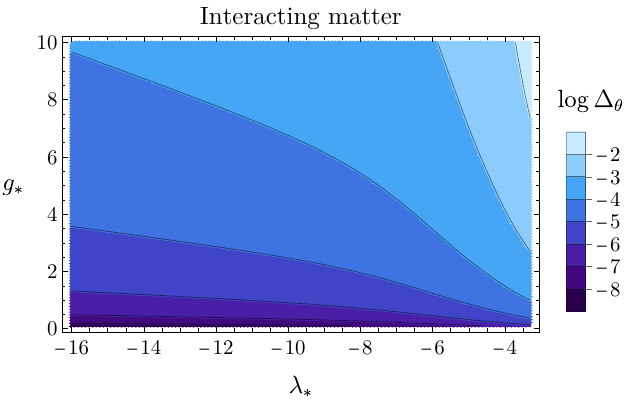}
    \caption{ 
   \label{fig:nonpert}
   The measure $\Delta_\theta$, cf. Eq.~\eqref{eqn:pert_measure}, measuring how close the critical exponents are to the canonical scaling dimension for the fixed point that is free (interacting) in the matter sector on the left (right). Both are interacting in the gravity sector. Note the different $\lambda_\ast$-axes in the two plots.
   }
\end{figure*}

 We use the deviation from canonical scaling as a measure of (near-)perturbativity, see also \cite{Falls:2013bv}. Specifically, we define
 \be
 \label{eqn:pert_measure}
 \Delta_{\theta} = \sqrt{\frac{\sum_i ({\rm Re}(\theta^{(i)})-d_{\bar{g}_i})^2}{\sum_i}},
 \ee
 where the sum runs over all matter couplings in our truncation and $d_{\bar{g}_i}$ denotes the canonical dimension of the $i$th coupling.
 Fig.~\ref{fig:nonpert} showcases that the matter system is near-perturbative at the interacting fixed points induced by its coupling to quantum gravity.
 This finding is well in line with previous indications for a near-perturbative nature of the Reuter fixed point in gravity-matter systems, e.g., \cite{Eichhorn:2018nda,Eichhorn:2018ydy,Eichhorn:2018akn}. It further supports our choice of truncation, and finally makes plausible that the Standard Model (extended by a few extra fields), that remains perturbative up to the Planck scale, could conceivably be UV extended by its coupling to asymptotically safe gravity. In particular, the mechanism that we explore here relies on a finite fixed-point value of the Yukawa coupling and is therefore not available in the region of parameter space which is generically tied to larger $\Delta_{\theta}$, lying at $\lambda_\ast>-3.3$. In fact, in this region, only a fixed point with vanishing matter couplings exist. Its critical exponents can depart more significantly from canonical scaling, see Fig.~\ref{fig:nonpert}. Thus, the region preferred on phenomenological grounds appears to coincide with the near-perturbative fixed-point regime.

 \subsection{Connection to the IR}
 \emph{...where we connect the UV fixed point to corresponding predictions for IR values and explore the quantitative relation between the various coupling parameters of the dark and visible sectors. In particular, we highlight the strong enhancement of predictivity, that asymptotic safety could entail.}\\

 To extract physical predictions, one needs to translate the UV fixed-point values
 into values at $k \to 0$. Within a standard EFT treatment that would start at some finite cutoff scale $k_{\rm UV}$, all  higher-order couplings would take arbitrary values. In addition, all couplings in our truncation correspond to canonically relevant or marginal couplings, and could also be specified freely. Thus, a large parameter-space of potential dark-matter properties would open up in the IR. 
 In contrast, an asymptotically safe UV completion (or extension) could impose a very high degree of predictivity on the setting. In fact,  in our truncation,   the two masses are the only two relevant couplings in the matter sector. Thus, these correspond to the only two parameters that can start to deviate from the fixed-point values at an arbitrary scale, such that their IR values can be adjusted. 
 Their values at $k \to 0$ are hence the only two free parameters in the scalar-Yukawa sector of the model in the asymptotically safe setting. 
 All canonically marginal couplings are irrelevant at FP3; their IR values are fixed as a function of the scalar masses. 

 \begin{figure*}
   \includegraphics[width=0.32\textwidth]{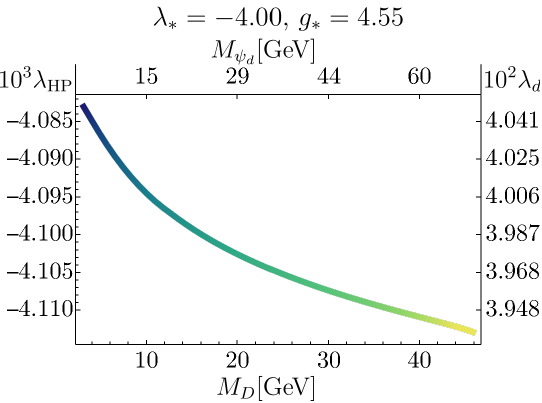}
   \includegraphics[width=0.32\textwidth]{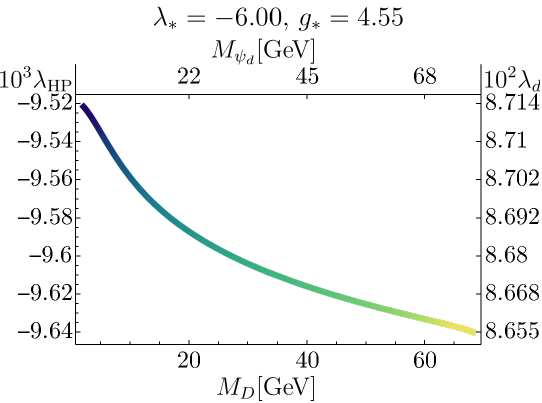}
   \includegraphics[width=0.32\textwidth]{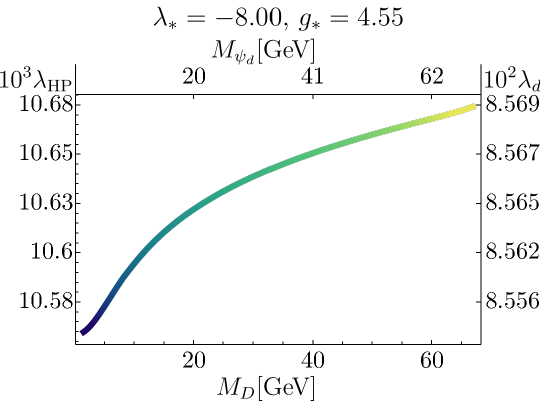}
   \includegraphics[width=0.20\textwidth]{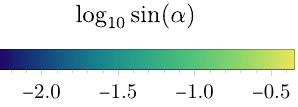}
   \caption{ \label{fig:portal_over_mass}
   Infrared relation between the dark scalar mass $M_D$, the portal coupling $\lambda_\text{HP}$, the dark quartic coupling $\lambda_d$ and the dark fermion mass $\psi_d$ for various fixed-point values of the gravitational couplings. The color coding indicates the mixing angle $\alpha$, see Eq.~\eqref{eqn:mixing_angle}. The scales on the right and top are not linear. In addition, the non-minimal coupling is also predicted.
   }
 \end{figure*}

 We break the accidental symmetry between the dark and visible sector through the RG flow to the IR by choosing different IR values for the two relevant mass parameters of the dark/visible scalar, thereby generating a distinction between the dark and visible sector at low energies. In an extension of the Standard Model, this exchange symmetry is broken explicitly due to a different set of degrees of freedom (four real scalars for the Higgs field), as well as the contribution of gauge interactions, which couple only to the visible, but not the dark, scalar and fermion.\\
To find the IR values corresponding to a fixed point trajectory, we proceed as follows: 
(i) We initialize the RG flow for the model at some scale $k_\text{UV}\gg M_\text{Pl}$ by perturbing the relevant directions of FP3.
(ii) We confirm that our choice of $k_\text{UV}$ does not influence our results by reversing the flow, i.e., flowing further ``into the UV", and confirming that typical initial conditions for the flow remain at the fixed point for a sufficiently large range of scales.  (iii) We then solve the beta functions numerically to flow towards small $k$ and extract the coupling values.\\
 At $k\approx M_\text{Pl}$, the gravitational degrees of freedom decouple dynamically. Since this decoupling occurs near-instantaneously (due to quadratic scaling $G_N \sim k^{-2}$ below the Planck scale), we can simplify the numerical flows such that we only have two instead of four relevant directions: Instead of solving the full system of flow equations, including the gravitational couplings, we decouple those degrees of freedom by hand at $k=M_\text{Pl}$, i.e., we set  $g(k)=g_\ast$ for $k>M_\text{Pl}$ and $g(k)=0$ for $k < M_{\rm Pl}$.\\
 Flowing further into the IR, $\phi_v$ or $\phi_d$ can undergo a spontaneous breaking of the $\mathbb{Z}_2$ symmetry. We do not study trajectories for which the fixed-point values already lie in the symmetry-broken regime, and leave the exploration of the corresponding phenomenology to future work.
 In order to make the spontaneous symmetry breaking in the IR explicit, it is advantageous to rewrite the potential as a Taylor expansion around the minimum:
 \bea
  V(\phi_v, \phi_d)&=& \frac{\lambda_\text{HP}}{4}\left(\phi_v^2-v_v^2\right)\left(\phi_d^2-v_d^2\right) \nonumber \\
   & & + \sum_{i=v,d} \frac{\lambda_i}{8}\left(\phi_i^2-v_i^2\right)^2
 .\eea
 For the flow in the symmetry-broken regime, we extend results of \cite{Gies:2009hq} in the fermionic sector and perform a computation around non-vanishing vacuum expectation value to obtain the additional contributions to beta functions in our system. Once the scale $k$ drops below the mass of both scalars, the flow freezes out and one obtains the infrared values for all couplings.
 \\

 Anticipating the application of our toy model to the full SM, we  focus on the regime where $\phi_v$ undergoes spontaneous symmetry breaking and fix the vacuum expectation value for $\phi_v$ to match the Higgs vacuum expectation value, $\expval{\phi_v}\equiv v_v \approx 246\,\text{GeV}$. We are then left with a single free parameter. We emphasize the marked  contrast to an EFT analysis of the same model, which would feature seven additional canonically marginal couplings as free parameters.\\
 Depending on  the IR value of the dark scalar mass $m_d^2$, three different scenarios can be distinguished: (i) The second scalar $\phi_d$ could remain in the unbroken regime in the IR, rendering the corresponding fermion massless. (ii) The vacuum expectation value of $\phi_d$ could be smaller than the one of $\phi_v$, or (iii) it could be larger. Case i) results in the dark scalar being unstable and decaying into the massless fermion. Due to its relativistic nature, it would be unlikely to provide a dark-matter candidate in agreement with constraints from structure formation. This leaves us with cases ii) and iii).
 Once $\phi_d$ also undergoes spontaneous symmetry breaking, the physical masses are given by the eigenvalues of the Hessian of the corresponding symmetry-broken potential,
 \bea
 M_{V/D}^2 &=& \frac{1}{2}\Bigl(\lambda_{v}v_v^2+ \lambda_d v_d^2\nonumber\\
 &{}& \quad \quad  \pm \sqrt{(\lambda_v v_v^2-\lambda_d v_d^2)^2+ 4 \lambda_{\rm HP}^2 v_v^2\, v_d^2}\Bigr).
 \eea 
 Both scalars mix with a mixing angle, see, e.g., \cite{Dupuis:2016fda,Arcadi:2019lka}
 \be
  \label{eqn:mixing_angle}
  \tan 2 \alpha = \frac{-2 v_v v_d \lambda_\text{HP}}{\lambda_v v_v^2 - \lambda_d v_d^2}
 .\ee
 For $M_D > M_V$, i.e., scenario (iii), the portal coupling in the SM to a dark scalar needs to be $\sim\mathcal{O}(0.1)$ in order to thermally produce the correct relic density, see, e.g., \cite{Beniwal:2015sdl}. For smaller portal couplings, an excess of dark matter will be produced. In comparison, the portal coupling $\lambda_\text{HP}$  in our toy model is typically very small at the fixed point FP3, cf.~Fig.~\ref{fig:portal_parameter_space}. Assuming that a quantitative comparison between the toy model and the SM is meaningful, we conclude that $M_D > M_V$ is likely disfavored as it might overclose the universe.
 Thus, we focus on scenario (ii) with $M_D<M_V$. 

 \begin{figure}
   \includegraphics[width=0.45\columnwidth]{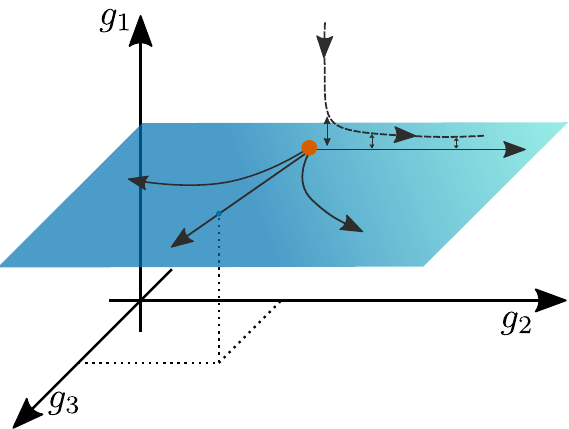}
   \includegraphics[width=0.45\columnwidth]{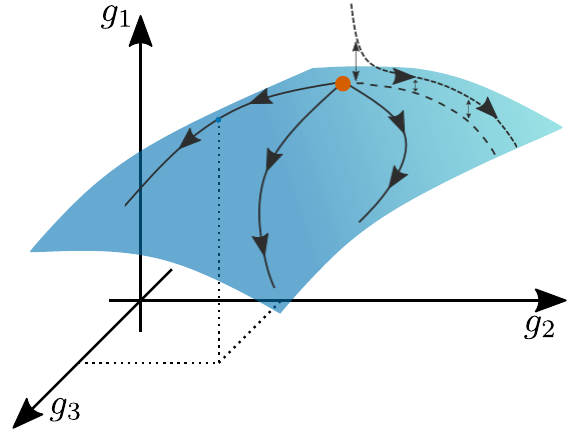}
   \caption{
   \label{fig:crit_surface}
   Illustration of the critical surface in a coupling space spanned by three couplings $g_i$. The first coupling $g_1$ is fixed in terms of $g_2$ and $g_3$. For a flat hypersurface $g_1$ is independent of $g_2$ and $g_3$. If the critical surface is curved it is fixed as a function of the other couplings.
   }
 \end{figure}

In Fig.~\ref{fig:portal_over_mass}, we vary the fixed-point values for $g_{\ast}$ and $\lambda_{\ast}$, since these depend on, e.g., the number of additional matter fields in the system. At a given choice of $g_{\ast}, \lambda_{\ast}$, a unique relationship between the portal coupling and the dark scalar mass emerges. 
The quantitative relationship also depends on the relevant parameters in the matter sector of the system, $m_{v}$ and $m_d$. This is a consequence of the curvature of the critical hypersurface of the fixed point: For a critical hypersurface that is not curved, each irrelevant coupling must take one uniquely fixed value in the IR as a consequence of asymptotic safety in the UV. In contrast, for a critical hypersurface with curvature, the IR value of an irrelevant coupling is a function of the values of the relevant ones, see illustration in Fig.~\ref{fig:crit_surface}. In particular, once  the relevant coupling $v_{v}(k_{\rm IR})$ is fixed by observations, one relevant direction remains in the matter sector, namely $v_d$. Due to the curvature of the critical hypersurface, the IR values of all irrelevant couplings depend on the value of this relevant direction. Translating it into the scalar dark-matter mass then results in a unique curve in the $\lambda_{\rm HP}, \lambda_d, M_{\psi_d}, \xi_d$-parameter space. In \cite{Eichhorn:2020kca}, we kept $g_{\ast}$ and $\lambda_{\ast}$ fixed, showing just one such curve.  Computing the gravitational fixed-point value with a particular set of beta functions and assuming a fixed matter content allowed us to obtain a single quantitative relation. Here, we instead explore the predictive relationship as a function of $g_{\ast}, \lambda_{\ast}$, accounting for systematic uncertainties in the gravitational beta functions.  Going beyond the gravitational fixed point considered in \cite{Eichhorn:2020kca}, the existence and qualitative nature of the predictive relation remains unchanged, see Fig.~\ref{fig:portal_over_mass}. Quantitatively, it changes upon varying the gravitational fixed point values. The portal coupling is always of order $\abs{\lambda_\text{HP}} \sim 10^{-3}$.
The existence of a predictive relation beyond the fixed point considered in \cite{Eichhorn:2020kca} illustrates, how asymptotic safety might identify interesting regions in the matter parameter space independent of the particular gravitational fixed point values.

To connect to existing and upcoming observational constraints, both from direct searches \cite{Wagner:2010mi,Hooper:2010mq,Ahnen:2016qkx,Aartsen:2016pfc,Akerib:2016vxi,Amole:2017dex,Agnese:2017jvy,Cui:2017nnn,Agnes:2018fwg,Aprile:2018dbl,Abdelhameed:2019hmk,Leane:2019xiy,Aguilar-Arevalo:2019wdi,Braine:2019fqb,Arnaud:2020svb,Andrianavalomahefa:2020ucg}, as well as from cosmological probes \cite{Gluscevic:2019yal,Drlica-Wagner:2019xan}, a crucial next step is to extend the visible sector to a more realistic representation of the SM Higgs sector. In this context it will also be crucial to extend the understanding of the role of the dark fermion in the thermal history of the universe.

 In summary, our scalar-Yukawa model provides a concrete example of a model with a non-vanishing portal coupling. It illustrates the high degree of predictivity that can emerge in asymptotically safe dark-matter models.
 Beyond our portal model, asymptotic safety allows to classify dark matter models with regard to whether they can be UV completed/extended through the existence of an asymptotically safe gravity-matter fixed point. In doing so, asymptotic safety could identify models and parameter regions that are of special interest for experimental dark matter searches.

\section{Conclusions and outlook}
\label{sec:conclusions}
 \subsection{Summary of key results}\label{sec:summary}
In this paper, we have studied the interplay of scalar fields with quantum gravity. We have mostly focused on asymptotically safe gravity, while including some comments from a broader, effective-field-theory perspective. Scalar fields play a key role in particle physics, constituting the Higgs sector in and beyond the Standard Model. Further, they could be relevant in a cosmological context, driving an early inflationary phase as well as late-time expansion of the universe. 
At the interface of particle physics and cosmology, scalars constitute dark-matter candidates. In all three contexts, a decoupling argument is typically invoked to neglect quantum gravity, based on a separation of the relevant scales from the Planck scale.
Here, we challenge this view, by pointing out that the interplay of scalar fields with quantum gravity could strongly constrain scalar systems. In fact, quantum gravity is non-negligible at the microscopic scales which determine the fundamental properties of scalar sectors. Hence, while quantum gravity is not \emph{dynamically} relevant at the scales at which we make observations in particle physics and cosmology today, it may be \emph{inherently} relevant by fixing the fundamental parameters of these systems which determine their observable properties. 

Within the asymptotic-safety approach to quantum gravity, the underlying quantum scale symmetry requires a balance between the microscopic interaction strength of quantum gravity with that of the matter fields. This balance relates various couplings in the system to each other at the microscopic level. 
The RG flow maps the corresponding relations to the values of couplings at macroscopic scales.  
Consequences of asymptotic safety are thereby (in principle) imprinted on the (observable) macrophysics. Therefore, free parameters of the SM, dark matter and dark energy as well as the inflationary sector could become calculable in asymptotic safety, see, e.g., \cite{Shaposhnikov:2009pv,Harst:2011zx,Wetterich:2016uxm,Eichhorn:2017als,Rubio:2017gty,Eichhorn:2017ylw,Eichhorn:2017lry,Eichhorn:2018whv,Pawlowski:2018ixd,deBrito:2019umw} for earlier work along these lines and \cite{Eichhorn:2018yfc,Eichhorn:2019tcj} for reviews and further references therein. Potentially, even free parameters of the geometry, such as the spacetime dimensionality, might be constrained by asymptotic safety \cite{Eichhorn:2019yzm}.

The leading question for this research is, whether scenarios in particle physics and/or cosmology could be excluded based on their incompatibility with asymptotically safe gravity. For phenomenology and model-building, this could  serve as a powerful guiding principle to distinguish among different effective theories. For fundamental quantum-gravity research, it could provide an observational test that could rule out asymptotically safe gravity. To achieve this goal, it is important to identify no-go examples, i.e., models which cannot be consistently embedded into asymptotic safety \footnote{One might also call this the ``swampland" of asymptotic safety -- in analogy to the swampland of string theory.}.
In this work, we provisionally strengthen previous evidence for such no-go examples and find tentative indications for new ones. For instance, we identify Higgs inflation as one such potential no-go-example. 
Quite intriguingly, it is the combination of cosmological considerations and particle-physics considerations which provide hints against the viability of asymptotically safe Higgs inflation in agreement with observations. This highlights that the development of a useful phenomenology of quantum gravity might require a broader view that encompasses particle physics and cosmology, instead of the focus on just one of these sectors.\\

In this work, we explore simple models for the matter sector, starting with a single scalar coupled to quantum gravity, promoting it to a simple Yukawa system, then two coupled scalar sectors and finally two coupled Yukawa systems. These simple systems serve to elucidate the mechanisms by which asymptotic safety might impose its predictive power on particle physics and cosmology.

 Let us add an important disclaimer: all results hold within truncations of the Euclidean Renormalization Group flow; therefore they are affected by systematic uncertainties, the choice of Euclidean vs. Lorentzian signature, and additionally, our restriction to simple models which account for a subset of the degrees of freedom of the Standard Model only. We will discuss these limitations in more detail after reviewing our results. \\

 It is a main aim of our investigations to provide new and strengthen existing hints for the potential consequences of asymptotic safety for IR physics.
 We mostly do so by leaving the fixed-point values for the gravitational couplings, the Newton coupling $g_\ast$ and the cosmological constant $\lambda_\ast$, unspecified. In fact, different consequences for low-energy physics follow from asymptotic safety in different regions within this parameter space. Note that this places no restriction on the measured low-energy values of these gravitational couplings. Exploring the full parameter space for these two couplings enables us to study which low-energy physics scenarios  could \emph{in principle} be compatible with asymptotic safety. Firstly, this allows us to identify whether a given scenario is ruled out in the entire region of the gravitational parameter space that we explore.  Secondly, the realization of a given scenario could single out a preferred region in the gravitational parameter space. Where in this parameter space the fixed point lies, is computable and depends on the number of matter fields, see, e.g., \cite{Dona:2013qba,Labus:2015ska,Meibohm:2015twa,Dona:2015tnf,Eichhorn:2016vvy,Biemans:2017zca,Christiansen:2017cxa,Alkofer:2018fxj,Eichhorn:2018nda,Wetterich:2019zdo,Daas:2020dyo}. Thus, within asymptotic safety, a given matter content (in the dark and visible sector) is (in principle) connected to a particular set of low-energy physics consequences, see also \cite{Eichhorn:2017ylw} for a discussion in the context of the top-quark mass.\\

\noindent \emph{Particle physics} \\
We find that the inclusion of the non-minimal coupling leaves the scenario from \cite{Eichhorn:2016esv,Eichhorn:2017ylw,Alkofer:2020vtb} intact, where the fixed-point structure of gravity-Yukawa systems could result in a calculation of the top and bottom quark masses. In our study, adding the non-minimal coupling to the setting, the physics of simple Yukawa systems is still determined by the interplay of two fixed points. These result in an upper bound on the fermion mass compatible with asymptotic safety. Quite intriguingly, there is only a narrow range of values for the non-minimal coupling $\xi$ for which this is the case. The fixed-point value $\xi_{\ast}$ falls right into this range in our truncation, constituting a rather nontrivial test of the mechanism proposed in \cite{Eichhorn:2016esv,Eichhorn:2017ylw,Alkofer:2020vtb}. \\
Moreover, we find that under the impact of quantum gravity, the non-minimal coupling could push simple Yukawa systems into a regime of spontaneous symmetry breaking already at the UV fixed point, see \cite{Eichhorn:2019dhg} for related considerations in grand unified theories. Within our approximation, this gravity-induced spontaneous symmetry breaking occurs only for sufficiently negative fixed-point values of the cosmological constant. In order to properly investigate the resulting low-energy phenomenology, global flows for the Higgs potential, see, e.g., \cite{Borchardt:2016xju}, are called for.\\
Finally, we find that the value of the non-minimal coupling $\xi$ is a prediction of asymptotic safety within our truncation. If this property persists for the full SM coupled to quantum gravity, the prediction for $\xi$ could become highly relevant for considerations of vacuum stability in the early universe, see \cite{Markkanen:2018pdo}.\\
\\

\noindent \emph{Cosmology}\\
 We strengthen previous results indicating that quantum gravity tends to drive scalar potentials towards flatness,  see, e.g., \cite{Narain:2009fy,Narain:2009gb,Percacci:2015wwa,Labus:2015ska,Wetterich:2016uxm, Eichhorn:2017als,Pawlowski:2018ixd, deBrito:2019umw}. This occurs in an appropriate part of the gravitational parameter space. This flattening-mechanism also reduces the number of free parameters in a scalar potential compared to the canonically relevant and marginal set. Within a cosmological setting, these results might play a role both for inflation as well as dynamical dark energy, and might potentially set asymptotic safety apart from other proposals for fundamental physics.\\

 We re-express our result in terms of scale-dependent slow-roll-inspired parameters which lend themselves to a study in an inflationary setting. We find that the predictive power of asymptotic safety could fix the values of these parameters in an appropriate regime of the gravitational parameter space. The gravitational flattening-mechanism appears encouraging in view of the observational indications for near-scale invariance. Yet, matching the amplitude of scalar fluctuations within the  potentially rather constraining setting of asymptotic safety might be a challenge.\\

The next decade in observational cosmology is expected to produce exciting novel insights into the value of the equation-of-state parameter for dark energy \cite{Abell:2009aa,Amendola:2012ys,Aghamousa:2016zmz,Bacon:2018dui,Bull:2018lat}. Within asymptotic safety, realizing a cosmological constant of the correct magnitude appears possible, since it corresponds to a relevant coupling. Additionally, scenarios with a very flat scalar potential also appear viable -- and even preferred -- within asymptotic safety in our truncation. Therefore, we tentatively conclude that an equation-of-state-parameter $w \approx -1$ might be preferred in asymptotic safety.\\

The degree of predictivity for scalar potentials depends on the gravitational fixed-point values. The latter in turn depend on the number of matter degrees of freedom -- both dark and visible ones  \cite{Dona:2013qba,Labus:2015ska,Meibohm:2015twa,Dona:2015tnf,Eichhorn:2016vvy,Biemans:2017zca,Christiansen:2017cxa,Alkofer:2018fxj,Eichhorn:2018nda,Wetterich:2019zdo,Daas:2020dyo}. Accordingly, a given set of matter fields could translate into strong constraints on inflation and dynamical dark energy. Conversely, for alternative BSM settings, matter fields might push the gravitational fixed-point values into a regime, where fewer theoretical constraints on scalar potentials exist. In the former case, a confrontation with observational data is more promising and might even allow to rule out asymptotic safety with a given matter content in the future. \\

\noindent \emph{At the interface}\\
To make a beginning step in the investigation of Higgs inflation, we use a simple Yukawa system as a toy model for the Higgs-top-sector of the SM. We find that the ratio of quartic coupling and non-minimal coupling could reach small enough values to match observational data on the amplitude of scalar fluctuations for a sufficiently small fixed-point value $g_{\ast}$. Due to the predictive nature of the asymptotically safe fixed point, this requirement simultaneously fixes the IR value of the Yukawa coupling in our truncation. Sizable values of the Yukawa coupling can only be accommodated at large values of $\lambda_4/\xi^2$. Hence, we conclude that within our truncation and toy model, a sizable top mass cannot be reconciled with a realistic amplitude of primordial scalar fluctuation anywhere in the gravitational parameter space. These studies clearly call for an extension of the model and the truncation; an investigation of the value of the Higgs mass is then also required. Nevertheless, using observations from particle physics and cosmology concertedly already provides us with a first tentative hint for the incompatibility of Higgs inflation with asymptotic safety. \\

Dark matter is another mystery at the interface of cosmology and particle physics. Here, we strengthen the decoupling result from \cite{Eichhorn:2017als} for a single, uncharged scalar as the only field in the dark sector: Under the impact of asymptotic safety, the Higgs portal coupling vanishes at all scales, ruling out the scalar field as a thermal relic, also in the presence of a non-minimal coupling. This could constitute a compelling example for a dark-matter model in the asymptotically safe swampland.\\

 We further expand on the dark-matter model we proposed in \cite{Eichhorn:2020kca} that consists of two Yukawa sectors. These represent a dark sector consisting of a dark scalar and a dark fermion and  a toy model for the Higgs-top sector of the SM. A non-vanishing fixed-point value for both Yukawa couplings induces finite non-minimal couplings and non-vanishing portal coupling; therefore the dynamical dark fermion is instrumental in lifting the decoupling property of the dark sector.\\
 The resulting fixed point is highly predictive, featuring only two free parameters in the matter sector within our truncation: Both Yukawa couplings, all three quartic couplings and both non-minimal coupling are predicted by asymptotic safety as a function of the two free parameters.
 Fixing one of them by requiring to match the vacuum expectation value of the Higgs field, we remain with just a single free parameter. This allows us to compute a unique relation between the dark-matter mass and all canonically marginal couplings in the infrared that holds in our truncation and toy model. In contrast, a canonical EFT treatment features four free parameters in the dark sector only, namely the dark Yukawa coupling, dark non-minimal coupling, dark quartic coupling and portal coupling, providing no relation between them. 
 This exemplifies that asymptotic safety might be a powerful paradigm to severely constrain the theoretically viable parameter space for dark-matter searches.\\

As emphasized previously, these results are subject to systematic uncertainties which need to be addressed in the future, and which we will discuss in more detail now.

\subsection{Outlook: improving on systematic uncertainties} 
Several sources of systematic uncertainties are affecting the current state-of-the-art in asymptotic-safety research and have been discussed extensively in \cite{Bonanno:2020bil}, see also \cite{Donoghue:2019clr}. More recently, progress in one of the key open questions, that of unitarity, has been made in \cite{Draper:2020bop,Draper:2020knh,Platania:2020knd}. In particular, a proposal for an asymptotically safe action which satisfies a perturbative notion of unitarity (tied to the poles of the graviton propagator about a flat background) has been put forward.

Here, we will  highlight some of the open questions that are most relevant for our discussion.\\

\noindent\emph{Systematic uncertainties: Choice of truncation}\\
Our results are achieved within a truncation of the effective dynamics. It follows the assumption that canonical power counting remains a good guiding principle at the interacting fixed point, which is supported by a number of explicit results \cite{Eichhorn:2018nda,Eichhorn:2018ydy,Eichhorn:2018akn,Falls:2013bv,Falls:2014tra,Falls:2017lst,Falls:2018ylp,Niedermaier:2009zz,Niedermaier:2010zz}, see Sec.~\ref{subsec:FRG} for more details.  We find further evidence in support of the corresponding near-perturbative nature of asymptotic safety, in that the critical exponents at the fixed point do typically not differ from the canonical dimensions by large amounts. This also constitutes an a-posteriori viability-check of our choice of truncation. 
\\
Fixed-point structures in simple Yukawa systems \cite{Eichhorn:2017ylw} and a dark sector with portal coupling \cite{Eichhorn:2017als} remain stable under the inclusion of the non-minimal coupling, lending further support to an asymptotically safe UV completion of such systems with an enhanced predictive power.
\\
 Based on the principle of near-perturbativity, the addition of $R^2$ and $R_{\mu\nu}^2$ to the gravitational dynamics appears indicated. This has partially been done in \cite{Hamada:2017rvn,Eichhorn:2017eht} for scalar systems and Yukawa systems coupled to gravity, see also \cite{deBrito:2019umw} for the unimodular asymptotic safety case. In short, the higher-order couplings modify the propagator for metric fluctuations, which is still proportional to $g$, but with a denominator that depends on the higher-order couplings in addition to the cosmological constant. Accordingly, the parameter space explored, e.g., in Fig.~\ref{fig:fv_nr_relv_dir}, \ref{fig:glambda_GFP_rel_with_yukawa}, \ref{fig:yukawa_contours} and \ref{fig:xi_g_lambda} is enlarged by two additional directions. At finite values of higher-order couplings, the boundaries in the $\lambda-g$ plane are deformed. Therefore, the mechanisms we explored here are expected to persist in this larger parameter space.

In the future, going beyond polynomial truncations for the scalar potential, e.g., with grid methods \cite{Reichert:2017puo} or spectral methods \cite{Borchardt:2015rxa,Borchardt:2016xju}, is also indicated, in particular to properly investigate the stability of the Higgs potential.\\

\noindent\emph{Systematic uncertainties: Lorentzian signature}\\
Our results are obtained within a fully Euclidean setting. While the RG flow at momentum scales below the Planck scale can be analytically continued, this is not necessarily the case for the transplanckian flow. As long as the gravitational fixed point is near-perturbative, and the vacuum of the theory is flat, or at least maximally symmetric, an analytic continuation might remain feasible even for the quantum gravity regime. This pressing question deserves further investigation in the future.\\

\noindent\emph{Systematic uncertainties: Missing degrees of freedom}\\
Here, we have worked in toy models with few matter fields. The RG flows of such simplified matter systems share many characteristics with that of the Higgs-top-sector of the Standard Model, and are therefore useful to explore possible structures that might arise in the Standard Model coupled to quantum gravity. 
Nevertheless, accounting for the missing degrees of freedom is an important future task.\\

 In summary, we have strengthened existing and found novel indications that certain models for dark matter as well as inflation might lie in the ``swampland" of asymptotic safety, i.e., be incompatible with an asymptotically safe UV sector. Such negative results are key in order to identify potential observational routes to ruling out this proposal for quantum gravity. Conversely, we have also identified potential phenomenological implications of asymptotic safety which could be compatible with observations, identifying matter theories in the ``landscape" of asymptotic safety.
In both cases,  their source is the predictive power of asymptotic safety, which conversely implies that asymptotic safety could serve as a very strong selection principle for particle physics, cosmology and at the interface of the two fields.

\section*{Acknowledgments}
We would like to thank Aaron Held for insightful discussions, Jan Kwapisz for a discussion on eternal inflation and the speakers and participants of the Online-Workshop Series on Quantum Gravity for discussions on the interplay of dark energy and quantum gravity.
 This work is supported by a research grant (29405) from VILLUM FONDEN. M.~P.~is  supported by  a  scholarship  of  the  German  Academic  Scholarship Foundation and gratefully acknowledges the extended hospitality at CP3-Origins during various stages of this work.

 \appendix
 \section{Formal aspects}
 \label{sec:methods}
 \emph{...where we specify the gauge-fixing procedure and the regulator with our focus on an auxiliary background-scalar field dependence related to the non-minimal coupling, which we then analyze. Further, we study the anomalous dimension of the scalar as a measure of robustness of our truncation.}\\

 While obtaining the beta functions from the effective action, one encounters subtleties related to the non-minimal coupling. We discuss these by focusing on the following truncation for scalar-gravity for the remainder of this section
 \be
  \Gamma_k = \Gamma_k^\text{scal+grav} + S_{\rm gf} + S_{\rm gh}. \label{eq:scalargravityaction}
  \ee
Here, we use a split of the  scalar into 
\be
\phi=\bar{\phi}+\varphi. 
\ee
In contrast to the metric, there is actually no need to introduce this split, since no gauge-fixing is required and the coarse-graining procedure is based on $\bar{g}_{\mu\nu}$. Nevertheless, it can be useful to introduce $\bar{\phi}$ for calculational convenience in the gauge-fixing and regularization term for the metric, see, e.g., \cite{Narain:2009fy}.

 \subsection{Gauge fixing}
 \label{sec:gauge_fixing}
 \emph{...where we specify the background-field-dependent gauge fixing we employ and detail how the scalar background-field appears in the construction of the gravitational gauge-fixing term.}\\

 To invert $\Gamma_k^{(2)}$, it is computationally advantageous to decompose the metric fluctuation $h_{\mu\nu}$ into a transverse-traceless tensor, a divergence-free vector and two scalar modes according to the York decomposition \cite{York:1973ia}, as follows
 \be
 \label{eqn:york_decomposition}
 h_{\mu\nu} = h_{\mu\nu}^{\rm TT} 
 + \bar{D}_{\mu}v_{\nu}^{\rm T} + \bar{D}_{\nu}v_{\mu}^{\rm T} + \bar{D}_{\mu}\bar{D}_{\nu}\sigma - \bar{g}_{\mu\nu}\bar{D}^2 \sigma - \frac{1}{4}\bar{g}_{\mu\nu}h.
 \ee
 Herein, $\bar{D}_{\mu}$ is the covariant derivative with respect to the background metric $\bar{g}_{\mu\nu}$. $h_{\mu\nu}^{\rm TT}$ is transverse, $\bar{D}^{\nu}h_{\mu\nu}^{\rm TT}=0$ and traceless $\bar{g}^{\mu\nu}h_{\rm \mu \nu}^{\rm TT}=0$, and $v_{\mu}^{\rm T}$ is transverse, $\bar{D}^{\mu}v_{\mu}^{\rm T}=0$.

 Due to the gauge symmetry in the gravitational sector, $\Gamma_k^{(2)}$ contains zero modes, requiring gauge fixing.
 We choose a linear, background-covariant gauge fixing
 \be
  \label{eqn:gauge_fixing}
  S_\text{gf} = \frac{1}{2\alpha} \int \dd[4]{x} \sqrt{\bar{g}}\, \bar{g}^{\mu\nu} \mathcal{F}_\mu \mathcal{F}_\nu ,
 \ee
 with 
 \be
  \mathcal{F}_\mu = \sqrt{\frac{1}{16\pi G_N} +  \alpha_\text{GF} \xi \bar{\phi}^2} \left(\bar{D}^\rho h_{\rho\mu} - \frac{1+\beta}{4} \bar{D}_\mu h^\rho_\rho \right).
 \ee
 For all results reported in the following, we  specify to $\beta=\alpha=0$, see, e.g., \cite{Gies:2015tca} for a study of the gauge dependence in the pure-gravity case. Our choice decouples one of the two scalar modes and leaves only contributions from $h_{\mu\nu}^{\rm TT}$ and $h$ in the beta functions of matter couplings. 

 In choosing the gauge fixing \eqref{eqn:gauge_fixing}, we have introduced a term dependent on the scalar background field into the gauge fixing. It is convenient to do so in truncations that include non-minimal gravity-scalar couplings, as the gauge-fixed propagator simplifies for this choice \cite{Narain:2009fy}. We point out that it is not consistent to disregard this background-scalar-field dependence in the Faddeev-Popov determinant
 \bea
 && \det(\fdv{\mathcal{F}_\mu}{\zeta_\lambda}) = \det \Bigg( \sqrt{\frac{1}{16\pi G_N} +  \alpha_\text{GF} \xi \bar{\phi}^2} \cdot \\ \nonumber
  && \quad \cdot \Big(\bar{D}^\rho \bar{g}_\mu^\sigma g_\sigma^\lambda D_\rho + \bar{D}^\sigma \bar{g}_\mu^\rho g_\sigma^\lambda D_\rho - \frac{1+\beta}{2} \bar{D}_\mu \bar{g}^{\rho\sigma} g_\sigma^\lambda D_\rho \Big) \Bigg). 
  \label{eq:FPdet}
 \eea
  In the above, $\zeta_\lambda$ is the parameter describing an infinitesimal diffeomorphism.

   As usual, Faddeev-Popov ghosts can be used to exponentiate the determinant.   A redefinition of the ghost fields absorbing a factor of $\sqrt{16 \pi G_N}$ that is present even at $\alpha_{\rm GF}=0$ allows to normalize the ghost term conveniently.
 We write the ghost term in the form
 \bea
  S_\text{gh} &=& -\int \dd[4]{x} \sqrt{ \bar{g}} \sqrt{1+\alpha_\text{GF} (1-\alpha_\text{ghost}) 16\pi G_N \xi \bar{\phi}^2  \cdot
  } \nonumber\\
 &{}&\quad   \cdot \bar{c}_{\mu} 
  \Bigl(\bar{D}^{\rho}\bar{g}^{\mu\kappa}g_{\kappa \nu}D_{\rho}  + \bar{D}^{\rho}\bar{g}^{\mu \kappa}g_{\rho\nu}D_{\kappa}\nonumber\\
  &{}& \quad \quad\quad - \frac{1+\beta}{2}\bar{D}^{\mu}\bar{g}^{\rho\sigma}g_{\rho\nu}D_{\sigma} \Bigr) c^{\nu}.
 \eea
 Here, the parameter $\alpha_\text{ghost}$ is an auxiliary parameter, that we introduce to track the background-scalar dependence of the ghost term. 
 Exponentiating the Faddeev-Popov determinant \eqref{eq:FPdet} actually results in $\alpha_\text{ghost} = 0$. With this choice,  we fully compensate the additional background- scalar dependence of the gauge fixing term.\\
In contrast, for $\alpha_\text{ghost}=1$ the background-scalar field dependence in the Faddeev-Popov determinant has been dropped. By comparing $\alpha_{\rm ghost}=0$ to $\alpha_{\rm ghost}=1$, we quantify the impact of such a treatment.
 \subsection{Choice of regulator}
 \label{sec:reg_choices_intro}
 \emph{...where we provide details of the regulator function we use and discuss how the scalar background field appears in the construction.}\\

 For the regulator, we utilize  
 \be
 \label{eqn:litim_reg}
  R_k(\bar{\mathcal{D}}) =\bar{ \mathcal{D}}\, r\left(\frac{k^2}{\bar{\mathcal{D}}}\right),
 \ee
 with the shape function $r(x) = (x-1) \Theta(x-1)$, where $\Theta(x)$ is the Heaviside distribution \cite{Litim:2001up}. The operator $\bar{\mathcal{D}}$ denotes a generalized background-covariant Laplacian. In the simplest case of a flat background, it reduces to $p^2$. 
More general choices for the argument of the regulator function in Eq.~\eqref{eqn:litim_reg} are possible and can be advantageous for practical calculations. There must not be an additional dependence on the dynamical fields in the regulator, since this would destroy the one-loop structure of the flow equation. In contrast, the background-field dependence is a priori unrestricted with unbroken background gauge invariance the only strict requirement.

 For simplicity, one would like the regulator to entirely remove the appearance of Laplacians in the denominator of the flow equation. This strategy has been implemented, e.g., in \cite{Narain:2009fy}. For a system with higher-order curvature invariants and/or non-minimal couplings, this strategy entails a choice of regulator that introduces a scalar background-field dependence in addition to the background-metric dependence contained in the background Laplacians. 
 In \cite{Bridle:2013sra}, it has been demonstrated in the context of a non-gravitational theory that such additional background-scalar field dependencies have to be treated with care within approximations, where they can lead to deformations of fixed points or even introduce new zeros of beta functions. Therefore, we will explore the dependence on the background scalar field that arises through the choice of regulator as in \cite{Narain:2009fy} carefully. We also refer to \cite{Labus:2015ska} for a brief discussion of the same point using the exponential split.

 In order to test the effect of additional background-scalar-field dependencies we introduce an additional parameter $\alpha_\text{reg}$ accompanying the background-scalar-field-dependent part of the regulator.
 Specifically, we choose 
 \bea
  R_{k\, \rm TT}^{\mu\nu\rho\sigma}&=&\frac{1}{4} \left(\frac{1}{16\pi G_N} + \xi \alpha_\text{reg} \bar{\phi}^2 \right)\cdot \nonumber \\
  &&\cdot \left(k^2 + 
  {(-\bar{D}}^2)\right) \: \Theta\!\left(k^2 + 
  {(-\bar{D}}^2)\right) \, \mathbbm{1}_\text{TT}^{\mu\nu\rho\sigma}\\ 
  R_{k\, v v}^{\mu\nu}&=&\frac{1}{2\alpha} \left(\frac{1}{16\pi G_N} + \xi \alpha_\text{GF} \bar{\phi}^2 \right) \cdot\nonumber \\
  && \cdot\left(k^2 + 
  {(-\bar{D}}^2)\right) \: \Theta\!\left(k^2 + 
  {(-\bar{D}}^2)\right) \, \mathbbm{1}_v^{\mu\nu}\\ 
  R_{k\, hh}&=& -\frac{3}{32} \left(\frac{1}{16\pi G_N} + \xi \alpha_\text{reg} \bar{\phi}^2 \right)\cdot \nonumber \\
  &&\cdot\left(k^2 + 
  {(-\bar{D}}^2) \right) \: \Theta\!\left(k^2 + 
  {(-\bar{D}}^2)\right)\\ 
  R_{k\, \sigma\sigma} &=& \frac{3}{32 \alpha}\left(\frac{(3-\alpha)}{16 \pi G_N}-(\alpha \alpha_\text{reg}-3\alpha_\text{GF}) \xi\bar{\phi}^2\right) \cdot \nonumber\\
  &&\quad \cdot\left(k^2 + 
  {(-\bar{D}}^2)\right) \: \Theta\!\left(k^2 +
  {(-\bar{D}}^2)\right) \\
  R_{k\, \phi\phi} &=& \frac{1}{2}\left(k^2 +
  {(-\bar{D}}^2)\right) Z_\phi \; \Theta\!\left(k^2 + 
  {(-\bar{D}}^2)\right)\\
  R_{k\, h \sigma} &=& -\frac{1}{16} \left(\frac{1}{16\pi G_N} + \xi \alpha_\text{reg} \bar{\phi}^2 \right) \; \Theta\!\left(k^2 + 
  {(-\bar{D}}^2)\right)\cdot \nonumber \\
  &&\cdot \Bigl(\sqrt{3 k^2(3k^2 -\bar{R})}\nonumber\\
  &{}&\quad \quad-  \sqrt{3
  {(-\bar{D}}^2)(3
  {(-\bar{D}}^2) + \bar{R})} \Bigr) \\ 
  R_{k\, h\phi} &=& - \frac{3}{2}\left(k^2 + 
  {(-\bar{D}}^2)\right) \sqrt{\alpha_\text{reg}} \xi \bar{\phi}  \; \Theta\!\left(k^2 + 
  {(-\bar{D}}^2)\right)\nonumber\\
  &{}&\\
  R_{k\, \sigma\phi}&=&-\frac{1}{2}\Bigl(\sqrt{3 k^2(3k^2 -\bar{R})} \nonumber\\
  &{}&\quad- \sqrt{3
  {(-\bar{D}}^2)(3
  {(-\bar{D}}^2) + \bar{R})} \Bigr)\cdot \nonumber \\ 
  &&\cdot \sqrt{\alpha_\text{reg}} \xi \bar{\phi} \; \Theta\!\left(k^2 + 
  {(-\bar{D}}^2)\right),
 \eea
 with $\mathbbm{1}_\text{TT}$ and $\mathbbm{1}_v$ the identity in the space of transverse-traceless tensors and transverse vectors, respectively. 
 The choice $\alpha_\text{reg} = 1$ corresponds to the choice in \cite{Narain:2009fy} with a background-scalar dependent regulator. 
 The choice $\alpha_\text{reg} = 0$ corresponds to a computation with a background-scalar independent regulator.\\
 In the sections \ref{sec:yukawa_model} and \ref{sec:portal}, where fermions are included, we choose a fermionic regulator of the form
 \be
 \label{eqn:reg_fermion}
 R_{k{\bar{\psi} \psi}}= i Z_\psi \bar{\slashed{\nabla}}\left(1-\sqrt{k^2/(-\bar{\slashed{\nabla}})^2} \right)\Theta \left(k^2-(-\bar{\slashed{\nabla}})^2  \right).
 \ee
 \subsection{Background field dependence}
 \label{sec:background_field_dependence}
 \emph{...where we study quantitatively, whether the auxiliary background-scalar-field dependence of regulator and gauge fixing changes key aspects of the results, such as, e.g., the signs of critical exponents within our truncation. This can also be interpreted as a study of the robustness of our approximation.}\\

 For clarity, we list the three parameters that track the background-scalar-field dependence in our calculation:
 \begin{itemize}
 \item $\alpha_{\rm GF}$ tracks the background-scalar-field dependence of the gauge-fixing term, which vanishes for $\alpha_{\rm GF}=0$,
 \item $\alpha_{\rm ghost}$ allows to independently vary the background-scalar-field dependence of the corresponding Faddeev-Popov ghost term.  In principle, one needs to choose $\alpha_\text{ghost}=0$ (if $\alpha_\text{GF}\neq 0$) for consistency. Here we treat $\alpha_\text{ghost}$ as an independent parameter to compare to previous work in the literature.
 \item $\alpha_{\rm reg}$ tracks the background-scalar-field dependence of the regulator for the metric fluctuations.
 \end{itemize}
 Choosing the values $\alpha_{\rm GF}=0= \alpha_{\rm reg}$ results in an absence of the extra background-scalar-field dependence from these terms.

\begin{figure}
   \includegraphics[width=0.45\textwidth]{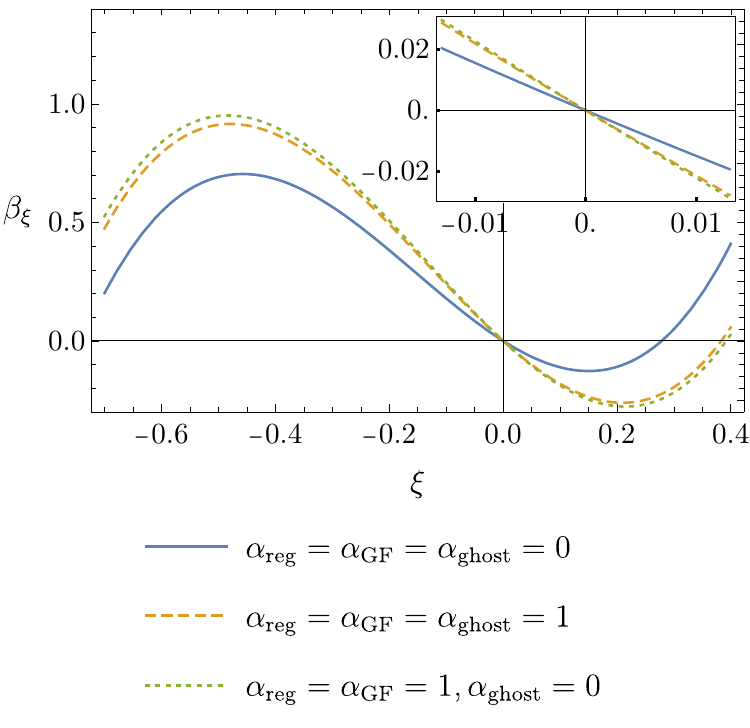}
   \caption{
   \label{fig:beta_xi_reg}
   The $\beta$-function for the non-minimal coupling $\xi$ as a function of $\xi$  for varying background-scalar dependence.
   We set the quartic coupling to its fixed-point value $\lambda_{4\,\ast}=0$.
   }
 \end{figure}

 The fixed-point values are independent of $\alpha_{\rm GF}, \alpha_{\rm reg}$ and $\alpha_{\rm ghost}$ and given by
 \footnote{ Quantitative deviations of this fixed point from \eqref{eq:FPpurescalar} arise due to the differing approximations for the graviton anomalous dimension $\eta_h$.}
 \be
  \label{eqn:ngfp}
  \lambda_\ast = 0.171, \quad g_\ast =0.843 \quad m^2_\ast = \lambda_{4\,\ast} = \xi_\ast = 0
 .\ee
In contrast, the critical exponents are affected by these parameters, as would the fixed-point coordinates be if $\xi_{\ast}$ was not equal to zero.  
Modifying $\alpha_\text{GF}$, $\alpha_\text{ghost}$ and $\alpha_\text{reg}$ can affect the $\beta$-functions for the individual couplings, as can also be seen in Fig.~\ref{fig:beta_xi_reg}. 
  As an example of the differences between the background-scalar dependent computation ($\alpha_\text{reg} = \alpha_\text{GF}= \alpha_\text{ghost} = 1$) and a background-scalar independent one ($\alpha_\text{reg} = \alpha_\text{GF}=0$), the difference for $\beta_{m^2}$ and $\beta_\xi$ in the transverse traceless sector reads
 \bea
  \beta_{m^2}^\text{dep}-\beta_{m^2}^\text{indep} &=& \frac{5 g (1-4 \lambda) \xi }{2 \pi  (1-2 \lambda )^2}, \\
  \beta_{\xi}^\text{dep} - \beta_{\xi}^\text{indep} &=& - \frac{5 g \lambda  (3 - 2 \lambda) \xi }{3 \pi  (1-2 \lambda)^3}\label{eq:diffxi}
 .\eea
 The additional terms arise from the field dependence of the regulator as well as the gauge fixing term. 
 These changes under a relatively small change of an unphysical parameter do not necessarily signify much if they occur in non-observable quantities like beta functions.
 At the same time, critical exponents of a physical fixed point are universal, i.e., must be independent of such a technical choice. This statement holds, as long as they are calculated exactly; an approximation introduces a dependence on such unphysical parameters.
 In the following we use this to probe the fixed point \eqref{eqn:ngfp} and test the quality of the approximation. 

If we choose $\alpha_\text{GF} = \alpha_\text{reg}= 0$ (with arbitrary $\alpha_\text{ghost}$) and our gauge choice $\beta=0$, 
 the beta functions we obtain for the potential terms at $\xi=0$ agree with the beta functions reported in \cite{Eichhorn:2017als}.
 For this case, the critical exponents are given by
 \begin{align}
 \theta^{(1/2)}&=2.72\pm 1.70i, & \theta^{(3/4)} &= 0.63\pm 2.55i, \nonumber \\
 \theta^{(5)} &= -2.26.
\label{eq:critexpbsfdvanishing}
 \end{align}
  The first two relevant directions overlap with $\lambda$ and  $g$ and the third and fourth align roughly with $m^2$ and  $\xi$. 

 Setting $\alpha_\text{GF}=\alpha_\text{reg}=\alpha_\text{ghost}=1$ and choosing the gauge $\beta=1$, our $\beta$-function for the scalar potential agrees with the one found in \cite{Narain:2009fy}. If we evaluate the traces with heat-kernel methods, also the $\beta$-function for the non-minimal coupling agrees with that reference.  
 To compare to those results, let us choose $\alpha_\text{GF} = \alpha_\text{reg} = \alpha_\text{ghost} = 1$ and take into account the effect of $\beta_g$ and $\beta_{\xi}$ terms on the rhs, cf.~discussion below in Sec.~\ref{sec:anomalous_dimension}.  At the fixed point \eqref{eqn:ngfp} the critical exponents read
 \begin{align}
  \label{eqn:ngfp_crit_exps}
  \theta^{(1/2)} &= 2.36 \pm 2.81 i, & \theta^{(3/4)} = 0.36 \pm 2.81 i, \nonumber\\
   \theta^{(5)} &= -2.26. & &
 \end{align}
 As expected, the characteristic shift of the third and fourth critical exponent by $-2$ compared to the leading critical exponents, first observed in \cite{Narain:2009fy}, is present in our results as well. 
 The minor numerical differences with respect to that reference arise due to a different gauge choice.

We now study the scalar-background dependence by first varying $\alpha_\text{GF}$ and $\alpha_\text{ghost}$ continuously and then varying $\alpha_\text{reg}$.

  \begin{figure}
   \includegraphics[width=0.45\textwidth]{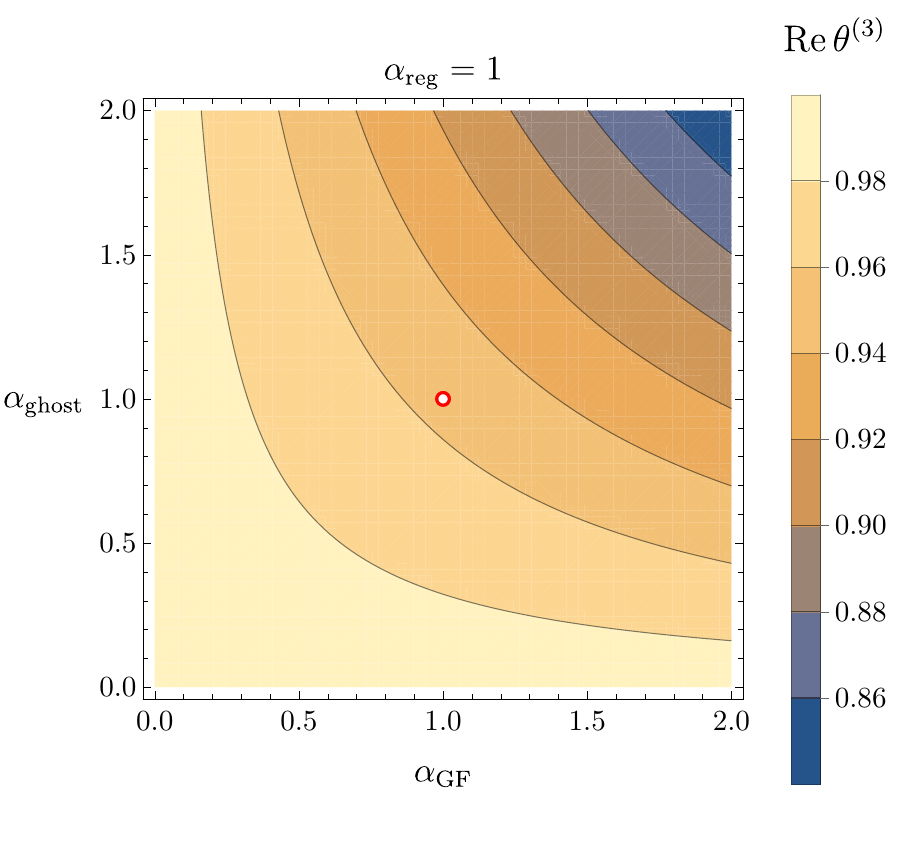}
   \caption{
   \label{fig:crit_exp_of_alpha_2}
   Real part of the critical exponent $\theta^{(3)}$ as a function of the parameters $\alpha_\text{GF}$ and $\alpha_\text{ghost}$.  
   The red dot marks the point corresponding to the computation with field dependent regulator and gauge-fixing term as in \cite{Narain:2009fy}. For $\alpha_\text{ghost} = 0$, the critical exponent becomes independent of $\alpha_\text{GF}$ as the contributions from the ghost term cancel those from the gauge fixing term.  The critical exponent is evaluated at the fixed point \eqref{eqn:ngfp}. For different values of the gravitational couplings as well as for $\alpha_\text{reg} = 0$, i.e., a background-scalar independent regulator, the dependence on $\alpha_\text{GF}$ and $\alpha_\text{ghost}$ is qualitatively similar.   }
 \end{figure}

 First, fixing $\alpha_\text{reg}=1$, we observe that as long as the background-scalar dependence of the gauge-fixing term is not fully accounted for by the ghost term ($\alpha_\text{ghost}\neq 0$), the critical exponents vary considerably under changes in $\alpha_\text{GF}$, cf.~Fig.~\ref{fig:crit_exp_of_alpha_2}. This can go as far as changing the number of relevant directions for $\alpha_\text{GF} = \alpha_\text{ghost}\approx 5.2$.   The choice $\alpha_\text{GF} = \alpha_\text{ghost} = 1$ used in \cite{Narain:2009fy} could be criticized based on the extra field-dependence it introduces, but does not yield qualitatively different results from the case without the additional field dependence. In fact, the two second-largest critical exponents are $\theta^{(3/4)} = 0.99 \pm 2.06 i$, once the scalar-background dependence is fully taken into account in the ghost action ($\alpha_\text{ghost}=0$). In this case, the critical exponents become independent of $\alpha_\text{GF}$, as they should.

 \begin{figure}
   \includegraphics[width=0.45\textwidth]{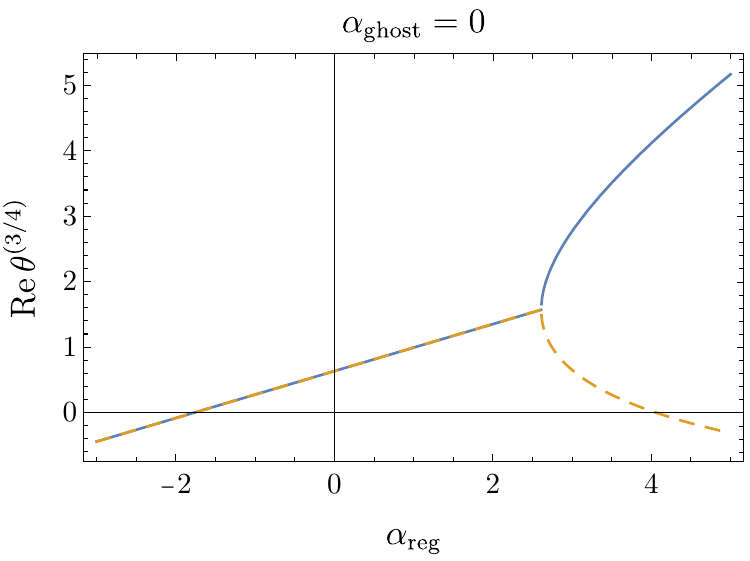}
   \caption{\label{fig:crit_exp_of_alpha_reg}
   Real part of the critical exponents $\theta^{(3/4)}$ as a function of $\alpha_\text{reg}$ at $\alpha_\text{ghost}=0$. The critical exponent is evaluated at the fixed point \eqref{eqn:ngfp}. For smaller $\lambda_\ast$, the curve flattens, while the value of $\alpha_{\rm reg}$ at which $\theta^{(3/4)}=0$, shifts to the left.}
 \end{figure} 	

\begin{figure}
   \includegraphics[width=0.45\textwidth]{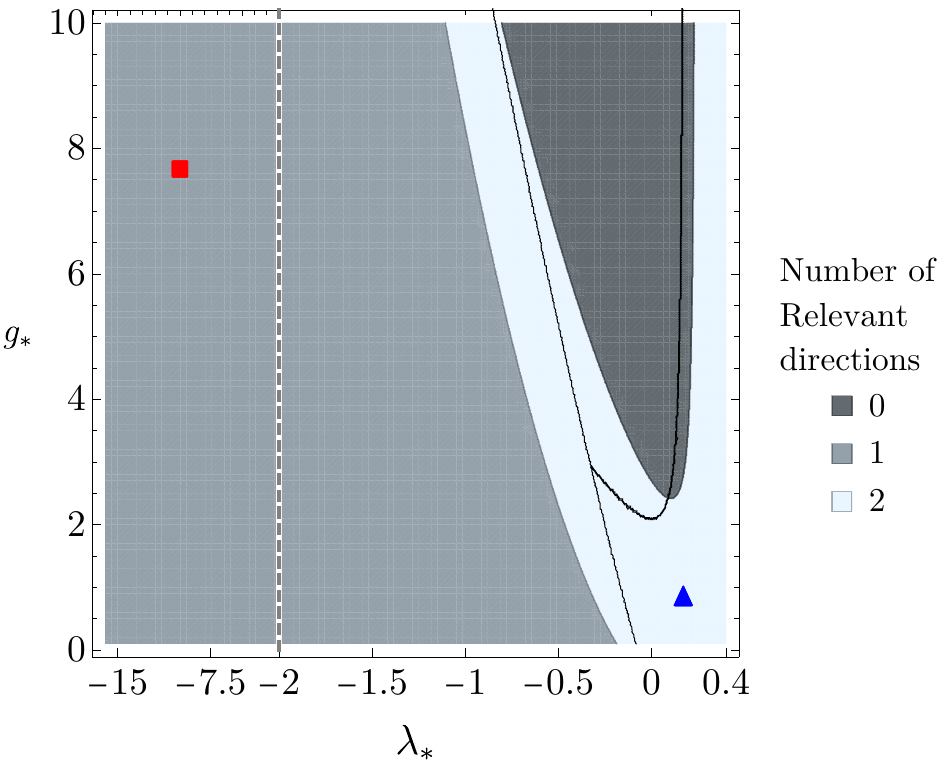}
   \includegraphics[width=\linewidth]{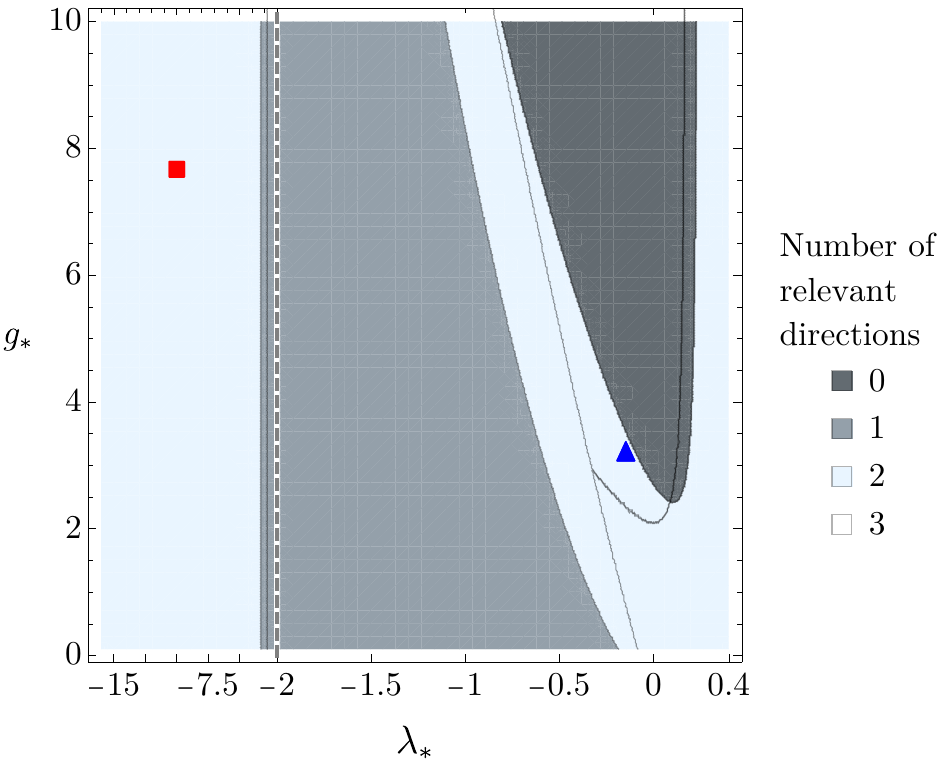}
   \caption{\label{fig:fv_nr_relv_dir_app}
   Number of relevant directions at the free fixed point as a function of the gravitational couplings $\lambda_\ast$ and $g_\ast$ for $\alpha_\text{reg}=\alpha_\text{GF} = 0$ in the purely scalar (Yukawa) system, for the three (four) scalar couplings $(m^2, \lambda_4, \xi(,y))$ in the top (bottom) panel. 
   The thin lines indicate the position of the corresponding areas using a scalar-background dependent regulator ($\alpha_\text{reg}=\alpha_\text{GF}=\alpha_\text{ghost}=1$)  and the background field approximation $\eta_h = -2$. Reduced background-field dependence reduces the sensitivity to a change in gravitational fixed-point values. The vertical dashed line highlights a change in the scaling of the $\lambda_\ast$ axis at $\lambda_\ast=-2$. For the Yukawa system the number of relevant directions changes. The blue triangle corresponds to the fixed point \eqref{eqn:ngfp}/\eqref{eqn:fp_grav_with_yukawa}, the red square to fixed point \eqref{eqn:fp_grav_canonical}.
   }
 \end{figure}

 Second, fixing $\alpha_\text{ghost} = 0$, such that the dependence on $\alpha_{\rm GF}$ vanishes, we vary $\alpha_\text{reg}$ in Fig.~\ref{fig:crit_exp_of_alpha_reg}. 
  For moderate values  of $|\alpha_\text{reg}| \leq 1.5$, the critical exponents depend quantitatively on $\alpha_\text{reg}$, with no value for $\alpha_\text{reg}$ preferred by the principle of minimum sensitivity. For large absolute values of $\alpha_\text{reg}$, the qualitative picture changes: For $\alpha_\text{reg}<-1.8$ and $\alpha_\text{reg}>4.1$, the number of relevant directions decreases. Similarly to the effect observed in \cite{Bridle:2013sra}, the single-field approximation for the scalar field can thus change the qualitative picture, and $\alpha_{\rm reg}=0$ appears preferred for this reason. For this choice one finds 
  the critical exponents in Eq.~\eqref{eq:critexpbsfdvanishing}.
  As long as the background-scalar dependence is not too strong, the number of relevant directions remains stable under technical modifications of the regulator and gauge-fixing term.  \\
 The number of relevant directions as a function of the gravitational fixed point parameters is shown in Fig.~\ref{fig:fv_nr_relv_dir_app}. Here we also display the number of relevant directions for the Yukawa system considered in Sec.~\ref{sec:yukawa_model}. In that system the number of relevant directions changes at the fixed point \eqref{eqn:fp_grav_with_yukawa} due to differing technical choices, elucidating the differing results in \cite{Oda:2015sma} and \cite{Hamada:2017rvn}. The qualitative change under a relatively minor technical choice exemplifies why we consider the full gravitational parameter space.

 \subsection{Momentum-dependent interactions}
 \label{sec:anomalous_dimension}
 \emph{...where we explore the size of the scalar anomalous dimension and its impact on the fixed-point results as a measure for the robustness of our truncation.}\\

  \begin{figure}
  \includegraphics[width=0.45\textwidth]{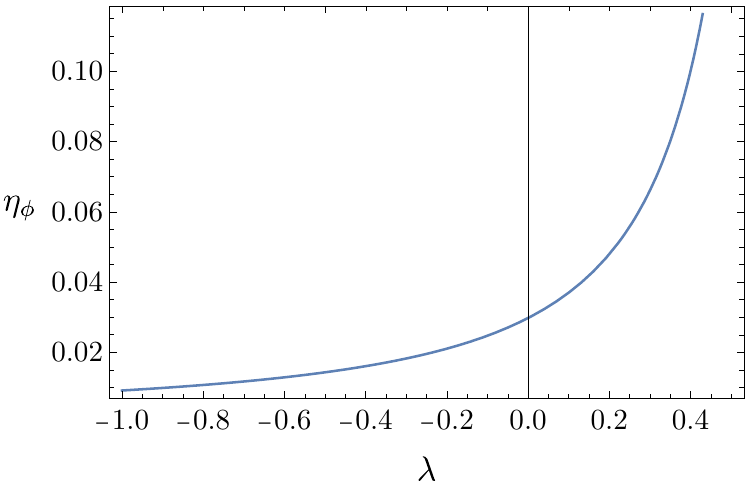}
  \includegraphics[width=0.45\textwidth]{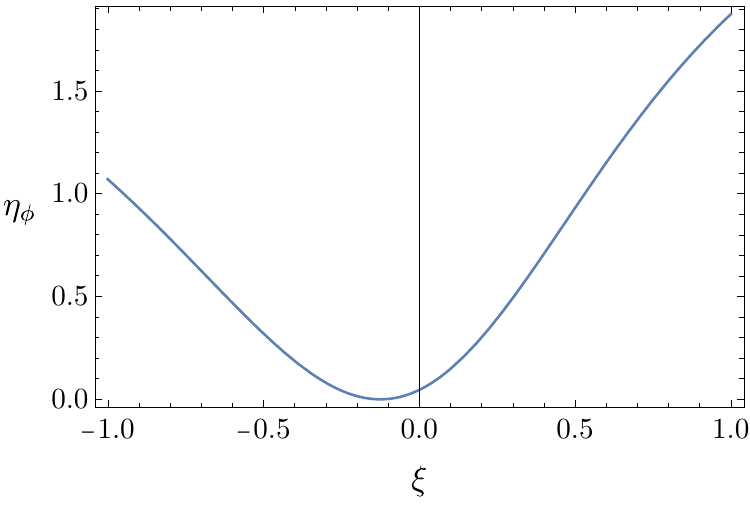}
  \caption{
    \label{fig:anomalous_dim}
    The anomalous dimension $\eta_\phi$ for a varying value of the cosmological constant $\lambda$ (upper panel) and non-minimal coupling $\xi$ (lower panel) around the fixed point $\lambda_*=0.171$ ($\xi_\ast=0$). All other couplings are held constant. The anomalous dimension  varies mildly with varying $\lambda$ and increases towards the pole at $\lambda=1/2$.
    It varies more strongly for varying $\xi$ and is slightly negative for a very small range of negative $\xi$.
  }
 \end{figure}
 As a test for the stability of the truncation \eqref{eqn:truncation}, we compute the scalar anomalous dimension $\eta_\phi = - \partial_t \ln Z_\phi(k^2)$ in the presence of a non-minimal coupling to explore how much this impacts the fixed-point results.
 We obtain 
\bea
  \eta_\phi &=& 
  g\frac{(8-\eta_h)(1+48(\xi-\frac{1}{2}m^2)) + 8\eta_h m^2}{16\pi(3-4\lambda)^2(1+m^2)} \nonumber \\
  &+& g\frac{(8-\eta_\phi)(1-24\xi+432\xi^2-144 \xi m^2)+48\eta_\phi\xi m^2}{48\pi(3-4\lambda)(1+m^2)^2} \nonumber \\
  &-& g\frac{108(\xi-\frac{1}{3}m^2)^2}{\pi (3-4\lambda)^2 (1+m^2)^2}
  .\eea
In the limit $\xi \rightarrow 0$ and  $m^2 \rightarrow 0$, this reduces to the result from \cite{Eichhorn:2017eht}, as it should.
\\
 Calculating the anomalous dimension with the fixed-point values \eqref{eqn:ngfp}, one obtains a small value of $\eta_\phi \approx 0.044$.
 In Fig.~\ref{fig:anomalous_dim} we explore how this result depends on the particular fixed-point values
 by studying the value of the anomalous dimension as a function of the non-minimal coupling and the cosmological constant treated as external parameters. 
 Away from the pole $\lambda = 1/2$ and for $|\xi|\lesssim 0.5$, the anomalous dimension remains small, tentatively indicating that momentum-dependent extensions of the truncation might be subleading. 
 For larger values of the non-minimal coupling, it appears indicated to extend the truncation to include higher-order momentum dependence.

 The shift due to the anomalous dimension modifies the fixed point \eqref{eqn:ngfp} at the sub-percent level. The critical exponents exhibit slightly larger modifications. The third/fourth critical exponents is shifted to $\theta^{(3/4)} = 0.58 \pm 2.55i$.
 In particular a scenario in which the anomalous dimension shifts these parameters into the irrelevant region is not realized.
 Given that the anomalous dimension is numerically small in a part of the parameter space, we will neglect it in some sections; but always indicate clearly whether we work in the approximation $\eta_{\phi}=0$.

 So far, in line with most of the existing literature, we employed the background-field approximation, in which the anomalous dimension for the metric fluctuation field $h_{\mu\nu}$ equals the anomalous dimension for the background Newton coupling, $\eta_h=\eta_N=-2$. In the main body of the paper, instead, we set $\eta_h=0$. Indeed, results in the literature indicate that this is a better approximation than $\eta_N=-2$ \cite{Codello:2013fpa, Dona:2013qba,Christiansen:2014raa,Dona:2015tnf,Christiansen:2015rva,Denz:2016qks,Eichhorn:2018ydy}. \\

 \section{Flow of the Potential}
 \label{app:flow_potential}
  The flow of the dimensionless potential in terms of the parameters \eqref{eqn:slow_roll_parameters} is given by
 \begin{widetext}
 \bea
  \partial_t v &=& \frac{1}{96 \pi^2 g (1-16\pi g v)\left(3(1+\epsilon_2)-32\pi g(v-2 \epsilon_1^2+ v \epsilon_2)\right)}\Big( 3g\left(19+16\epsilon_2-384\pi^2 v (1+\epsilon_2)\right)  \nonumber \\
  &+& 16\pi g^2 \left(48\epsilon_1^2 + 1920 \pi^2 v^2 (1+\epsilon_2) - 3 v(9 + 512\pi^2\epsilon_1^2+ 4\epsilon_2) \right)   \nonumber \\
  &+& 512 \pi^2 g^3 v \left( 16 \epsilon_1^2 - 384 \pi^2 v^2 (1+\epsilon_2) - v (5-768\pi^2 \epsilon_1^2 + 8 \epsilon_2)\right)
  \Big)
 .\eea
 \end{widetext}
 Taking derivatives of this equation with respect to the field $\phi$, one obtains the flow of the $\epsilon_i$.

  \section{Beta functions}
  \label{app:beta_functions}
  In the following we report the beta functions for the three couplings $m^2, \xi, \lambda_4$ in a scalar gravity theory with gauge choice $\beta=0$ and a background-scalar independent regulator. 
  \begin{widetext}
  \bea
    \beta_{m^2}
     &=& - (2 - \eta_\phi) m^2
    -\frac{36 g \xi ^2 \eta _{\phi }}{5 \pi  (3-4 \lambda) \left(1+m^2\right)^2}
    +\frac{4 g m^2 (12\xi  - \eta_\phi (m^2+6\xi))}{3 \pi  (3-4 \lambda) \left(1+m^2\right)^2}
    -\frac{3 g \xi  (6\xi - \eta_\phi(2m^2+3\xi))}{\pi  (3-4 \lambda) \left(1+m^2\right)^2} \nonumber \\
    &-&\frac{2 g m^4 \left(2-\eta _{\phi }\right)}{\pi  (3-4 \lambda) \left(1+m^2\right)^2}
    -\frac{108 g \xi ^2 \eta _h}{5 \pi  (3-4 \lambda )^2 \left(1+m^2\right)}
    -\frac{9 g \xi  (6\xi - \eta_h (2m^2+3\xi))}{\pi  (3-4 \lambda )^2 \left(1+m^2\right)}
    -\frac{5 g \left(\xi  \left(8-\eta _h\right)-m^2 \left(6-\eta _h\right)\right)}{12 \pi  (1-2 \lambda )^2}  \nonumber \\
    &-&\frac{g \left(3 \xi  \left(8-\eta _h\right)-2 m^2 \left(6-\eta _h\right)\right)}{4 \pi  (3-4 \lambda )^2}
    +\frac{4 g m^2 (12\xi - \eta_h (m^2+6\xi))}{\pi  (3-4 \lambda )^2 \left(1+m^2\right)}
    -\frac{6 g m^4 \left(2-\eta _h\right)}{\pi  (3-4 \lambda )^2 \left(1+m^2\right)}
    -\frac{\lambda _4 \left(6-\eta _{\phi }\right)}{64 \pi ^2 \left(1+m^2\right)^2},
  \eea
  \bea
    \beta_{\lambda_4} &=& 2 \lambda _4 \eta _{\phi }
    - \frac{64 g^2 \left(81 m^2 \xi ^2 \left(10-\eta _{\phi }\right)-45 m^4 \xi  \left(8-\eta _{\phi }\right)+10 m^6 \left(6-\eta _{\phi }\right)-54 \xi ^3 \left(12-\eta _{\phi }\right)\right)}{15 (3-4 \lambda )^2 \left(1+m^2\right)^2}  \nonumber \\
    &+&\frac{512 g^2 \left(-756 m^2 \xi ^3 \left(12-\eta _{\phi }\right)+567 m^4 \xi ^2 \left(10-\eta _{\phi }\right)-210 m^6 \xi  \left(8-\eta _{\phi }\right)+35 m^8 \left(6-\eta _{\phi }\right)+405 \xi ^4 \left(14-\eta _{\phi }\right)\right)}{105 (3-4 \lambda )^2 \left(1+m^2\right)^3}  \nonumber \\
    &+&\frac{8 g^2 \left(-10 m^2 \xi  \left(8-\eta _h\right)+5 m^4 \left(6-\eta _h\right)+6 \xi ^2 \left(10-\eta _h\right)\right)}{3 (1-2 \lambda)^3}
    -\frac{16 g^2 \left(-30 m^2 \xi  \left(\eta _h-8\right)+10 m^4 \left(\eta _h-6\right)+27 \xi ^2 \left(\eta _h-10\right)\right)}{5 (3-4 \lambda)^3}  \nonumber \\
    &+&\frac{128 g^2 \left(-81 m^2 \xi ^2 \left(10-\eta _h\right)+45 m^4 \xi  \left(8-\eta _h\right)-10 m^6 \left(6-\eta _h\right)+54 \xi ^3 \left(12-\eta _h\right)\right)}{5 (3-4 \lambda)^3 \left(1+m^2\right)}  \nonumber \\
    &+&\frac{512 g^2 \left(-756 m^2 \xi ^3 \left(12-\eta _h\right)+567 m^4 \xi ^2 \left(10-\eta _h\right)-210 m^6 \xi  \left(8-\eta _h\right)+35 m^8 \left(6-\eta _h\right)+405 \xi ^4 \left(14-\eta _h\right)\right)}{35 (3-4 \lambda)^3 \left(1+m^2\right)^2}  \nonumber \\
    &+&\frac{4 g \lambda _4 \left(-30 m^2 \xi  \left(8-\eta _{\phi }\right)+10 m^4 \left(6-\eta _{\phi }\right)+27 \xi ^2 \left(10-\eta _{\phi }\right)\right)}{5 \pi  (3-4 \lambda) \left(1+m^2\right)^3}
    +\frac{4 g \lambda _4 \left(3 \xi  \left(8-\eta _{\phi }\right)-2 m^2 \left(6-\eta _{\phi }\right)\right)}{3 \pi  (3-4 \lambda) \left(1+m^2\right)^2}  \nonumber \\
    &+&\frac{6 g \lambda _4 \left(-30 m^2 \xi  \left(8-\eta _h\right)+10 m^4 \left(6-\eta _h\right)+27 \xi ^2 \left(10-\eta _h\right)\right)}{5 \pi  (3-4 \lambda )^2 \left(1+m^2\right)^2}
    -\frac{4 g \lambda _4 \left(2 m^2 \left(6-\eta _h\right)-3 \xi  \left(8-\eta _h\right)\right)}{\pi  (3-4 \lambda )^2 \left(1+m^2\right)}
    +\frac{5 g \lambda _4 \left(6-\eta _h\right)}{12 \pi  (1-2 \lambda )^2}  \nonumber \\
    &+&\frac{g \lambda _4 \left(6-\eta _h\right)}{2 \pi  (3-4 \lambda )^2}
    +\frac{3 \lambda _4^2 \left(6-\eta _{\phi }\right)}{32 \pi ^2 \left(1+m^2\right)^3},
  \\
    \beta_\xi &=& \xi  \eta _{\phi }
    +\frac{2 g \left(18 \xi ^2 \left(2-\eta _h\right)+m^6 \eta _h-m^4 (12\xi - \eta_h(1+6\xi))-6 m^2 \xi (2 + 3 \xi (2-\eta_h)) \right)}{3 \pi  (3-4 \lambda )^2 \left(1+m^2\right)^2}   \nonumber \\
    &-&\frac{5 g \left(\xi (6+20\lambda-\eta_h(1+2\lambda))-m^2 \left(15-6\lambda + \eta _h (2 \lambda -3)\right)\right)}{36 \pi  (1-2 \lambda)^3}
    +\frac{g \left(12\xi - m^2 \left(6-\eta _h\right) \right)}{12 \pi  (3-4 \lambda )^2}   
    +\frac{9 g \xi ^2 \eta _{\phi } \left(5 m^2+32 \xi +5\right)}{20 \pi  (3-4 \lambda) \left(1+m^2\right)^3}   \nonumber \\
    &-&\frac{g \xi  \left(-3 \xi  (4 + 18\xi +\eta_h(1-9\xi))-m^2 (12\xi - \eta_h (4+15\xi+4m^2))\right)}{\pi  (3-4 \lambda )^2 \left(1+m^2\right)^2} 
    + \frac{27 g \xi ^2 \eta _h \left(5+5 m^2+16 \xi\right)}{20 \pi  (3-4 \lambda )^2 \left(1+m^2\right)^2}    \nonumber \\
    &+&\frac{g m^2 \left(2-\eta _h\right) \left(m^2(1+m^2)-6 \xi \right)}{\pi  (3-4 \lambda )^2 \left(1+m^2\right)^2}  
    -\frac{g m^2 \left(2-\eta _{\phi }\right) \left(m^4+m^2 (1+6 \xi)-6 \xi \right)}{3 \pi  (4 \lambda -3) \left(1+m^2\right)^3}    \nonumber \\
    &+& \frac{g \xi  \left(3 \xi  (4+36\xi+\eta_\phi(1-18\xi))+m^2 (12\xi - \eta_\phi (4+4m^2+33\xi))\right)}{3 \pi  (3-4 \lambda) \left(1+m^2\right)^3}    \nonumber \\
    &+&\frac{2 g \left(18 \xi ^2 \left(2-\eta _{\phi }\right)+m^6 \eta _{\phi }-m^4 \left(12\xi - \eta _{\phi } (1+12\xi)\right)-6 m^2 \xi  \left(2+9 \xi  \left(2-\eta _{\phi }\right)\right)\right)}{9 \pi  (3-4 \lambda) \left(1+m^2\right)^3}
    +\frac{\lambda _4 \left(6-\eta _{\phi }\right) \left(1+m^2+12 \xi\right)}{384 \pi ^2 \left(1+m^2\right)^3}.
  \eea
  The beta function for the Yukawa coupling is
  \bea
    \beta_y &=& y\, \eta _{\psi }+y \frac{\eta _{\phi }}{2} + y^3 \left( \frac{5-\eta _{\psi }}{80 \pi ^2 \left(1+m^2\right)}+\frac{6-\eta _{\phi }}{96 \pi ^2 \left(1+m^2\right)^2} \right) 
    \nonumber \\
      &-&\frac{3y\,g}{16} \left(\frac{16 \left(6-\eta _{\psi }\right)}{15 \pi  (3-4 \lambda)}+\frac{192 (7-\eta_h)}{35 \pi  (3-4 \lambda )^2}\right)
      +\frac{9y\,g}{256} \left(\frac{32 \left(7-\eta _{\psi }\right)}{21 \pi  (3-4 \lambda)}+\frac{4 (8-\eta_h)}{\pi  (3-4 \lambda )^2}\right)
      +\frac{5y\,g (6-\eta_h)}{12 \pi  (1-2 \lambda )^2}
      +\frac{y\,g(6-\eta_h)}{2 \pi  (3-4 \lambda )^2} \nonumber \\
&-&\frac{y\,g \left(45 \xi  \left(8-\eta _{\psi }\right)-28 m^2 \left(6-\eta _{\psi }\right)\right)}{140 \pi  (3-4 \lambda) \left(1+m^2\right)}
+\frac{y\,g \left(36 m^2 \left(7-\eta _{\phi }\right)-60 \xi  \left(9-\eta _{\phi }\right)\right)}{105 \pi  (3-4 \lambda) \left(1+m^2\right)^2}
+\frac{y\,g \left(3 \xi  \left(8-\eta _{\phi }\right)-2 m^2 \left(6-\eta _{\phi }\right)\right)}{3 \pi  (3-4 \lambda) \left(1+m^2\right)^2} \nonumber \\
&-&\frac{y\,g \left(2 (6-\eta_h) m^2-3 (8-\eta_h) \xi \right)}{\pi  (3-4 \lambda )^2 \left(1+m^2\right)}
+\frac{y\,g \left(36 (7-\eta_h) m^2-60 (9-\eta_h) \xi \right)}{35 \pi  (3-4 \lambda)^2 \left(1+m^2\right)},
  \eea
  \end{widetext}
  where parts of the Yukawa beta function is taken from \cite{Eichhorn:2017eht}.
  The portal system described in Sec.~\ref{sec:portal} consists of two copies of the these beta functions. Additionally, the dark sector contributes
  \begin{widetext}
  \bea
    \Delta \beta_{m_v^2} &=& - \frac{(6-\eta_{\phi_d}) \lambda_\text{HP}}{192\pi^2 (1+m_d^2)^2}, \\
    \Delta \beta_{\lambda_{4,v}} &=& \frac{(6-\eta_{\phi_d}) \lambda_\text{HP}^2}{96\pi^2 (1+m_d^2)^3}, \\
    \Delta \beta_{\xi_{v}} &=& \frac{(6-\eta_{\phi_d}) \lambda_\text{HP} (1 + m_d^2 + 12 \xi_d)}{1152 \pi^2 (1+m_d^2)^3},
  \eea
  to the beta functions for the visible couplings and vice versa. The beta function for the Higgs portal coupling reads
  \bea
    \beta_{\lambda_\text{HP}} &=&  \left(\eta _{\phi _d}+\eta _{\phi _v}\right) \lambda _{\text{HP}}  \nonumber \\
    &+& \frac{41472 g^2 \xi _d^2 \left(14-\eta _h\right) \xi _v^2}{7 (3-4 \lambda)^3}
     + \frac{16 g^2 \xi _d \left(10-\eta _h\right) \xi _v}{(1-2 \lambda)^3}
     + \frac{432 g^2 \xi _d \left(10-\eta _h\right) \xi _v}{5 (3-4 \lambda)^3}  \nonumber \\
    &+&\frac{6912 g^2 \xi _d^2 \xi _v^2 \left(28-\eta _{\phi _d}-\eta _{\phi _v}\right)}{7 (3-4 \lambda )^2}
     + \frac{36 g \xi _d \lambda _{\text{HP}} \xi _v \left(20-\eta _{\phi _d}-\eta _{\phi _v}\right)}{5 \pi  (3-4 \lambda)}
     + \frac{108 g \xi _d \left(10-\eta _h\right) \lambda _{\text{HP}} \xi _v}{5 \pi  (3-4 \lambda )^2}  \nonumber \\
    &+&\frac{5 g \left(6-\eta _h\right) \lambda _{\text{HP}}}{12 \pi  (1-2 \lambda )^2}
     + \frac{g \left(6-\eta _h\right) \lambda _{\text{HP}}}{2 \pi  (3-4 \lambda )^2}  \nonumber \\
    &+&\frac{3456 g^2 \xi _v^2 \xi _d \left(12-\eta _h\right)}{5 (3-4 \lambda)^3}
     + \frac{3456 g^2 \xi _v \xi _d^2 \left(12-\eta _h\right)}{5 (3-4 \lambda)^3}
     + \frac{576 g^2 \xi _d \xi _v^2 \left(12-\eta _{\phi _v}\right)}{5 (3-4 \lambda )^2}
     + \frac{576 g^2 \xi _d^2 \xi _v \left(12-\eta _{\phi _d}\right)}{5 (3-4 \lambda )^2}  \nonumber \\
    &+&\frac{6 g \xi _d \left(8-\eta _h\right) \lambda _{\text{HP}}}{\pi  (3-4 \lambda )^2}
     + \frac{6 g \xi _v \left(8-\eta _h\right) \lambda _{\text{HP}}}{\pi  (3-4 \lambda )^2}
     + \frac{27 g \xi _d^2 \left(10-\eta _h\right) \lambda _{\text{HP}}}{5 \pi  (3-4 \lambda )^2}
     + \frac{27 g \xi _v^2 \left(10-\eta _h\right) \lambda _{\text{HP}}}{5 \pi  (3-4 \lambda )^2}  \nonumber \\
    &+&\frac{18 g \xi _d^2 \lambda _{\text{HP}} \left(10-\eta _{\phi _d}\right)}{5 \pi  (3-4 \lambda)}
     + \frac{18 g \lambda _{\text{HP}} \xi _v^2 \left(10-\eta _{\phi _v}\right)}{5 \pi  (3-4 \lambda)}
     + \frac{2 g \xi _d \lambda _{\text{HP}} \left(8-\eta _{\phi _d}\right)}{\pi  (3-4 \lambda)}
     + \frac{2 g \xi _v \lambda _{\text{HP}} \left(8-\eta _{\phi _v}\right)}{\pi  (3-4 \lambda)}  \nonumber \\
    &+&\frac{\lambda _{\text{HP}}^2 \left(6-\eta _{\phi _v}\right)}{48 \pi ^2}
     + \frac{\lambda _{\text{HP}}^2 \left(6-\eta _{\phi _d}\right)}{48 \pi ^2}
     + \frac{\lambda _{\text{HP}} \lambda _{4 v} \left(6-\eta _{\phi _v}\right)}{32 \pi ^2}
     + \frac{\lambda _{4 d} \lambda _{\text{HP}} \left(6-\eta _{\phi _d}\right)}{32 \pi ^2}.
  \eea
  where in the last expression we set the masses to zero for better readability. 
    \end{widetext}
   \bibliography{references}
 \end{document}